\documentclass[10pt]{article}
\usepackage{epsfig}
\usepackage{axodraw}
\usepackage{epsfig}
\usepackage{graphicx}
\usepackage{rotate}
\usepackage{latexsym}
\usepackage{amssymb}
\newcommand\pubnumber{}
\newcommand\pubdate{\today}
\newcommand\hepnumber{hep-ph/0612123}

\def\csuma{Deutsches Electronen - Synchrotron, DESY, Platanenallee 6, 15738
Zeuthen, Germany}
\def\csumb{Dipartimento di Fisica Teorica, Universit\`a di Torino, Italy\\
INFN, Sezione di Torino, Italy}
\def\support{\footnote{Work supported by MIUR under contract
2001023713$\_$006 and by  the European Community's Marie-Curie Research 
Training Network under contract MRTN-CT-2006-035505
`Tools and Precision Calculations for Physics Discoveries at Colliders'.}}
\def\Title#1{\begin{center} {\Large\bf #1 } \end{center}}

\def\Author#1{\begin{center}{ \sc #1} \end{center}}
\def\Address#1{\begin{center}{ \it #1} \end{center}}

\newcommand\pubblock{\rightline{\begin{tabular}{l} \pubnumber\\
         \pubdate\\ \hepnumber \\ DESY 06-224 \\  SFB/CPP-06-55\end{tabular}}}
\newenvironment{Abstract}{\begin{quotation}  }{\end{quotation}}

\def\Acknowledgments{\bigskip  \bigskip \begin{center}
          \large\bf Acknowledgments\end{center}}
\def\email#1{\footnote{#1}}
\makeatletter
\def\section{\@startsection{section}{0}{\z@}{5.5ex plus .5ex minus
 1.5ex}{2.3ex plus .2ex}{\large\bf}}
\def\subsection{\@startsection{subsection}{1}{\z@}{3.5ex plus .5ex minus
 1.5ex}{1.3ex plus .2ex}{\normalsize\bf}}
\def\subsubsection{\@startsection{subsubsection}{2}{\z@}{-3.5ex plus
-1ex minus  -.2ex}{2.3ex plus .2ex}{\normalsize\sl}}

\renewcommand{\@makecaption}[2]{%
   \vskip 10pt
   \setbox\@tempboxa\hbox{\small #1: #2}
   \ifdim \wd\@tempboxa >\hsize     % IF longer than one line:
       \small #1: #2\par          %   THEN set as ordinary paragraph.
     \else                        %   ELSE  center.
       \hbox to\hsize{\hfil\box\@tempboxa\hfil}
   \fi}
 \def\citenum#1{{\def\@cite##1##2{##1}\cite{#1}}}
\def\citea#1{\@cite{#1}{}}
\newcount\@tempcntc
\def\@citex[#1]#2{\if@filesw\immediate\write\@auxout{\string\citation{#2}}\fi
  \@tempcnta\z@\@tempcntb\m@ne\def\@citea{}\@cite{\@for\@citeb:=#2\do
    {\@ifundefined
       {b@\@citeb}{\@citeo\@tempcntb\m@ne\@citea\def\@citea{,}{\bf }\@warning
       {Citation `\@citeb' on page \thepage \space undefined}}%
    {\setbox\z@\hbox{\global\@tempcntc0\csname b@\@citeb\endcsname\relax}%
     \ifnum\@tempcntc=\z@ \@citeo\@tempcntb\m@ne
       \@citea\def\@citea{,}\hbox{\csname b@\@citeb\endcsname}%
     \else
      \advance\@tempcntb\@ne
      \ifnum\@tempcntb=\@tempcntc
      \else\advance\@tempcntb\m@ne\@citeo
      \@tempcnta\@tempcntc\@tempcntb\@tempcntc\fi\fi}}\@citeo}{#1}}
\def\@citeo{\ifnum\@tempcnta>\@tempcntb\else\@citea\def\@citea{,}%
  \ifnum\@tempcnta=\@tempcntb\the\@tempcnta\else
  {\advance\@tempcnta\@ne\ifnum\@tempcnta=\@tempcntb \else\def\@citea{--}\fi
    \advance\@tempcnta\m@ne\the\@tempcnta\@citea\the\@tempcntb}\fi\fi}
\makeatother
\newcommand{\nl}{\nonumber\\}

\newcommand{\lpar}{\left(}                            % bracketing
\newcommand{\rpar}{\right)}

\newcommand{\bq}{\begin{equation}}                    % equationing
\newcommand{\eq}{\end{equation}}
\newcommand{\bqa}{\arraycolsep 0.14em\begin{eqnarray}}
\newcommand{\eqa}{\end{eqnarray}}
\newcommand{\ba}[1]{\begin{array}{#1}}
\newcommand{\ea}{\end{array}}
\newcommand{\ben}{\begin{enumerate}}
\newcommand{\een}{\end{enumerate}}
\newcommand{\bei}{\begin{itemize}}
\newcommand{\eei}{\end{itemize}}
\newcommand{\eqn}[1]{Eq.(\ref{#1})}
\newcommand{\eqns}[2]{Eqs.(\ref{#1})--(\ref{#2})}

\newcommand{\eqnsc}[2]{Eqs.(\ref{#1}) and (\ref{#2})}
\newcommand{\eqnst}[3]{Eqs.(\ref{#1}), (\ref{#2}) and (\ref{#3})}

\newcommand{\fig}[1]{Fig.~\ref{#1}}

\newcommand{\sect}[1]{Section~\ref{#1}}

\newcommand{\subsect}[1]{Subsection~\ref{#1}}

\newcommand{\appendx}[1]{Appendix~\ref{#1}}

\def\Re{\mathop{\operator@font Re}\nolimits}
\def\Im{\mathop{\operator@font Im}\nolimits}
\newcommand{\ord}[1]{{\cal O}\lpar#1\rpar}

\newcommand{\wb}{W}

\newcommand{\mw}{M_{_W}}

\newcommand{\mz}{M_{_Z}}

\newcommand{\mh}{M_{_H}}

\newcommand{\mt}{m_t}

\newcommand{\muq}{m_u}
\newcommand{\md}{m_d}

\newcommand{\mb}{m_b}
\newcommand{\mws}{M^2_{_W}}

\newcommand{\mhs}{M^2_{_H}}

\newcommand{\gf}{G_{\ssF}}

\newcommand{\stw}{s_{\theta}}             % bare, Lagrangian parameters
\newcommand{\ctw}{c_{\theta}}
\newcommand{\stws}{s_{\theta}^2}

\newcommand{\ctws}{c_{\theta}^2}

\newcommand{\spro}[2]{{#1}\cdot{#2}}
\newcommand{\li}[2]{\mathrm{Li}_{#1}\lpar\displaystyle{#2}\rpar} % polylog
\newcommand{\egam}[1]{\Gamma\lpar#1\rpar}               % Euler's Gamma
\newcommand{\intfx}[1]{\int_{\scriptstyle 0}^{\scriptstyle 1}\,d#1}

\newcommand{\MSB}{\overline{MS}}
\newcommand{\NMSB}{\overline{NMS}}

\newcommand{\ep}{\epsilon}
\newcommand{\Reb}{{\rm{Re}}}

\newcommand{\upar}[1]{u}

\newcommand{\ssA}{{\scriptscriptstyle{A}}}
\newcommand{\ssB}{{\scriptscriptstyle{B}}}
\newcommand{\ssC}{{\scriptscriptstyle{C}}}
\newcommand{\ssD}{{\scriptscriptstyle{D}}}
\newcommand{\ssE}{{\scriptscriptstyle{E}}}
\newcommand{\ssF}{{\scriptscriptstyle{F}}}
\newcommand{\ssG}{{\scriptscriptstyle{G}}}
\newcommand{\ssH}{{\scriptscriptstyle{H}}}
\newcommand{\ssI}{{\scriptscriptstyle{I}}}

\newcommand{\ssK}{{\scriptscriptstyle{K}}}
\newcommand{\ssL}{{\scriptscriptstyle{L}}}
\newcommand{\ssM}{{\scriptscriptstyle{M}}}
\newcommand{\ssN}{{\scriptscriptstyle{N}}}
\newcommand{\ssO}{{\scriptscriptstyle{O}}}

\newcommand{\ssQ}{{\scriptscriptstyle{Q}}}
\newcommand{\ssR}{{\scriptscriptstyle{R}}}
\newcommand{\ssS}{{\scriptscriptstyle{S}}}
\newcommand{\ssT}{{\scriptscriptstyle{T}}}
\newcommand{\ssU}{{\scriptscriptstyle{U}}}
\newcommand{\ssV}{{\scriptscriptstyle{V}}}
\newcommand{\ssW}{{\scriptscriptstyle{W}}}
\newcommand{\ssX}{{\scriptscriptstyle{X}}}
\newcommand{\ssY}{{\scriptscriptstyle{Y}}}
\newcommand{\ssZ}{{\scriptscriptstyle{Z}}}

\newcommand{\bqas}{\begin{eqnarray*}}
\newcommand{\eqas}{\end{eqnarray*}}
\def\app#1#2 {{\it Acta. Phys. Pol.} {\bf#1},#2}
\def\cpc#1#2 {{\it Computer Phys. Comm.} {\bf#1},#2}
\def\np#1#2 {{\it Nucl. Phys.} {\bf#1},#2}
\def\pl#1#2 {{\it Phys. Lett.} {\bf#1},#2}
\def\prep#1#2 {{\it Phys. Rep.} {\bf#1},#2}
\def\prev#1#2 {{\it Phys. Rev.} {\bf#1},#2}
\def\prl#1#2 {{\it Phys. Rev. Lett.} {\bf#1},#2}
\def\zp#1#2 {{\it Zeit. Phys.} {\bf#1},#2}
\def\sptp#1#2 {{\it Suppl. Prog. Theor. Phys.} {\bf#1},#2}
\def\mpl#1#2 {{\it Modern Phys. Lett.} {\bf#1},#2}
\def\jetp#1#2 {{\it Sov. Phys. JETP} {\bf#1},#2}
\def\fpj#1#2 {{\it Fortschr. Phys.} {\bf#1},#2}
\def\afp#1#2 {{\it Acta.Phys. Polon.} {\bf#1},#2}
\def\err#1#2 {{\it Erratum} {\bf#1},#2}
\def\ijmp#1#2 {{\it Int. J. Mod. Phys} {\bf#1},#2}
\def\nc#1#2 {{\it Nuovo Cimento} {\bf#1},#2}
\def\ap#1#2 {{\it Ann. Phys.} {\bf#1},#2}
\def\cmp#1#2 {{\it Comm. Math. Phys.} {\bf#1},#2}
\def\el#1#2 {{\it Europhys. Lett.} {\bf#1},#2}
\def\hpa#1#2 {{\it Helv. Phys. Acta} {\bf#1},#2}
\def\yf#1#2 {{\it Yad. Fiz.} {\bf#1},#2}
\def\nim#1#2 {{\it Nucl. Instrum. Meth.} {\bf#1},#2}
\def\spz#1#2 {{\it Sov. Pisma Zhetf} {\bf#1},#2}
\def\jetpl#1#2 {{\it JETP Lett.} {\bf#1},#2}
\def\sjnp#1#2 {{\it Sov. J. Nucl. Phys.} {\bf#1},#2}
\def\ptp#1#2 {{\it Progr. Theor. Phys. (Kyoto)} {\bf#1},#2}
\def\rmp#1#2  {{\it Rev. Mod. Phys.} {\bf#1},#2}
\def\zhetf#1#2 {{\it ZhETF} {\bf#1},#2}
\def\prs#1#2 {{\it Proc. Roy. Soc.} {\bf#1},#2}
\def\phys#1#2 {{\it Physica} {\bf#1},#2}

\def\bfi{\begin{figure}}
\def\efi{\end{figure}}

\newcommand{\LB}{{\cal L}oop{\cal B}ack}
\newcommand{\GS}{{\cal G}raph{\cal S}hot}
\newcommand{\bmid}{\Bigr|}
\newcommand{\DUV}{{\Delta}_{\ssU\ssV}}

\newcommand{\ext}{\,;\,{\rm ext}}

\newcommand{\uv}{\,;\,\ssU\ssV}
\newcommand{\Fin}{\,;\,\ssF}

\newcommand{\gpAZ}{\xi_{\ssA\ssZ}}
\newcommand{\gpA}{\xi_{\ssA}}
\newcommand{\gpZ}{\xi_{\ssZ}}
\newcommand{\gpW}{\xi_{\ssW}}
\newcommand{\Gred}{G_{\rm red}}
\newcommand{\Girr}{G_{\rm irr}}
\newcommand{\rre}{\,;\,{\rm red}}
\newcommand{\irr}{\,;\,{\rm irr}}
\newcommand{\bos}{\,;\,{\rm bos}}
\newcommand{\lep}{\,;\,{\rm lep}}
\newcommand{\fer}{\,;\,{\rm fer}}
\newcommand{\ren}{\,;\,{\rm ren}}
\newcommand{\ct}{\,;\,{\rm ct}}

\newcommand{\cL}{{\cal L}}

\begin{document}
\begin{titlepage}
\pubblock
\vfill
\def\thefootnote{\fnsymbol{footnote}}
\Title{Two-Loop Renormalization in the Standard Model\\[5mm]
Part II: Renormalization Procedures and\\[5mm]
Computational Techniques\support}
\vfill
\Author{Stefano Actis\email{Stefano.Actis@desy.de}}
\Address{\csuma}
\Author{Giampiero Passarino\email{giampiero@to.infn.it}}
\Address{\csumb}
\vfill
\begin{Abstract}
\noindent 
In part I general aspects of the renormalization of a spontaneously broken
gauge theory have been introduced. Here, in part II, two-loop renormalization 
is introduced and discussed within the context of the minimal Standard Model. 
Therefore, this paper deals with the transition between bare parameters and 
fields to renormalized ones. The full list of one- and two-loop counterterms 
is shown and it is proven that, by a suitable extension of the formalism 
already introduced at the one-loop level, two-point functions suffice in 
renormalizing the model. The problem of overlapping ultraviolet divergencies 
is analyzed and it is shown that all counterterms are local and of polynomial 
nature. The original program of 't Hooft and Veltman is at work.
Finite parts are written in a way that allows for a fast and reliable numerical
integration with all collinear logarithms extracted analytically. Finite
renormalization, the transition between renormalized parameters and physical
(pseudo-)observables, will be discussed in part III where numerical results,
e.g.\ for the complex poles of the unstable gauge bosons, will be shown.
An attempt will be made to define the running of the electromagnetic coupling 
constant at the two-loop level.
\end{Abstract}
\vfill
\begin{center}
\\[5mm]
\end{center}
\end{titlepage}
\def\thefootnote{\arabic{footnote}}
\setcounter{footnote}{0}
%--
\small
\thispagestyle{empty}
\tableofcontents
%\setcounter{page}{1}
%\normalsize
%--
\clearpage
\setcounter{page}{1}
\normalsize
%--
\section{Introduction}
\label{intro}
%--
After the end of the Lep era it became evident that including estimates
of higher-order radiative corrections into one-loop calculations for
physical (pseudo-)observables could not satisfy anymore the need for precision
required by the forthcoming generation of experiments. 
Since LHC will be an arena for discovery physics, high precision will not be 
mandatory during its first phase. However, according to some predestinate 
design, hadron machines are alternating with electron-positron ones and,
hopefully, ILC will come into operation: here the highest available 
theoretical precision will play a fundamental role.

As a matter of fact, it is not clear what kind of scenario will arise after 
the first few months of running at LHC. Any evidence of new physics will lead 
to a striking search for new theoretical models and their Born predictions; 
the quake could be so strong to remove any interest in quantum effects of the 
Standard Model. On the contrary, we could be back to the familiar landscape: 
effects of new physics hidden inside loops.

Since we have no firm opinion, we decided to follow the old rule 
\emph{si vis pacem para bellum}, building the environment that allows for 
a complete two-loop analysis of a spontaneously broken gauge field theory. 
This construction requires several elements, and it is difficult to 
characterize our strategy with a single acronym; although our work implies 
a lot of analytical aspects, the final step (computing arbitrary two-loop 
diagrams) can only be done with an \emph{algebraic-numerical} approach.  

If one thinks for a while, everything is in the old papers of 't Hooft and
Veltman~\cite{'tHooft:1972fi}, but translating few formal properties 
into a working scheme is far from trivial. Most of the times it is not a 
question of {\em How do I do it?}, rather it is a question of bookkeeping: 
{\em Can I do it without exhausting the memory of my computer?}, or, 
{\em Is there any practical way of presenting my results besides making my 
codes public?}.

We devoted a first paper~\cite{partI} (hereafter I) to deal with general 
aspects of a spontaneously broken gauge theory. First of all, we showed how 
to treat tadpoles; although everybody knows how to do it, general results are 
rarely presented in a way that everyone can use them.
In addition, we analyzed how to perform an order-by-order diagonalization 
of the neutral sector of a theory of fundamental interactions; once again, 
one needs a comprehensive collection of results which allows for practical 
applications. 

Alternative solutions to solving these problems exist, noticeably in
the background-field formalism~\cite{Denner:1994xt} (compare also with
Ref.~\cite{Jegerlehner:1985ch}); our claim here is restricted to the 
construction of a set of procedures which do not rely on other sources and 
cover broadly all the aspects, from generation of diagrams and
renormalization to evaluation (mostly numerical) of physical processes.
In particular, our results show that the structure of the counterterms
at the two-loop level, as well as of the whole set of renormalized Green
functions and the Ward-Slavnov-Taylor identities~\cite{Taylor:ff}, 
has (in the {\em conventional} approach) a degree of simplicity comparable
to the one obtained at one loop in the background-field approach.

Another perennial question is: what about renormalization, with or without 
counterterms? 
In a way, it is a fake question, since the two approaches are fully equivalent 
as far as $S$-matrix elements are concerned. In this paper we focus
on the transition from bare parameters to renormalized ones, 
and in a third paper (hereafter III) we will then discuss the ultimate step 
in any renormalization procedure: the transition from renormalized parameters 
to a set of physical (pseudo-)observables.
Perhaps, one should try to make a clear vocabulary of renormalization in 
quantum field theory.
A renormalization procedure is designed to bring you from a Lagrangian to
theoretical predictions; it includes regularization (nowadays dimensional
regularization~\cite{'tHooft:1972fi} is easy to understand), a renormalization
scheme and the choice of an input parameter set. The scheme, being a 
transitory step, is almost irrelevant; it can be on-mass-shell or $\MSB$ or 
based on complex poles, but unless you do something illegal (resummations 
that are not allowed or similar things) it really does not matter. 
Admittedly, one can define $\MSB$ quantities as convenient tools, but it is 
the last step that matters, at least as long as we have a convenient 
subtraction point (which we miss in QCD). 

Renormalized quantities should always be expressed in terms of a set of 
physical quantities. One may indulge to the introduction of an $\MSB$ running 
electromagnetic coupling constant (importing a concept from QCD to QED, 
which sounds strange anyway) but, at the end of the day, only cross sections 
matter.

In this paper, we have done one thing: all the two-loop Green's functions 
of the theory are made finite by introducing non-logarithmic counterterms and 
respecting unitarity. 
In addition, one can easily check that renormalized Ward-Slavnov-Taylor 
identities are satisfied. Actually, we have done more, since all 
ultraviolet-finite parts have been classified and an algorithm has been 
designed for their evaluation at any scale. What is innovative in our 
approach, as well as in other modern approaches~\cite{Freitas:2002ja}, is the 
idea that everything can be generated (is generated) by a set of automatized 
procedures which deals satisfactorily with the somewhat greater complexity of 
a two-loop calculation. 

Furthermore, classification of ultraviolet divergencies is dynamically linked 
to a well-defined computational scheme. In other words, in our approach, the 
ultraviolet-finite parts are back in their privileged position where they can 
play the role of creating the predictive power of the theory. There are, of 
course, preliminary steps (not always the easy ones), but it is only the 
full control on the multi-scale level that pays off.

The outline of this paper is as follows. After introducing our notation and 
conventions in~\sect{conve}, we outline the strategy of our calculation 
in~\sect{strategy}, where we classify all the ultraviolet-divergent parts of 
the needed loop integrals.
Next, in~\sect{sec:renormalization}, we choose the gauge-fixing Lagrangian 
and we define our renormalization scheme. In~\sect{olR} we review one-loop 
renormalization, and in~\sect{sec:one:beta2} we show explicit results for 
tadpole renormalization and neutral-sector diagonalization at two loops.
Finally, we analyze the ultraviolet structure of two-loop self-energies 
in~\sect{sec:one:twoloop} and we show explicit results for the two-loop 
counterterms in~\sect{sec:one:ct2}. \sect{conclu} contains the summary, and
several technical details connected with the relevant kinematical limits,
which represent the backbone of our computational techniques, are discussed 
in the appendices.
%--
\section{Notation and conventions}
\label{conve}
%--

\noindent
\underline{\emph{Regularization.}} We employ dimensional 
regularization~\cite{'tHooft:1972fi}, 
denoting the number of the space-time dimensions by $n= 4-\ep$.  
In addition, we use a short-hand notation for regularization-dependent factors,
%--
\bq
\DUV\, =\, \gamma\, +\, \ln\pi\, +\, \ln \frac{M^2}{\mu^2},
\qquad\qquad
\DUV(x)\,=\, \DUV\, -\, \ln\frac{M^2}{x},
\label{defMSB}
\eq
%--
where $\gamma= 0.5772156\cdots$ is the Euler constant, $\mu$ is the 't Hooft 
unit of mass, $M$ stands for the bare (renormalized) $W$-boson mass
(we do not distinguish unless strictly needed) and $x$ is a positive-definite 
kinematical variable. In our conventions the logarithm has a cut along the 
negative real axis and it is understood that for all masses:
$M^2\, \to \, M^2 -\, i\,\delta$, with $\delta\, \to\, 0_+$.
\vspace{0.1cm}

\noindent
\underline{\emph{Masses.}} We introduce a compact notation for ratios of 
squared masses,
%--
\bq
x_\ssH\, =\, \frac{\mhs}{M^2},\qquad
x_{l,i}\,=\, \frac{m_{l,i}^2}{M^2},\qquad
x_{u,i}\,=\, \frac{m_{u,i}^2}{M^2},\qquad
x_{d,i}\,=\, \frac{m_{d,i}^2}{M^2}.
\eq
%--
Here $\mh$ is the Higgs-boson bare (renormalized) mass and  
$m_{l,i}$, $m_{u,i}$ and $m_{d,i}$ are the bare (renormalized) masses of the 
charged lepton and the up and down quarks of the $i$th fermion doublet, with 
$i=1,\ldots,3$.
We consider the minimal representation for the Standard-Model scalar sector,
defining $M_0=M\slash c_\theta$ for the $Z$-boson bare (renormalized) 
mass. Here $c_\theta$ ($s_\theta$) is the bare or renormalized cosine (sine) 
of the weak-mixing angle $\theta$.
For notational clarity we frequently employ the notation
%--
\bq
\{m\}_{12\ldots N}=m_1,m_2,\ldots,m_N.
\eq
%--
\section{Outline of the calculation}
\label{strategy}
%--
Our calculation builds upon an automatic strategy for generating Feynman 
diagrams and evaluating the necessary one- and two-loop integrals. 
In addition, it does not rely on any black-box tool: diagrams are generated 
through a set of FORM~\cite{Vermaseren:2000nd} routines implemented in the 
${\cal G}raph{\cal S}hot$ package~\cite{GraphShot} and loop integrals are 
computed through the FORTRAN/95 ${\cal L}oop{\cal B}ack$~\cite{LoopBack} code.

The present work uses a set of results derived in a series of previous 
papers. The general strategy for handling multi-loop multi-leg Feynman 
diagrams was designed in Ref.~\cite{Passarino:2001wv} and the whole body of 
results necessary for evaluating two-loop two-point integrals can be found
in Ref.~\cite{Passarino:2001jd}.
The calculation of two-loop three-point scalar integrals is considerably more
involved: infrared-convergent configurations are discussed 
in Ref.~\cite{Ferroglia:2003yj} and infrared- and collinear-divergent ones are 
analyzed in Ref.~\cite{Passarino:2006gv}.
Finally, our method for dealing with two-loop tensor integrals can be found 
in Ref.~\cite{Actis:2004bp} and results for one-loop multi-leg integrals are 
shown in Ref.~\cite{Ferroglia:2002mz}.

It is worth noting that in any spontaneously-broken gauge theory the large 
number of masses and external kinematical variables leads to difficulties 
for the familiar analytical methods. Therefore, we use an alternative 
approach (mostly numerical) which, however, does not imply a blind rejecting 
of analytical techniques; for instance, in this work, we devote a special 
attention to the analytical extraction of leading and sub-leading collinear 
logarithms, since they play a dominant role in any 
calculation~\cite{Denner:2006jr}.

If we focus on renormalization, we can safely state that all the necessary
ingredients are available. Here the crucial point is to connect a set of 
input experimental data (an input-parameter set, hereafter IPS) to the free 
parameters of the theory:
%--
\begin{itemize}

\item[--] mass renormalization involves the calculation of 
self-energies~\cite{Passarino:2001wv,Passarino:2001jd};

\item[--] renormalization of coupling constants requires additional elements, 
which depend on the choice of the (pseudo-)observables (hereafter POs) in 
the IPS.

\end{itemize}
%--
The most-obvious selection of an IPS is based on the choice of those data 
which are known with the best experimental precision: the 
electromagnetic coupling constant, $\alpha$, and the Fermi coupling constant, 
$\gf$.
>From the diagrammatic point of view, the whole set of renormalization
equations (including finite parts) requires the evaluation of two-loop 
vertex and box integrals, since $\alpha$ is defined by means of the 
Thomson-scattering amplitude and $\gf$ is extracted from the muon lifetime.
However, it is possible to show that:
%--
\begin{itemize}
\item[--] neglecting $\ord{m^2_{\mu}/M^2}$ terms, where $m_\mu$ is the
muon mass, the evaluation of the muon lifetime can be carried on through 
tadpole diagrams (zero external momenta);

\item[--] the relevant Ward-Slavnov-Taylor (hereafter WST) identities of the 
Standard Model imply a set of algebraic cancellations involving vertex 
diagrams and wave-function renormalization factors for the external legs. 
Here the role played by the parameter $\Gamma$ introduced in I (all-order
diagonalization of the neutral sector) plays an essential role.
As a result, one can write an electric-charge renormalization equation which 
depends only on the photon vacuum-polarization function, as in QED.
%--
\end{itemize}
%--
We defer a complete discussion of this computation to a forthcoming paper 
(hereafter III); for a short review, see Ref.~\cite{Actis:2006xf}.
For the present purposes, it is enough to say that the 
results collected in Refs.~\cite{Passarino:2001wv,Passarino:2001jd} allow to 
complete the whole renormalization program. Of course, properties of two-loop 
vertex diagrams have to be checked, and here we use the results 
of~Refs.\cite{Ferroglia:2003yj,Passarino:2006gv,Actis:2004bp}.

An important remark concerns tensor loop integrals, namely those configurations
with a non-trivial spin structure for the numerators.
On the one hand, they can be reduced to scalar integrals if one uses the 
standard methods of Ref.~\cite{Passarino:1978jh}, supplemented by the sub-loop 
techniques of Ref.~\cite{Weiglein:1993hd}. 
It is worth noting that a complete scalarization for two-loop vertex diagrams
requires the introduction of generalized scalar integrals, to be evaluated in 
shifted space-time dimensions (the corresponding propagators are raised to 
non-canonical powers).
On the other hand, the number of terms generated by the reduction procedure 
increases rapidly in any realistic computation.

In order to carry out a systematic check of our computation, we have used
both approaches: a reduction procedure followed by an extraction of the 
ultraviolet (hereafter UV) resides of the poles for the scalar configurations
and a direct computation of the UV divergencies for the tensor cases.
Reduction techniques can introduce negative powers of Gram determinants,
which can be dangerous for the numerical stability of the results.
The approach we take in the present paper avoids these problems since 
scalarization is only employed to prove that the relevant algebraic 
properties are satisfied.

One of the main goals of this paper is to show that UV divergencies at two 
loops can be removed through suitable polynomial subtraction terms.
Therefore, in this section we collect the results for the residues of the 
UV poles of all the relevant one- and two-loop integrals employed
in our calculation.
However, we stress that only a full control over the (less-trivial) UV-finite 
parts will finally allow for theoretical predictions. 
In \subsect{ss:uvdeco} we introduce general UV decompositions
for one- and two-loop integrals. In \subsect{oneloopFUNC} we present 
the relevant results for the one-loop ingredients.
In Subsections~\ref{ss:1}, \ref{ss:2} and \ref{ss:3} we show the necessary 
results for one-, two- and three-point two-loop integrals.
%--
\subsection{Ultraviolet decompositions}
\label{ss:uvdeco}
%--

\noindent
\underline{\emph{One-loop integrals.}}
Any one-loop integral $f^1$ can be decomposed 
as~\footnote{The UV decompositions of \eqns{dec1loop}{UVfactors} and the
corresponding classification of UV poles are due to S.~Uccirati.}
%--
\bq
\label{dec1loop}
f^1\lpar \,\{l\}\,\rpar = \sum_{k=-1}^{1}\,
f^1\lpar \,\{l\}\,;\,k\,\rpar \, F^1_k(x),
\eq
%--
where $\{l\}$ denotes a given set of arguments: powers of the inverse 
propagators, external kinematical variables, masses of internal particles.
$x$ is some kinematical variable (usually $M$ for tadpoles or a squared 
external momentum) and the dependence on the dimensional regulator $\ep$ and
the regularization-dependent factors introduced in \eqn{defMSB} 
is entirely transferred to the \emph{universal} UV factors,
%--
\bq
F^1_{-1}(x) = \frac{1}{\ep} - \frac{1}{2}\,\DUV(x) + 
\frac{1}{8}\,\DUV^2(x)\,\ep, \qquad
F^1_0(x) = 1 - \frac{1}{2}\,\DUV(x)\,\ep, \qquad
F^1_1(x) = \ep.
\label{OLUVF}
\eq
%--
It is worth noting that, because of overlapping divergencies (UV-divergent 
one-loop sub-diagrams), we include $\ord{\ep}$ terms in all one-loop results.
%--
\vspace{0.2cm}

\noindent
\underline{\emph{Two-loop integrals.}}
A generic two-loop integral $f^2$ can be written as
%--
\bq
f^2\lpar \,\{l\}\,\rpar = 
\sum_{k=-2}^{0}\,f^2\lpar \,\{l\}\,;\,k\rpar\,F^2_k(x).
\label{TLUVD}
\eq
%--
Here the two-loop UV factors read as follows:
%--
\bq
F^2_{-2}(x) = \frac{1}{\ep^2} - \frac{\DUV(x)}{\ep} + \frac{1}{2}\,\DUV^2(x),
\qquad
F^2_{-1}(x) = \frac{1}{\ep} - \DUV(x),
\qquad
F^2_0(x) = 1.
\label{UVfactors}
\eq
%--
Note that the product of two one-loop integrals can be written through the 
same UV decomposition of a two-loop integral.
%--
%\bqa
%f^1_a(\{a\})\,f^1_b(\{b\}) &=& 
%\sum_{k=-2}^{0}\,f^2_{ab}(\{a\},\{b\};\,k)\,F^2_k(x),
%\nl
%f^2_{ab}(\{a\},\{b\};k_1+k_2) &=& 
%\sum_{k_1,k_2=-1}^1 f^1_a(\{a\};\,k_1)\,f^1_b(\{b\};\,k_2).
%\eqa
%--
Finally, in the following we collect two overall factors for one- and 
two-loop integrals,
%--
\bq
\mu_1(\ep)\, =\, \frac{\mu^{\ep}}{i\,\pi^2}, \qquad \qquad
\mu_2(\ep)\, =\, \frac{\mu^{2\ep}}{\pi^4} = -\,\mu_1^2(\ep).
\eq
%--
%--
\subsection{One-loop integrals \label{DefOLF}}
\label{oneloopFUNC}
%--
\noindent
\textbf{Tadpoles}
\vspace{0.2cm}
%--

\noindent
A scalar one-loop tadpole, with the inverse propagator raised to an integer 
power $\alpha$, is defined by
%--
\bq
A_0(\alpha,m) \, =\, 
\mu_1(\ep)\,\int\,\frac{d^nq}{(q^2 + m^2)^\alpha}.
\eq
%--
Integration-by-parts (IBP) identities~\cite{Tkachov:1981wb} allow to reduce 
any configuration with $\alpha>1$ to the case $\alpha=1$ by means of 
a recursive application of the identity
%--
\bq
A_0(\alpha,m) \, =\,  -\,\frac{1}{\alpha - 1}\,
\left( \frac{n}{2} - \alpha + 1\right)\,\frac{A_0(\alpha-1,m)}{m^2}.
\label{defgAfun}
\eq
%--
We factorize everywhere the scale dependence in order to deal with 
dimensionless loop integrals. Here we introduce the function $a_0(\alpha,m)$,
%--
\bq
A_0(\alpha,m) \, =\,  
 (m^2)^{2-\alpha} \,a_0(\alpha,m),
\eq
%--
and apply the UV decomposition defined in \eqn{dec1loop} to the case 
$a_0(m)=a_0(1,m)$,
%--
\bq
a_0(m) = \sum_{k=-1}^{1}\,a_0(m\,;\,k)\,F^1_k(M^2).
\label{cUV}
\eq
%--
The coefficients of the UV factors in \eqn{cUV} read as
%--
\bqa
a_0(m\,;\,-1) &=& -2, \qquad
a_0(m\,;\,0) = - 1 + \ln\frac{m^2}{M^2},
\nl
a_0(m\,;\,1) \, &=& \frac{1}{2}\,\Bigl[ - 1 - \frac{1}{2}\,\zeta(2)
          + \,\Bigl( 1 - 
         \frac{1}{2}\,\ln \frac{m^2}{M^2}\Bigr)\, \ln \frac{m^2}{M^2} \Bigr],
\label{defAfun}
\eqa
%--
where $\zeta(z)$ is the Riemann zeta function.
%--
\vspace{0.2cm}

\noindent
\textbf{Two-point integrals}
\vspace{0.2cm}

\noindent
\underline{\emph{Ultraviolet decompositions.}}
For two-point integrals we introduce $s= -p^2$, where $p$ is the external 
momentum. We extract form factors for tensor configurations and introduce 
scaled quantities according to the following definitions:
%--
\bqa
\label{1loop2p}
B_{\mu\nu}(\alpha,\beta,s,\{m\}_{12}) &=& \mu_1(\ep)\,\int\,
d^n q\,\frac{q_{\mu}\,q_{\nu}}
{(q^2 + m^2_1)^{\alpha}\,[(q+p)^2 + m^2_2]^{\beta}} 
\nl
{}&=&
s^{2-\alpha-\beta}\,b_{21}(\alpha,\beta,s,\{m\}_{12})\,p_{\mu}\,p_{\nu} +
s^{3-\alpha-\beta}\,b_{22}(\alpha,\beta,s,\{m\}_{12})\,\delta_{\mu\nu},
\nl
%--
B_{\mu}(\alpha,\beta,s,\{m\}_{12}) &=& \mu_1(\ep)\,\int\,
d^nq\,\frac{q_{\mu}}{(q^2 + m^2_1)^{\alpha}\,[(q+p)^2 + m^2_2]^{\beta}} 
= s^{2-\alpha-\beta}\,b_1(\alpha,\beta,s,\{m\}_{12})\,p_{\mu},
\nl
B_0(\alpha,\beta,s,\{m\}_{12}) \,&=&\, \mu_1(\ep)\,\int\,
\frac{d^nq}{(q^2 + m^2_1)^{\alpha}\,[(q+p)^2 + m^2_2]^{\beta}}\,  =
\, s^{2-\alpha-\beta}\, b_0(\alpha,\beta,s,\{m\}_{12}).
\eqa
%--
Note that scaled one-loop form factors will often appear in the residues
of the UV poles for two-loop integrals. Reduction of higher-rank form factors 
to the scalar case can be obtained by means of well-established techniques.
UV decompositions read as
%--
\bq
b_i(\alpha,\beta,s,\{m\}_{12}) = 
\sum_{k=-1}^{1}\,b_i(\alpha,\beta,s,\{m\}_{12};\,k)\,F^1_k(s),
\eq
%--
where $i=0,1,21,22$. The coefficients of the UV factors for the canonical 
scalar case ($\alpha=\beta=1$) read as
%--
\bqa
b_0(1,1,s,\{m\}_{12};\,-1) &=& 2, \qquad
b_0(1,1,s,\{m\}_{12};\,0) = b_0^{\rm fin}(1,1,s,\{m\}_{12}),
\nl
b_0(1,1,s,\{m\}_{12};\,1) &=& 
 \frac{1}{4}\,\Bigl[ \zeta(2) - b_0^{\ep}(1,1,s,\{m\}_{12})\Bigr].
\label{B0oo}
\eqa
%--
We will need to consider additional cases, where one propagator is raised to a 
non-canonical power, i.e. $\alpha=2, \beta=1$. Here results are given by 
%--
\bqa
b_0(2,1,s,\{m\}_{12};\,-1) &=& 0, \qquad
b_0(2,1,s,\{m\}_{12};\,0) = b_0^{\rm fin}(2,1,s,\{m\}_{12}),
\nl
b_0(2,1,s,\{m\}_{12};\,1) &=& -\, \frac{1}{4}\,b_0^{\ep}(2,1,s,\{m\}_{12}).
\label{B0to}
\eqa
%--
Note that \eqn{B0to} holds strictly only when $m_1\neq 0$, otherwise
an infrared (IR) divergency arises and we need a different decomposition.
The finite parts and the $\ord{\ep}$ terms introduced in
\eqnsc{B0oo}{B0to} have the following integral representations:
%--
\bqa
b^{\rm fin}_0(1,1,s,\{m\}_{12}) &=& -\,\intfx{x}\,\ln\frac{\chi(x)}{s},
\qquad
b^{\rm fin}_0(2,1,s,\{m\}_{12}) = \intfx{x}\,\frac{(1-x)\,s}{\chi(x)}, 
\nl
b^{\ep}_0(1,1,s,\{m\}_{12}) &=& -\,\intfx{x}\,\ln^2\frac{\chi(x)}{s},
\qquad
b^{\ep}_0(2,1,s,\{m\}_{12}) = 2\,\intfx{x}\,
\frac{(1-x)\,s}{\chi(x)}\,\ln\frac{\chi(x)}{s},
\label{defb0ep}
\eqa
%--
where the quadratic form $\chi$ reads as
%--
\bq
\chi(x) = s\,x^2\, -\, (\,s\, +\, m^2_1\, -\, m^2_2\,)\,x \,
+\, m^2_1\, -\, i\,\delta.
\eq
%--
\noindent
\underline{\emph{Finite components.}}
%--
For the finite parts we compute explicitly the integrals introduced in 
\eqn{defb0ep}. For the case $\alpha= \beta= 1$ we obtain
%--
\bqa
b_0^{\rm fin}(1,1,s,0,0) &=& 2 - L_{\rm an},
\nl
b^{\rm fin}_0(1,1,s,0,m) &=& %- \DUV(s)
-
\ln\frac{m^2}{s} + 2 - \lpar 1 - \frac{m^2}{s} \rpar\,
\ln \lpar 1 - \frac{s}{m^2}\rpar,
\nl
b^{\rm fin}_0(1,1,s,m,m) &=& %- \DUV(s)
-
\ln\frac{m^2}{s} + 2 -\beta(\frac{m^2}{s})\,
\ln\frac{\beta(m^2/s)+1}{\beta(m^2/s)-1},
\nl
b_0^{\rm fin}(1,1,s,m_1,m_2) &=& 
-\frac{1}{2}\,\Bigl[ \ln\frac{m^2_1}{s} + \ln \frac{m^2_2}{s}
+ \frac{m_1^2-m_2^2}{s}\,
\Bigl(\ln\frac{m^2_1}{s} - \ln \frac{m^2_2}{s}\Bigr)\Bigr]
\nl
{}&+& 2 - \frac{1}{s}\,\lambda^{1/2}(s,m_1^2,m_2^2)\,L(s,m_1^2,m_2^2),
\eqa
%--
where $\lambda(x,y,z)$ is the usual K\"allen function and we introduced
%--
\bqa
\beta^2(\frac{m^2}{s}) &=& 1 - 4\,\frac{m^2}{s\, +\, i\, \delta}, 
\nl
L(s,m^2_1,m^2_2)&=& \ln\frac
{- s + m^2_1 + m^2_2 - \lambda^{1/2}(s,m^2_1,m^2_2)}
{2\,m_1\,m_2},
\nl
L_{\rm an} &=& 
\ln ( - 1 - i\,\delta), \qquad \delta \to 0_+.
\label{deflambda}
\eqa
For the regular ($m_1\neq 0$) case, where $\alpha=2$ and $\beta=1$, we obtain
%--
\bqa
b^{\rm fin}_0(2,1,s,m,0) &=&
- \ln \lpar 1-\frac{s}{m^2}\rpar,
\nl
b^{\rm fin}_0(2,1,s,m,m) &=& -\,\frac{1}{\beta(m^2/s)}\,
\ln\frac{\beta(m^2/s)+1}{\beta(m^2/s)-1},
\nl
b_0^{\rm fin}(2,1,s,m_1,m_2) &=&
\frac{m_1^2-m_2^2-s}{\lambda^{1/2}(s,m_1^2,m_2^2)}\,L(s,m_1^2,m_2^2)
+ \frac{1}{2}\, \Bigl( \ln \frac{m^2_1}{s} \, - \, \ln \frac{m^2_2}{s} \Bigr).
\eqa
%--
%--
\noindent
\underline{\emph{Infrared-divergent configurations.}}
Two special cases requires a separate discussion: when $m_1=0$ in 
\eqn{1loop2p} and $\alpha= 2$ or $\beta= 1$, an IR divergency shows up.
Here we regularize IR divergencies by setting $n=4+{\hat{\ep}}$,
and then we use the relation between the UV regulator $\ep$ and the IR
one ${\hat{\ep}}$, i.e. $\ep= -{\hat{\ep}}$. The results read as follows:
%--
\bqa
b_0(2,1,s,0,0) &=& \frac{2}{\ep} - \DUV(s) - L_{\rm an}
   + \frac{\ep}{4}\,\Bigl[  
\DUV^2(s) + 2\,\DUV(s)\,L_{\rm an}
+ L^2_{\rm an} - \zeta(2)\Bigr],
\nl
b_0(2,1,s,0,m) &=& \frac{s}{s - m^2}\,\Bigl\{
     \frac{2}{\ep} - \DUV(s) - \ln \frac{m^2}{s}
    - \lpar 1 + \frac{m^2}{s}\rpar\,\ln \lpar 1-\frac{s}{m^2}\rpar
\label{irone}\nl
{}&+& \frac{\ep}{2}\,\Bigl[\,
        \frac{1}{2}\,\zeta(2) + \frac{1}{2}\,\DUV^2(s)
    + \ln \frac{m^2}{s}\,\lpar \DUV(s) + \frac{1}{2} \ln \frac{m^2}{s}\, \rpar
\nl
{}&+& \lpar 1 + \frac{m^2}{s}\rpar\,\Bigl(
      \ln\lpar 1 - \frac{s}{m^2}\rpar\,\DUV(s) 
      + \ln \frac{m^2}{s}\,\ln\lpar 1 - \frac{s}{m^2}\rpar 
\nl
{}&+& \frac{1}{2}\,\ln^2\lpar 1 - \frac{s}{m^2}\rpar 
      - \li{2}{\frac{s}{s-m^2}} \, \Bigl)\Bigr]\Bigr\}. 
\eqa
%--
Here we used generalized Nielsen polylogarithms,
\bq
S_{n,p}(z) = \frac{(-1)^{n+p-1}}{(n-1)\,!\;p\,!}\,
\int_0^1\,\frac{dx}{x}\,\ln^{n-1} x\;\ln^p (1 - z\,x), \qquad
S_{n-1,1}(z) = \li{n}{z}.
\eq
%--

\noindent
\textbf{Three-point integrals}
\vspace{0.2cm}
%--

\noindent
\underline{\emph{Ultraviolet decompositions.}}
For three-point integrals we extract form factors for tensor configurations 
and then we introduce scaled quantities. Our notation for a generic tensor 
structure with $l$ Lorentz indices is
%--
\bq
C_{\mu_1\ldots\mu_l}(\alpha_1,\alpha_2,\alpha_3,p_1,p_2,P,\{m\}_{123}) =
\mu_1(\ep)\,\int\,
d^nq\,\frac{q_{\mu_1}
\ldots q_{\mu_l}}{(q^2 + m^2_1)^{\alpha_1}\,[(q+p_1)^2 + m^2_2]^{\alpha_2}
[(q+P)^2 + m^2_3]^{\alpha_3}}, 
\eq
%--
where $P=p_1+p_2$. Scaled quantities read as (we omit list of arguments)
%--
\bqa
%--
C_{\mu\nu}\,&=&\,
s^{2-\alpha_1-\alpha_2-\alpha_3}\,\Bigl[\,
c_{21}\,p_{1\mu} p_{1\nu}\, +\, 
c_{22}\,p_{2\mu} p_{2\nu} \,
+\, 
c_{23}\,\{p_1 p_2\}_{\mu\nu}\,\Bigr]\, +\, 
s^{3-\alpha_1-\alpha_2-\alpha_3}\,
c_{24}\,\delta_{\mu\nu}, 
\nl
C_{\mu}\,&=&\, 
s^{2-\alpha_1-\alpha_2-\alpha_3}\,\Bigl(\,
c_{11}\,p_{1\mu} \,+\, 
c_{12}\,p_{2\mu}\,\Bigr),
\qquad
C_{0}\,=\, 
s^{2-\alpha_1-\alpha_2-\alpha_3}\,
c_{0},
\label{GCfunV0}
\eqa
%--
where we have also introduced the symmetrized combination $\{p\,k\}_{\mu\nu} = 
p_{\mu}\,k_{\nu} + p_{\nu}\,k_{\mu}$.
Three-point, scaled, one-loop form factors will often appear in the residues
of the UV poles for two-loop vertex integrals. Reduction of higher-rank form 
factors to the scalar case can be obtained by means of standard methods,
and UV decompositions read as
%--
\bq
c_i(\alpha_1,\alpha_2,\alpha_3,p_1,p_2,P,\{m\}_{123}) = 
\sum_{k=-1}^{1}\,
c_i(\alpha_1,\alpha_2,\alpha_3,p_1,p_2,P,\{m\}_{123};\,k)\,F^1_k(s),
\eq
%--
where $i$ contains the information on the spin structure and $s=-P^2$.
Examples of UV decomposition are
%--
\bq
c_0(1,1,1,p_1,p_2,P,\{m\}_{123};\,-1) = 0,
\quad \dots \quad
c_{24}(1,1,1,p_1,p_2,P,\{m\}_{123};\,-1) = \frac{1}{2}, 
\quad \mbox{etc.}
\eq
%--
\subsection{Two-loop tadpoles \label{vacB}}
\label{ss:1}
%--
A special class of two-loop integrals is represented by tadpoles, defined by
%--
\bqa
T_{ijk}\lpar \{m\}_{123} \rpar &=&
\mu_2(\ep)\,\int \frac{d^n q_1\,d^n q_2}
{\lpar q^2_1 + m^2_1\rpar^{i}\,
\lpar q^2_2 + m^2_2\rpar^{j}\,
[ (q_1 - q_2)^2 + m^2_3]^{k}}.
\eqa
%--
Any two-loop tadpole can be reduced to a single two-loop master tadpole
(we choose the configuration with $i=k=1,\ j=2$) and to products
of one-loop tadpoles through iterated application of IBP identities.
A few examples are given in \appendx{ibpvb}. The UV decomposition for the 
master tadpole reads as
%--
\bq
T_{121}\lpar \{m\}_{123}\rpar \, = \, \sum_{k=-2}^0\,
T_{121}\lpar \{m\}_{123};k\rpar\, F_{k}^2(M^2),
\label{masterB}
\eq 
%--
where the coefficients of the UV-divergent factors are given by
%--
\bqa
T_{121}(\{m\}_{123};-2)\, &=&\, -2, \qquad
T_{121}(\{m\}_{123};-1)\, =\,
-\, \Bigl(\, 1\, -\, 2\, \ln \frac{m_2^2}{M^2} \,\Bigr).
\label{defTf}
\eqa
%--
These relations hold if $m_2 \neq 0$, since a collinear logarithm is
present when $m_2 \to 0$. 
When $m_2 = 0$ we can change integration variables and apply again
IBP identities in order to use \eqn{defTf}, i.e.
%--
\bq
T_{121}(m_1,0,m_3)\, = \, T_{112}(m_1,m_3,0)\, 
\underbrace{\to}_{IBP}\, T_{121}(m_1,m_3,0),
\eq
%--
and the $(m_1,m_3,0)$ configuration, which is collinear-free, can be computed 
as shown in \eqn{asusual} of \appendx{ibpvb}.
%--
\subsection{Two-loop two-point integrals \label{TLtlF}}
\label{ss:2}
%--
\noindent
\textbf{Notation.}
We follow the notation of Section~7 of Ref.~\cite{Actis:2004bp} and we 
introduce scaled irreducible two-loop two-point integrals.
Note that when the squared external momentum vanishes, $p^2=0$,
the tadpole-reduction methods of \subsect{ss:1} can be easily employed.
Otherwise, we extract $s= -p^2$ and accordingly define
%--
\bqa
S_0^{\ssA}= s\, s_0^{\ssA},\qquad
S_0^{\ssC}= s_0^{\ssC}, \qquad
S_0^{\ssD}=\frac{1}{s}\,s_0^{\ssD},\qquad
S_0^{\ssE}=\frac{1}{s}\,s_0^{\ssE}.
\label{scale2}
\eqa
%--
The alphameric classification of graphs can be found in Subsection~2.2 of 
Ref.~\cite{Actis:2004bp}.
For tensor configurations we employ form factors 
according to Section~7 of Ref.~\cite{Actis:2004bp},
%--
\bqa
\texttt{rank 1}\quad \to \quad S^{\ssI}(\mu\,|\,0) &=& S^{\ssI}_1\,p_{\mu},
\qquad \qquad \qquad\qquad \
S^{\ssI}(0\,|\,\mu) = S^{\ssI}_2\,p_{\mu},
\nl
\texttt{rank 2} \  \to \ \
S^{\ssI}(\mu,\nu\,|\,0) &=& S^{\ssI}_{112}\,\delta_{\mu\nu} +
S^{\ssI}_{111}\,p_{\mu}\,p_{\nu},
\qquad
S^{\ssI}(\mu\,|\,\nu) = S^{\ssI}_{122}\,\delta_{\mu\nu} +
S^{\ssI}_{121}\,p_{\mu}\,p_{\nu},
\nl
S^{\ssI}(0\,|\,\mu,\nu) &=& S^{\ssI}_{222}\,\delta_{\mu\nu} +
S^{\ssI}_{221}\,p_{\mu}\,p_{\nu},
\label{generalabg}
\eqa
%--
with $I= A,C,D,E$, and we extract the scale dependence in complete analogy 
with \eqn{scale2}. 
%--
\vspace{0.2cm}
%--

\noindent
\textbf{Ultraviolet decompositions.}
The UV decomposition introduced in \eqnsc{TLUVD}{UVfactors} reads as
%--
\bq
s^{\ssI}_j\lpar s, \{m\} \rpar\, \, = \,
\sum_{k=-2}^0\, s^{\ssI}_j\lpar s, \{m\} ; k \rpar\,\, F_k^2(s), 
\eq
%--
where $I=A,C,D,E$ classifies the family of integrals and $j$ contains the 
information about the tensor structure. It is worth noting that the following 
results can be written in a compact form through the one-loop, two-point 
functions introduced in \subsect{oneloopFUNC}.
We shall be quoting the tensor configurations relevant for our calculation.
%--
\vspace{0.2cm}

\noindent
\underline{\emph{A family}}
%--
\bq
s^{\ssA}_0\lpar s,\{m\}_{123}\,;\,-2\rpar \,=\, 
2\,\sum_{i=1}^3\, \frac{m^2_i}{s},
\eq
%--
\bq
s^{\ssA}_0\lpar s,\{m\}_{123}\,;\,-1\rpar \,=\,
\sum_{i=1}^{3}\,\frac{m^2_i}{s}\,
\Bigl[\, - \, 2\, \ln\lpar \frac{m_i^2}{s}\rpar \,+\, 3\,\Bigr]\, -\, 
\frac{1}{2}.
\eq
%--
\noindent
\underline{\emph{C family}}
\bq
s^{\ssC}_0\lpar s,\{m\}_{1234}\,;\,-2\rpar \,=\, -\,2, \qquad
s^{\ssC}_1\lpar s,\{m\}_{1234}\,;\,-2\rpar \,=\, \frac{1}{2}, \qquad
s^{\ssC}_2\lpar s,\{m\}_{1234}\,;\,-2\rpar \,=\, 1,
\label{dpoleC}
\eq
%--
\bqa
s^{\ssC}_0\lpar s,\{m\}_{1234}\,;\,-1\rpar \,&=&\,
   - \,2\,b_0(1,1,s,\{m\}_{34}\,;\,0)\, -\, 1,
\nl
s^{\ssC}_1\lpar s,\{m\}_{1234}\,;\,-1\rpar \,&=&\,
   - \,b_1(1,1,s,\{m\}_{34}\,;\,0) \,+\, \frac{1}{8},
\nl
s^{\ssC}_2\lpar s,\{m\}_{1234}\,;\,-1\rpar \,&=&\,
   - \,2\,b_1(1,1,s,\{m\}_{34}\,;\,0) \,+\, \frac{1}{4}.
\label{spoleC}
\eqa
%--
\noindent
\underline{\emph{D family}}
%--
\bqa
s^{\ssD}_0\lpar s,\{m\}_{12345}\,;\,-2\rpar &=& 0, \qquad\quad
s^{\ssD}_i\lpar s,\{m\}_{12345}\,;\,-2\rpar = 0,
\nl
s^{\ssD}_{112}\lpar s,\{m\}_{12345}\,;\,-2\rpar &=& -\,\frac{1}{2}, \qquad
s^{\ssD}_{122}\lpar s,\{m\}_{12345}\,;\,-2\rpar = 0, \qquad
s^{\ssD}_{222}\lpar s,\{m\}_{12345}\,;\,-2\rpar = -\,\frac{1}{2},
\nl
s^{\ssD}_{ij1}\lpar s,\{m\}_{12345}\,;\,-2\rpar &=& 0,
\eqa
%--
\bqa
s^{\ssD}_0\lpar s,\{m\}_{12345}\,;\,-1\rpar \,&=& 
\,0, \qquad\qquad\qquad\qquad\qquad
\qquad\qquad\, \,
s^{\ssD}_i\lpar s,\{m\}_{12345}\, ;\,-1\rpar \,=\, 0,
\nl
s^{\ssD}_{112}\lpar s,\{m\}_{12345}\,;\,-1\rpar \,&=&\,
   - \,\frac{1}{2}\,b_0(1,1,s,\{m\}_{45}\,;\,0) \,-\, \frac{3}{8},\qquad
s^{\ssD}_{122}\lpar s,\{m\}_{12345}\,;\,-1\rpar \,=\,
   - \,\frac{1}{4},
\nl
s^{\ssD}_{222}\lpar s,\{m\}_{12345}\,;\,-1\rpar \,&=&\,
   - \,\frac{1}{2}\,b_0(1,1,s,\{m\}_{12}\,;\,0) \,-\, \frac{3}{8},
\nl
s^{\ssD}_{ij1}\lpar s,\{m\}_{12345}\,;\,-1\rpar \,&=&\, 0.
\eqa
%--
\noindent
\underline{\emph{E family}}
%--
\bqa
s^{\ssE}_0\lpar s,\{m\}_{12343}\,;\,-2\rpar \,&=& \,0, \qquad
s^{\ssE}_i\lpar s,\{m\}_{12343}\,;\,-2\rpar \,=\, 0, 
\nl
s^{\ssE}_{112}\lpar s,\{m\}_{12343}\,;\,-2\rpar \,&=&\, 0,\qquad 
s^{\ssE}_{122}\lpar s,\{m\}_{12343}\,;\,-2\rpar \,=\, -\,\frac{1}{4}, \qquad
s^{\ssE}_{222}\lpar s,\{m\}_{12343}\,;\,-2\rpar \,=\, -\,\frac{1}{2},
\nl
s^{\ssE}_{ij1}\lpar s,\{m\}_{12343}\,;\,-2\rpar \,&=&\, 0, 
\label{Edouble}
\eqa
\bqa
s^{\ssE}_0\lpar s,\{m\}_{12343}\,;\,-1\rpar \,&=&\,
   - \,2 \, \, b_0(2,1,s,\{m\}_{34}\,;\,0),
\nl
s^{\ssE}_1\lpar s,\{m\}_{12343}\,;\,-1\rpar \,&=&\,
   - \,b_1(2,1,s,\{m\}_{34}\,;\,0),
\nl
s^{\ssE}_2\lpar s,\{m\}_{12343}\,;\,-1\rpar \,&=&\,
   - \,2\,b_1(2,1,s,\{m\}_{34}\,;\,0),
\nl
s^{\ssE}_{112}\lpar s,\{m\}_{12343}\,;\,-1\rpar \,&=&\,
     \frac{1}{2}\,\frac{m^2_1\,+\,m^2_2}{s}\,b_0(2,1,s,\{m\}_{34}\,;\,0)
   \,+\, \frac{1}{6}\,b_{21}(2,1,s,\{m\}_{34}\,;\,0),
\nl
s^{\ssE}_{111}\lpar s,\{m\}_{12343}\,;\,-1\rpar \,&=&\,
   - \,\frac{2}{3}\,b_{21}(2,1,s,\{m\}_{34}\,;\,0),
\nl
s^{\ssE}_{122}\lpar s,\{m\}_{12343}\,;\,-1\rpar \,&=&\,
   - \,b_{22}(2,1,s,\{m\}_{34}\,;\,0) \,-\, \frac{1}{16},
\nl
s^{\ssE}_{121}\lpar s,\{m\}_{12343}\,;\,-1\rpar \,&=&\,
   - \,b_{21}(2,1,s,\{m\}_{34}\,;\,0),
\nl
s^{\ssE}_{222}\lpar s,\{m\}_{12343}\,;\,-1\rpar \,&=&\,
   - \,2\,b_{22}(2,1,s,\{m\}_{34}\,;\,0) \,-\, \frac{1}{8},
\nl
s^{\ssE}_{221}\lpar s,\{m\}_{12343}\,;\,-1\rpar \,&=&\,
   - \,2\,b_{21}(2,1,s,\{m\}_{34}\,;\,0).
\eqa
%--
The results shown for the $s^\ssE$ family hold when $m_3\neq 0$.
\vspace{0.2cm}
%--

\noindent
\textbf{Finite components.}
The UV-finite parts of two-point two-loop functions can be found 
in Ref.~\cite{Passarino:2001wv}, Eq.~(89) and Eqs.~(146-147), for 
$S^{\ssA}_0 = S^{111}_0$; in Ref.~\cite{Passarino:2001jd}, Subsection~5.8 for 
$S^{\ssC}_0 = S^{121}_0$, Subsection~7.3 for $S^{\ssE}_0 = S^{131}_0$ and 
Subsection~7.8 for $S^{\ssD}_0 = S^{221}_0$.  
\vspace{0.2cm}
%--

\noindent
\textbf{Special configurations.}
Some configurations where one or more masses vanish show overlapping UV and IR
divergencies and deserve a separate discussion.

As a non trivial example, we consider the scalar function of the $C$ family 
evaluated at $p^2=0$ ($s_0^{\ssC}$), since here the residue of the single UV 
pole can be IR divergent.
Note that this function represents a tadpole, and it can be treated
by means of the methods of \subsect{vacB}.
The residue of the double UV pole of $s^{\ssC}_0$ can be found
in \eqn{dpoleC} and the residue of the single UV pole is given in \eqn{spoleC}.
The same UV components arise when $p^2=0$ and $m_3=m_4$ is not vanishing,
%--
\bqa
s_0^{\ssC}(0,m_1,m_2,m_3,m_3) &=& 
-2\,F^2_{-2}(M^2)
%-\,\frac{2}{\ep^2}
%+2\frac{\DUV}{\ep}
%-\DUV^2
- F^2_{_1}(M^2)
%- \Bigl(\frac{1}{\ep}-\DUV \Bigr)
\Bigl( 1-2\ln\frac{m_3^2}{M^2}\Bigr)
+ s^{\ssC}_{\rm fin},
\label{IR}
\eqa
%--
where $s^{\ssC}_{\rm fin}$ is finite when $\ep \to 0$.
However, reduction in sub-loops requires to evaluate cases where $m_3=m_4=0$. 
Here the residue of the single UV pole becomes IR singular, and the explicit 
expressions read as
%--
\bqa
%s_0^{\ssC}(0,0,m,0,0) &=& 2\,F^2_{-2}(M^2) + 
%2\,\Bigl( 1 - \ln\frac{m^2}{M^2}\Bigr)\,F^2_{-1}(M^2) 
%- \frac{1}{\ep}\,\Bigl( \frac{2}{\ep} - 2\,\DUV + 1\Bigr),
%
%s_0^{\ssC}(0,0,m,0,0) &=& \frac{2}{\ep^2} - \DUV 
% + \Bigl( \frac{1}{\ep} - \DUV\Bigr)\,\Bigl(
% 1 - 2\,\DUV - 2\,\ln\frac{m^2}{M^2}\Bigr),
\nl
s_0^{\ssC}(0,m_1,m_2,0,0) &=& 2\,F^2_{-2}(M^2) + 2\,
\Bigl[ 1 + \frac{1}{m^2_2 - m^2_1}\,\Bigl(
m^2_1\,\ln\frac{m^2_1}{M^2} - m^2_2\,\ln\frac{m^2_2}{M^2}\Bigr)\Bigr]\,
F^2_{-1}(M^2) + F_{\ssI\ssR},
\nl
F_{\ssI\ssR} &=& - \frac{1}{\ep}\,\Bigl( \frac{2}{\ep} - 2\,\DUV + 1\Bigr).
%
%s_0^{\ssC}(0,m_1,m_2,0,0) &=& \frac{2}{\ep^2} - \DUV  
% + \Bigl( \frac{1}{\ep} - \DUV\Bigr)\,\Bigl[
% 1 - 2\,\DUV + 2\,\frac{1}{m^2_2 - m^2_1}\,\Bigl(
%m^2_1\,\ln\frac{m^2_1}{M^2} - m^2_2\,\ln\frac{m^2_2}{M^2}\Bigr)\Bigr].
\label{IRaa}
\eqa
%--
Overlapping UV and IR divergencies appear also for special configurations of 
the scalar function of the $E$ family ($s_0^{\ssE}$), evaluated at arbitrary 
$s$, when $m_3=0$, i.e.
%--
\bqa
s_0^{\ssE}(s,0,m_2,0,M,0) &=&
 \frac{2}{1 - x}\,\Bigl[ \DUV^2(s) - \frac{2}{\ep^2}\Bigr]
+ 2\,\Bigl[ \frac{1}{\ep} - \DUV(s)\Bigr]\,\Bigl\{
\frac{1}{1 - x}\,\Bigl[ 2\,\DUV(s) - 1 
- \ln x
\nl
{}&+& 2\,\ln (x - 1) + \ln\frac{m^2_2}{s}\Bigr] +
\ln x - \ln (x - 1) \Bigr\} + 
s^{\ssE}_{\rm fin}(s,0,m_2,0,M,0),
\nl
s_0^{\ssE}(s,m_1,m_2,0,M,0) &=&
 \frac{2}{1 - x}\,\Bigl[ \DUV^2(s) - \frac{2}{\ep^2}\Bigr]
+ 2\,\Bigl[ \frac{1}{\ep} - \DUV(s)\Bigr]\,\Bigl\{
\frac{1}{1 - x}\,\Bigl[
2\,\DUV(s) - 1 
\nl
{}&-& \ln x + 2\,\ln (x-1)
+ \frac{m^2_2}{m^2_2-m^2_1}\,\ln\frac{m^2_2}{s} 
- \frac{m^2_1}{m^2_2-m^2_1}\,\ln\frac{m^2_1}{s}\Bigr] 
\nl
{}&+& \ln x -
\ln (x-1) \Bigr\} + 
s^{\ssE}_{\rm fin}(s,m_1,m_2,0,M,0),
\label{irtwo}
\eqa
%--
where $s^{\ssE}_{\rm fin}$ is finite when $\ep \to 0$ and $x= M^2/s$. 
Note that here double poles arise, whereas the residue of the double UV pole 
vanishes in \eqn{Edouble}.

These spurious singularities are a consequence of the sub-loop reduction 
of Ref.~\cite{Weiglein:1993hd}. This technique allows to handle integrals with 
irreducible scalar products in the numerator and introduces new propagators 
with a corresponding zero mass. IR poles are then generated by the related 
integrals behaving as
%--
\bq
\int_0\,d^nq \frac{1}{q^4}.
\eq
%--
The treatment of overlapping UV and IR poles is not free from ambiguities. 
Strictly speaking, the integrals of \eqnsc{IRaa}{irtwo} cannot be defined in 
any strip of the complex $n$-plane. UV regulation requires $n < 4$ and IR one 
should be performed for $n > 4$. Therefore, one should first of all
disentangle overlapping UV and IR singularities. This can be achieved by 
means of the IBP identities introduced in Ref.~\cite{Tkachov:1981wb}, which 
allow  to write the original integral as a combination of objects which are 
UV or IR singular but never both.
At the end one can use the well-known relation between UV and IR regulators, 
i.e. ${\hat{\ep}} = - \ep$. We verified that this recipe gives the same result
of the naive approach where one regularizes integrals for $n < 4$ without 
separating the UV and the IR domains.
%--
\subsection{Two-loop three-point integrals}
\label{ss:3}
%--
\noindent
\textbf{Notation.}
Our notation for three-point two-loop integrals can be found in Section~9 of 
Ref.~\cite{Actis:2004bp}. Here $P= p_1+p_2$ and we extract $s= -P^2$ in order 
to define scaled functions. For scalar configurations we introduce
%--
\bq
V_0^{\ssE} = v_0^{\ssE}, \qquad V_0^{\ssI,\ssG}= s^{-1}\,v_0^{\ssI,\ssG}, 
\qquad
V_0^{\ssM,\ssK,\ssH} = s^{-2}\,v_0^{\ssM,\ssK,\ssH}.
\label{scalars}
\eq
%--
For tensor integrals, first we introduce form factors according to 
Section~9 of Ref.~\cite{Actis:2004bp}. Then, we define dimensionless 
functions in complete analogy with \eqn{scalars}.
The (somehow redundant) definition of the rooting of momenta in two-loop 
vertices has been introduced for all families in Ref.~\cite{Actis:2004bp}; we
stick to the notation used there and one should only remember momentum
conservation, i.e. $P = p_1+p_2$.
\vspace{0.2cm}
%--

\noindent
\textbf{Ultraviolet decompositions.}
The UV decomposition introduced in \eqnsc{TLUVD}{UVfactors} reads as
%--
\bq
v^{\ssL}_j\lpar \{\hbox{momenta}\}\,,\,\{m\} \rpar\, \, = \,
\sum_{k=-2}^0\, v^{\ssL}_j\lpar \{\hbox{momenta}\}\,,\,\{m\} ; 
k \rpar\,\, F_k^2(s), 
\eq
where $L=E,I,G,M,K,H$ classifies the family of integrals under consideration
and $j$ summarizes the information about the tensor structure.
Here we show only UV-divergent tensor configurations relevant for our 
calculation.
\vspace{0.1cm}

%--
\noindent
\underline{\emph{E family}}
\bqa
v^{\ssE}_0\lpar \cdots\,;\,-2\rpar &=& -2, \qquad
v^{\ssE}_{11}\lpar \cdots\,;\,-2\rpar = \frac{1}{2}, \qquad
v^{\ssE}_{12}\lpar \cdots\,;\,-2\rpar = 1, 
\nl
v^{\ssE}_{21}\lpar \cdots\,;\,-2\rpar &=& 1, \qquad \ \ 
v^{\ssE}_{22}\lpar \cdots\,;\,-2\rpar = 2, 
\eqa
%--
\bqa
v^{\ssE}_0\lpar p_2,P,\{m\}_{1234}\,;\,-1\rpar &=&
   - 2\,b_0(1,1,p_1,\{m\}_{34}\,;\,0) - 1,
\nl
v^{\ssE}_{11}\lpar p_2,P,\{m\}_{1234}\,;\,-1\rpar &=&
   - b_1(1,1,p_1,\{m\}_{34}\,;\,0) + \frac{1}{8},
\nl
v^{\ssE}_{12}\lpar p_2,P,\{m\}_{1234}\,;\,-1\rpar &=&
   b_0(1,1,p_1,\{m\}_{34}\,;\,0) + \frac{1}{4},
\nl
v^{\ssE}_{21}\lpar p_2,P,\{m\}_{1234}\,;\,-1\rpar &=&
   - 2\,b_1(1,1,p_1,\{m\}_{34}\,;\,0) + \frac{1}{4},
\nl
v^{\ssE}_{22}\lpar p_2,P,\{m\}_{1234}\,;\,-1\rpar &=&
   2\,b_0(1,1,p_1,\{m\}_{34}\,;\,0) + \frac{1}{2}.
\eqa
%--
\noindent
\underline{\emph{I family}}
%--
\bqa
v^{\ssI}_0\lpar \cdots\,;\,-2\rpar &=& 0, \quad
v^{\ssI}_{ij}\lpar \cdots\,;\,-2\rpar = 0, 
\nl
v^{\ssI}_{114}\lpar \cdots\,;\,-2\rpar &=& 0, \quad
v^{\ssI}_{124}\lpar \cdots\,;\,-2\rpar = -\,\frac{1}{4}, \quad
v^{\ssI}_{224}\lpar \cdots\,;\,-2\rpar = -\,\frac{1}{2}, 
\nl
v^{\ssI}_{ijk}\lpar \cdots\,;\,-2\rpar &=& 0, 
%\quad i,j = 1,2, \, k= 1,2,3, 
\eqa
%--
\bqa
v^{\ssI}_0\lpar p_1,P,\{m\}_{12345}\,;\,-1\rpar &=&
   - 2\,c_0(1,1,1,p_1,p_2,P,\{m\}_{345}\,;\,0),
\nl
v^{\ssI}_{1i}\lpar p_1,P,\{m\}_{12345}\,;\,-1\rpar &=&
   - c_{1i}(1,1,1,p_1,p_2,P,\{m\}_{345}\,;\,0),
\nl
v^{\ssI}_{2i}\lpar p_1,P,\{m\}_{12345}\,;\,-1\rpar &=&
   - 2\,c_{1i}(1,1,1,p_1,p_2,P,\{m\}_{345}\,;\,0),
\nl
{}{} v^{\ssI}_{114}\lpar p_1,P,\{m\}_{12345}\,;\,-1\rpar &=&
     \frac{1}{2}\,\frac{m^2_1+m^2_2}{s}\,
      c_0(1,1,1,p_1,P,\{m\}_{345}\,;\,0)
\nl
+ \frac{1}{6}\,\frac{p^2_1}{s}\,
      c_{21}(1,1,1,p_1,p_2,P,\{m\}_{345}\,;\,0)
&+&  \frac{1}{6}\,\frac{P^2-p^2_1-p^2_2}{s}\,
      c_{23}(1,1,1,p_1,p_2,P,\{m\}_{345}\,;\,0)
\nl
&+&  \frac{1}{6}\,\frac{p^2_2}{s}\,
      c_{22}(1,1,1,p_1,p_2,P,\{m\}_{345}\,;\,0),
\nl
v^{\ssI}_{11i}\lpar p_1,P,\{m\}_{12345}\,;\,-1\rpar &=&
    - \frac{2}{3}\,c_{2i}(1,1,1,p_1,p_2,P,\{m\}_{345}\,;\,0),
\nl
v^{\ssI}_{124}\lpar p_1,P,\{m\}_{12345}\,;\,-1\rpar &=&
   - c_{24}(1,1,1,p_1,p_2,P,\{m\}_{345}\,;\,0) - \frac{1}{16},
\nl
v^{\ssI}_{12i}\lpar p_1,P,\{m\}_{12345}\,;\,-1\rpar &=&
    - c_{2i}(1,1,1,p_1,p_2,P,\{m\}_{345}\,;\,0),
\nl
v^{\ssI}_{224}\lpar p_1,P,\{m\}_{12345}\,;\,-1\rpar &=&
   - 2\,c_{24}(1,1,1,p_1,p_2,P,\{m\}_{345}\,;\,0) - \frac{1}{8},
\nl
v^{\ssI}_{22i}\lpar p_1,P,\{m\}_{12345}\,;\,-1\rpar &=&
    - 2\,c_{2i}(1,1,1,p_1,p_2,P,\{m\}_{345}\,;\,0).
\eqa
%--
\noindent
\underline{\emph{G family}}
%--
\bqa
v^{\ssG}_0\lpar \cdots\,;\,-2\rpar &=& 0, \quad
v^{\ssG}_{ij}\lpar \cdots\,;\,-2\rpar = 0, 
\nl
v^{\ssG}_{114}\lpar \cdots\,;\,-2\rpar &=& -\,\frac{1}{2}, \quad 
v^{\ssG}_{124}\lpar \cdots\,;\,-2\rpar = 0, \quad 
v^{\ssG}_{224}\lpar \cdots\,;\,-2\rpar = -\,\frac{1}{2}, 
\nl
v^{\ssG}_{ijk}\lpar \cdots\,;\,-2\rpar &=& 0, 
%\quad i,j = 1,2, \, k= 1,2,3.
\eqa
%--
\bqa
v^{\ssG}_0\lpar \cdots\,;\,-1\rpar &=& 0,
\qquad
v^{\ssG}_{1i}\lpar \cdots\,;\,-1\rpar = 0,
\nl
v^{\ssG}_{114}\lpar p_1,p_1,P,\{m\}_{12345}\,;\,-1\rpar &=&
   - \frac{1}{2}\,b_0(1,1,p_2,\{m\}_{45}\,;\,0) - \frac{3}{8},
\nl
v^{\ssG}_{124}\lpar p_1,p_1,P,\{m\}_{12345}\,;\,-1\rpar &=&
   - \frac{1}{4},
\nl
v^{\ssG}_{224}\lpar p_1,p_1,P,\{m\}_{12345}\,;\,-1\rpar &=&
   - \frac{1}{2}\,b_0(1,1,p_1,\{m\}_{12}\,;\,0) - \frac{3}{8}.
\eqa
%--
\noindent
\underline{\emph{M family}}
\bqa
v^{\ssM}_0\lpar p_1,P,\{m\}_{123453}\,;\,-1\rpar &=&
   - 2\,c_0(2,1,1,p_1,p_2,P,\{m\}_{345}\,;\,0),
\nl
v^{\ssM}_{1i}\lpar p_1,P,\{m\}_{123453}\,;\,-1\rpar &=&
   - c_{1i}(2,1,1,p_1,p_2,P,\{m\}_{345}\,;\,0),
\nl
v^{\ssM}_{2i}\lpar p_1,P,\{m\}_{123453}\,;\,-1\rpar &=&
   - 2\,c_{1i}(2,1,1,p_1,p_2,P,\{m\}_{345}\,;\,0),
\nl
v^{\ssM}_{114}\lpar p_1,P,\{m\}_{123453}\,;\,-1\rpar &=&
     \frac{1}{2}\,\frac{m^2_1+m^2_2}{s}\,
      c_0(2,1,1,p_1,p_2,P,\{m\}_{345}\,;\,0)
\nl
+ \frac{1}{6}\,\frac{p^2_1}{s}\,
      c_{21}(2,1,1,p_1,p_2,P,\{m\}_{345}\,;\,0)
&+& \frac{1}{6}\,\frac{P^2-p^2_1-p^2_2}{s}\,
      c_{23}(2,1,1,p_1,p_2,P,\{m\}_{345}\,;\,0)
\nl
&+& \frac{1}{6}\,\frac{p^2_2}{s}\,
       c_{22}(2,1,1,p_1,p_2,P,\{m\}_{345}\,;\,0),
\nl
v^{\ssM}_{11i}\lpar p_1,P,\{m\}_{123453}\,;\,-1\rpar &=&
    - \frac{2}{3}\,c_{2i}(2,1,1,p_1,p_2,P,\{m\}_{345}\,;\,0),
\nl
v^{\ssM}_{12i}\lpar p_1,P,\{m\}_{123453}\,;\,-1\rpar &=&
   - c_{2i}(2,1,1,p_1,p_2,P,\{m\}_{345}\,;\,0),
\nl
v^{\ssM}_{22i}\lpar p_1,P,\{m\}_{123453}\,;\,-1\rpar &=&
   - 2\,c_{2i}(2,1,1,p_1,p_2,P,\{m\}_{345}\,;\,0).
\eqa
%--

\noindent
\underline{\emph{H,K families}}
%--
\bqa
v^{\ssK}_0\lpar \cdots\,;\,-1\rpar &=& 0,
\quad
v^{\ssK}_{1i}\lpar \cdots\,;\,-1\rpar = 0,
\nl
v^{\ssK}_{114}\lpar P,p_1,P,\{m\}_{123456}\,;\,-1\rpar &=&
   - \frac{1}{2}\,c_0(1,1,1,p_1,p_2,P,\{m\}_{456}\,;\,0),
\nl
v^{\ssK}_{ijk}\lpar \cdots\,;\,-1\rpar &=& 0,
\qquad
v^{\ssH}_{ijk}\lpar \cdots\,;\,-1\rpar = 0.
\eqa
%--
\section{$\MSB$ renormalization at two loops}
\label{sec:renormalization}
%--
We now briefly summarize the notion of renormalization procedure.
Any renormalization procedure for a gauge theory requires three steps:
%--
\begin{itemize}
\item[--] the definition of the gauge-fixing term of the Lagrangian;
\item[--] the renormalization-scheme prescription;
\item[--] the choice of an IPS.
\end{itemize}
%--
In \subsect{gaugeF} we define the gauge-fixing part of the 
Standard-Model Lagrangian and in \subsect{renScheme} we discuss the 
renormalization scheme employed in this paper.
This is a rather technical subject, but one should realize that choosing a
renormalization scheme is equivalent to changing the variables which describe 
the degrees of freedom of the theory.
Bare quantities are traded for renormalized ones through multiplicative
renormalization constants, and the latter are expanded in perturbation theory
introducing  counterterms.
The prescription imposed for the counterterms defines the so-called 
renormalization scheme.

Whatever the renormalization scheme employed, a theory becomes predictive 
when renormalized quantities are expressed in terms of the chosen IPS. 
As long as the same data are employed, different renormalization 
schemes lead to the same theoretical predictions.
Special care should be devoted to the choice of the POs used as input: 
two predictions for the same PO differ by an amount proportional to the 
missing higher-order corrections if they refer to different IPS.
We will discuss this issue in details in III, where we will show that a 
consistent description of unstable particles at two loops can be achieved in 
the framework of complex poles~\cite{Jegerlehner:2003py}.
%--
\subsection{Gauge fixing \label{gaugeF}}
%--
We fix the gauge for the electroweak sector of the Standard-Model Lagrangian
introducing six gauge parameters $\xi_i$ ($i=A,Z,AZ,\varphi^{0},W,\varphi$),
%--
\bqa
\label{eq:th:gfun}
{\cal{L}}^{\ssE\ssW}_{gf}\, &=& 
\, -\,{\cal{C}}^{+}\,{\cal{C}}^{-}\,-\,\frac{1}{2}\,
\Bigl[\, \left({\cal{C}}^{\ssA}\right)^2\, +
\, \left({\cal{C}}^{\ssZ}\right)^2\,\Bigr],
\nl
{\cal{C}}^{\ssA} =
           -\frac{1}{\gpA} \partial_{\mu} A_{\mu} 
           - \gpAZ \,\partial_{\mu} Z_{\mu},
\qquad
{\cal{C}}^{\ssZ} &=&
           -\frac{1}{\gpZ} \partial_{\mu} Z_{\mu} 
           + \xi_{\varphi^{0}}\, M_0\, \varphi^{0},
\qquad
{\cal{C}}^{\pm} =
           -\frac{1}{\gpW} \partial_\mu W^{\pm}_{\mu} 
           + \xi_{\varphi} \, M\, \varphi^{\pm}.
\eqa
%--
Here $A_\mu$, $Z_\mu$ and $W_\mu^{\pm}$ are the fields for the photon and
the $Z$ and the $W$ bosons and $\varphi^0$ and $\varphi^{\pm}$ are
the fields for the neutral and charged unphysical Higgs-Kibble scalars.
For the QCD sector, we employ the usual 't Hooft-Feynman gauge.

Note that gauge parameters are usually introduced in order to test 
predictions for physical (pseudo-)observables (hereafter POs); $S$-matrix 
elements are gauge independent, and one can indeed check that gauge-parameter 
dependence cancels out. Here, instead, the gauge-fixing term which defines 
what we call the $R_{_{\xi\xi}}$ gauge is closely related to our 
renormalization strategy.
We will discuss this issue in \subsect{renScheme}, where we will 
introduce renormalized gauge parameters.
%--

\noindent
%--
\textbf{Dyson-resummed propagators in the $R_{\xi\xi}$ gauge.}
%--
In this Section we illustrate the method of Dyson resummation of higher-order 
corrections which plays a crucial role in any renormalization procedure, 
because mass renormalization and wave-function factors for external legs 
naturally arise in this context.
In the $R_{\xi\xi}$ gauge, in addition, transitions between the vector
gauge bosons and the Higgs-Kibble scalars, as well as a mixing
between the photon and the $Z$ boson, take place at the Born level.
Therefore, Dyson resummation is a mandatory step already at lowest
order in perturbation theory.

Here we show the Dyson-resummed propagators for the vector gauge bosons
and the Higgs-Kibble scalars, including all the allowed transitions.
Furthermore, we provide the explicit expressions for the propagators
of the Faddeev-Popov ghost fields.
%--
\vspace{0.2cm}
%--

\noindent
\underline{\emph{The charged sector:}}
The Dyson-resummed propagator for the $W$ boson reads as
%--
\bq
\overline{\Delta}_{\mu\nu}^{\ssW\ssW} \, =\,
\frac{1}{(2\pi)^4 i}\,\,
\frac{1}{p^2\, +\, M^2}\,
\Bigl( \, \delta_{\mu\nu}\, +\, 
\overline{\Delta}_{pp}^{\ssW\ssW}\,p_{\mu}\, p_{\nu}\,
\Bigr),
\eq
%--
where we introduced
%--
\bq
\overline{\Delta}_{pp}^{\ssW\ssW}\, = \,
\frac{1}{p^2 \,+\, \gpW^2\, M^2}\, \left[\,
\gpW^2\, -\, 1\, +\, \frac{(\gpW\, 
-\, \xi_{\varphi})^2\, \gpW^2\, M^2\, (p^2\, +\, M^2)}
{(p^2\,+\,\gpW \,\xi_{\varphi}\, M^2)^2}
\right].
\eq
%--
Moreover, for the propagator of the $\varphi$ boson we get
%--
\bq
\overline{\Delta}^{\varphi \varphi} \, =\,
\frac{1}{(2\pi)^4 i}\,\,
\frac{p^2\, +\, \gpW^2\, M^2}{(p^2\,+\,\gpW\, \xi_{\varphi}\, M^2)^2}.
\eq
%--
Finally, a $W$-$\varphi$ transition takes place,
%--
\bq
\overline{\Delta}_{\mu}^{\ssW \varphi} \,=\,
\frac{1}{(2\pi)^4 i}\,\,
\frac{M\, \gpW\, (\xi_{{\varphi}}\, 
-\, \gpW)}{(p^2 \,+\, \gpW\, \xi_{{\varphi}}\, M^2)^2}\,
i \, p_{\mu}.
\eq
%--
\noindent
\underline{\emph{The neutral sector:}}
Let us start with the Dyson-resummed propagators for the photon and the $Z$ 
boson,
%--
\bq
\overline{\Delta}_{\mu\nu}^{\ssA\ssA} \, =\,
\frac{1}{(2\pi)^4 i}\,\frac{1}{p^2}\,
\Bigl( \,\delta_{\mu\nu} \,+\, \overline{\Delta}_{pp}^{\ssA\ssA}\,
p_{\mu} p_{\nu}\, \Bigr),
\qquad
\overline{\Delta}_{\mu\nu}^{\ssZ\ssZ} \,=\,
\frac{1}{(2\pi)^4 i}\,\frac{1}{p^2\, +\, M^2_0}
\Bigl( \, \delta_{\mu\nu} \,+\, \overline{\Delta}_{pp}^{\ssZ\ssZ}\,
p_{\mu} p_{\nu}\, \Bigr),
\eq
%--
where we defined
%--
\bqa
\overline{\Delta}_{pp}^{\ssA\ssA}\, =\,
\frac{\gpA^2\, -\, 1}{p^2} &+& \frac{\gpA^2\, \gpAZ^2\, \gpZ^2\, 
(p^2\, +\, \xi_{\varphi^{0}}^2\,M^2_0)}
{(p^2\, +\, \gpZ\, \xi_{\varphi^{0}}\, M^2_0)^2},
\nl
\overline{\Delta}_{pp}^{\ssZ\ssZ} \,=\,
\frac{1}{p^2\, +\, \xi_{\ssC}^2\, M^2_0}\,\Bigl[\,
\xi_{\ssC}^2\, -\, 1\, &+&\, \xi_{\ssC}^2\, (p^2\, +\, M^2_0)
\frac{M^2_0\, (\gpZ\, -\, \xi_{\varphi^{0}})^2\, +\,
\gpAZ^2\, \gpZ^2\, (p^2\, +\, \xi_{\varphi^{0}}^2\, M^2_0)}
{(p^2\, +\, \xi_{\varphi^{0}} \,\gpZ \,M^2_0)^2}\Bigr],
\eqa
%--
and we introduced a combination of gauge parameters,
$\xi_{\ssC}^2 = \gpZ^2/(1 + \gpZ^2 \, \gpAZ^2)$.
Furthermore, the propagator for the $\varphi^{0}$ boson reads as
%--
\bq
\overline{\Delta}^{\varphi^{0} \varphi^{0} } \,=\,
\frac{1}{(2\pi)^4 i}\,
\frac{p^2\, +\, \gpZ^2\, M^2_0}
{(p^2\, +\, \gpZ\, \xi_{\varphi^{0}}\, M^2_0)^2}.
\eq
%--
In the neutral sector we have vector-scalar transitions,
%--
\bq
\overline{\Delta}_{\mu}^{\ssA \varphi^{0}} \,=\,
\frac{1}{(2\pi)^4 i}\,
\frac{M_0\, \gpZ\, \gpA \,\gpAZ\, (\xi_{{\varphi^{0}}}\, -\, \gpZ)}
{(p^2\, +\, \gpZ\, \xi_{{\varphi^{0}}}\, M^2_0)^2}\,i \,p_{\mu},
\quad
\overline{\Delta}_{\mu}^{\ssZ \varphi^{0}} \,=\,
\frac{1}{(2\pi)^4 i}\,
\frac{M_0\, \gpZ\, (\xi_{{\varphi^{0}}} \,-\, \gpZ)}
{(p^2\, +\, \gpZ\, \xi_{{\varphi^{0}}}\, M^2_0)^2}\,i \,p_{\mu},
\eq
%--
as well as a mixing between the photon and the $Z$ boson,
%--
\bq
\overline{\Delta}_{\mu\nu}^{\ssA\ssZ} \,=\,
\frac{1}{(2\pi)^4 i}\,
\frac{\gpA\, \gpAZ\, \gpZ^2\, (p^2\, +\, \xi_{\varphi^{0}}^2\,M^2_0)}
{p^2\,(p^2\, +\, \gpZ\, \xi_{{\varphi^{0}}} \, M^2_0)^2}\,p_{\mu}\, p_{\nu}.
\eq
%--
\noindent
\underline{\emph{Ghost fields:}}
Finally, the gauge-fixing Lagrangian defines the propagators for the charged 
and neutral ghost fields $X^\pm$, $Y^\ssZ$ and $Y^\ssA$,
%--
\bq
\Delta^{\ssX^{\pm}} \,=\,
\frac{1}{(2\pi)^4 i}\,
\frac{\gpW}{p^2\, +\, \xi_{\varphi} \,\gpW \,M^2},
\quad
\Delta^{\ssY^{\ssZ}} \,=\,
\frac{1}{(2\pi)^4 i}\,
\frac{\gpZ}{p^2 \,+\, \xi_{\varphi^{0}} \, \gpZ \, M^2_0},
\quad
\Delta^{\ssY^{\ssA}} \,=\,
\frac{1}{(2\pi)^4 i}\,\frac{\gpA}{p^2}.
\eq
%--
\subsection{Definition of the renormalization scheme: $\MSB$ and beyond}
\label{renScheme}
%--
The \emph{orthodox} approach to renormalization~\cite{Freitas:2002ja}
uses the language of \emph{counterterms}.
It is worth noting that this is not a mandatory step, since one could write 
directly renormalization equations that connect the \emph{bare} parameters of 
the Lagrangian to an IPS, skipping the introduction of intermediate 
renormalized quantities and avoiding any unnecessary reference to 
a given renormalization scheme.
In this approach, carried on at one loop in~\cite{Passarino:1989ta}, no 
special attention is paid to individual Green functions, and one is mainly 
concerned with:
%--
\begin{itemize}
\item[--] UV finiteness of $S$-matrix elements after the proper treatment
of external legs in amputated Green functions, which greatly reduces the
complexity of the calculation;
\item[--] gauge independence of predictions for (pseudo-)observables.
\end{itemize}
%--
However, renormalization equations are usually organized through different 
building blocks, where gauge-boson self-energies embed process-independent 
(universal) higher-order corrections and play a privileged role.
Therefore, their structure has to be carefully analyzed, and the language
of counterterms allows to disentangle UV overlapping divergencies
which show up at two loops.
In a renormalizable gauge theory, in fact, the UV poles of any Green function 
can be removed order-by-order in perturbation theory. 
In addition, the imaginary part of a Green function at a given order is fixed, 
through unitarity constraints, by the previous orders.
Therefore, UV-subtraction terms have to be at most polynomials in the 
external momenta (in the following, \emph{local} subtraction terms).
In this paper, we will express our results using the language of 
counterterms:
%--
\begin{itemize}
\item[--] we promote bare quantities (parameters and fields) to renormalized 
ones;
\item[--] we fix the counterterms at one loop in order to remove the 
UV poles from all one-loop Green functions;
\item[--]  at this stage, we check that two-loop Green functions develop local 
UV residues;
\item[--] our final task is to fix the counterterms at two loops and to 
remove the UV poles from two-loop Green functions.
\end{itemize}
%--
Obviously, the absorption of UV divergencies into local counterterms
does not exhaust the renormalization procedure, because we have
still to connect renormalized quantities to POs, thus making the theory 
predictive. This will be the subject of III. In the remainder of this section
we discuss renormalization constants for all parameters and fields.
\vspace{0.1cm}
%--

\noindent
\textbf{Renormalization constants}
\vspace{0.1cm}

\noindent
We relate bare quantities to renormalized ones introducing multiplicative 
renormalization constants $Z_i$ and (if not otherwise stated) we 
expand them through the renormalized $SU(2)$ coupling constant $g_\ssR$,
%--
\bq
Z_{i} = 1 + \sum_{n=1}^{\infty}\,\lpar \frac{g^2_{\ssR}}{16\,\pi^2}
\rpar^n\,\delta Z^{(n)}_{i},
\label{exppar}
\eq
%--
where $\delta Z^{(n)}_{i}$ are counterterms and the subscript $i$ refers to 
masses, couplings, gauge parameters and fields. 
Let us make more precise the notion of renormalization constants. 
%--
\bei
\item[--] for the bare masses of the $W$ boson ($M$), the Higgs boson 
($\mh$) and the fermions ($m_f$) we write $m=Z^{1/2}_{m}\,m_{\ssR}$, where 
$m=M,\mh,m_f$ and $m_{\ssR}$ corresponds to one of the associated renormalized
masses. $Z_m$ is expanded through \eqn{exppar}.
\eei
%--
\bei
\item[--] For the bare $SU(2)$ coupling constant $g$ and the bare cosine 
(sine) of the weak-mixing angle $c_\theta$ ($s_\theta$) we define
$p=Z_{p}\,p_{\ssR}$, where $p=g,c_\theta,s_\theta$
and $p_{\ssR}$ is one of the related renormalized parameters.
$Z_p$ is expanded through \eqn{exppar}.
\eei
%--
\bei
\item[--] For the bare gauge parameters introduced in \eqn{eq:th:gfun}
we use $\xi = Z_{\xi}\,\xi_{\ssR}$, where $\xi$ is one of the bare
gauge parameters and $\xi_{\ssR}$ is the associated renormalized quantity.
$Z_\xi$ is expanded by means of \eqn{exppar} except for the case 
$\xi=\xi_{\ssA\ssZ}$, where we use
%--
\bq
Z_{\xi_{\ssA\ssZ}} = \sum_{n=1}^{\infty}\,\lpar \frac{g^2_{\ssR}}{16\,\pi^2}
\rpar^n\,\delta Z^{(n)}_{\xi_{\ssA\ssZ}}.
\eq
%--
After the expansions, we use the freedom in choosing the values for
the renormalized gauge parameters and we set $\xi=1$, thus compromising 
between two alternative exigencies.
On the one hand, counterterms for the gauge parameters are required
to remove UV divergencies; on the other hand, we recover part of the 
simplicity of 't Hooft-Feynman gauge.
%--
\eei
%--
\bei
\item[--] For a given bare field $\phi$ it is convenient to write 
$\phi = Z^{1/2}_{\phi}\,\phi_{\ssR}$,
where $\phi_{\ssR}$ is a renormalized field, and we expand $Z_{\phi}$ through
\eqn{exppar}. The bare photon field $A^\mu$ represents an exception, 
and here we use
%--
\bq
A^\mu = Z^{1/2}_{\ssA\ssA}\,A^\mu_{\ssR} +
Z^{1/2}_{\ssA\ssZ}\,Z^\mu_{\ssR},
\qquad
Z^{1/2}_{\ssA\ssZ} =
\sum_{n=1}^{\infty}\,\lpar \frac{g^2_{\ssR}}{16\,\pi^2}
\rpar^n\,\delta Z^{(n)}_{\ssA\ssZ},
\eq
%--
where $A^\mu_\ssR$ and $Z^\mu_\ssR$ are the renormalized fields for the photon
and the $Z$ boson. Note that $Z_{\ssA\ssA}$ is expanded through \eqn{exppar}.

In addition, bare fermion fields $\psi$ (we omit flavour labels) are written 
by means of bare left-handed and right-handed chiral fields $\psi^\ssL$ and 
$\psi^\ssR$. The latter are traded for renormalized fields  
$\psi^\ssL_{\ssR}$ and $\psi^\ssR_{\ssR}$ expanding the renormalization 
constants through \eqn{exppar},
%--
\bq
\psi^{\ssL,\ssR} = \frac{1}{2}\,(1 \pm \gamma^5)\psi, \qquad
\psi^{\ssL,\ssR} = Z^{1/2}_{\psi_{_{\ssL,\ssR}}}\,\psi_{\ssR}^{\ssL,\ssR}.
\label{fermRENchi}
\eq
%--
Finally, Faddeev-Popov ghost fields are not renormalized.
\eei
%--
\vspace{0.2cm}
%--

\noindent
\textbf{UV decompositions for Green functions}
\vspace{0.2cm}

\noindent
Before defining our renormalization scheme, we introduce UV decompositions 
also for Green functions.
Given a one- ($i=1$) or two-loop ($i=2$) Green function with $N$ external 
lines carrying Lorentz indices $\mu_j$, $j=1,\ldots,N$, we introduce 
form factors,
%--
\bq
G^{i}_{\mu_1\,\,\dots\,\,\mu_{\ssN}}\, =\,
\sum_{a=1}^{\ssA}
G^{i}_a \,K^a_{\mu_1\,\,\dots\,\,\mu_{\ssN}},\qquad i=1,2.
\label{eq:one:ASDF}
\eq
%--
Here the set $K^a$, with $a = 1,\dots,A$, contains independent tensor 
structures made up of external momenta, Kronecker-delta functions, elements 
of the Clifford algebra and Levi-Civita tensors.
Next, we introduce UV decompositions for the form factors $G^{i}_a$ in analogy
with \eqns{dec1loop}{UVfactors},
%--
\bq
\texttt{1 loop}\,\to
\,G^{1}_a = \sum_{k=-1}^{1}\,G^{1}_{a\,;\,k}\,F^1_k(M_\ssR^2),
\qquad \quad  \texttt{2 loops}\,\to\,
G^{2}_a = \sum_{k=-2}^{0}\,G^{2}_{a\,;\,k}\,F^2_k(M_\ssR^2).
\label{eq:one:polect11}
\eq
%--
The UV factors $F^1_k$ ($k=-1,0,1$) and $F^2_k$ ($k=-2,-1,0$) can be read
in \eqnsc{OLUVF}{UVfactors}.
%--
\vspace{0.2cm}
%--

\noindent
\textbf{Renormalization prescriptions: $\MSB$ and beyond}
\vspace{0.2cm}
%--

\noindent
In the spirit of the UV decomposition of \eqn{eq:one:polect11}, one could
define a non-minimal ($\NMSB$) subtraction scheme where the one- and two-loop 
counterterms are defined by
%--
\bq
\texttt{1 loop}\,\to
\,\delta Z_i^{(1)} = \,\Delta Z^{(1)}_{i}\,F^1_{-1}(M_\ssR^2),
\qquad \quad  \texttt{2 loops}\,\to\,
\delta Z^{(2)}_i = \sum_{k=-2}^{-1}\,\Delta Z^{(2)}_{i\,;\,k}\,F^2_k(M_\ssR^2),
\label{eq:one:ct11}
\eq
%--
and are fixed in order to remove order-by-order the poles at
$\ep=0$ for any Green function. As a result, also the product of a one-loop 
counterterm with a one-loop diagram (i.e., a one-loop counterterm insertion) 
has the same UV decomposition of a two-loop function thus simplifying two-loop 
renormalized Green functions. Even if we scale non-tadpole diagrams with 
$s$ instead of $M_\ssR^2$, the difference is proportional to 
$\ln M_\ssR^2/(s\, \ep)$, a non-local term which cancels in the total.  

Although the $\NMSB$ scheme has the virtue of respecting a universal
UV decomposition, we adopt the conventional $\MSB$ choice, keeping the
counterterms as simple as possible. Let us accordingly define
%--
\bq
\delta Z^{(1)}_i =
\lpar - \frac{2}{\ep} + C \rpar\,\Delta\,Z^{(1)}_i,
\label{eq:one:pole1ctaaa}
\eq
%--
and we define the renormalization scheme choosing the explicit value for $C$.
We define a minimal $\MSB$ subtraction scheme,
%--
\bq
\delta Z^{(1)}_i =
\lpar - \frac{2}{\ep} + \DUV \rpar\,\Delta\,Z^{(1)}_i,
\label{eq:one:pole1ctaaaDUV}
\eq
%--
and we fix the counterterms in order to remove the poles at $\ep=0$ for any 
one-loop Green function. Next, we extend the scheme at two loops by introducing
%--
\bq
\delta Z^{(2)}_i =
\lpar \frac{1}{\ep} - \DUV\rpar\,\lpar
\frac{\Delta Z^{(2)}_{i;1}}{\ep} + \Delta Z^{(2)}_{i;2}\rpar + 
\DUV^2\,\Delta Z^{(2)}_{i;3}.
\label{eq:one:pole1ct2a}
\eq
%--
Accordingly, we decompose the two-loop form factors,
%--
\bq
G^{2}_a =
\lpar \frac{1}{\ep^2} - \frac{\DUV}{\ep} \rpar\,G^{2}_{a\uv\,;\,1}
+ \lpar  \frac{1}{\ep} - \DUV \rpar\,G^{2}_{a\uv\,;\,2} +
\DUV^2\,G^{2}_{a\uv\,;\,3} + G^{2}_{a\Fin},
\label{eq:one:polect12}
\eq
%--
and we fix the counterterms in order to remove the poles at $\ep=0$ for any 
two-loop Green function. Note the correspondence with the decomposition in 
\eqn{eq:one:polect11},
%--
\bqa
G^{2}_{a\uv\,;\,1} &=& G^{2}_{a\,;\,-2},
\qquad
G^{2}_{a\uv\,;\,2} = G^{2}_{a\,;\,-1},
\nl
G^{2}_{a\uv\,;\,3} &=& \frac{1}{2}\,G^{2}_{a\,;\,-2},
\quad
G^{2}_{a\Fin} = G^{2}_{a\,;\,0}.
\eqa
%--
\section{One-loop renormalization \label{olR}}
%--
In this section we collect all the one-loop results necessary for our
renormalization procedure. Tadpole renormalization ($\beta$ vertices) and 
neutral-sector diagonalization ($\Gamma$ vertices) are shown in 
\subsect{subsec:bt1} and counterterms for parameters and fields are presented 
in \subsect{olCT}.

In \subsect{FiniteR} we discuss the relevance of one-loop finite
renormalization through some examples and in \subsect{sec:one:LABEL}
we analyze how counterterms affect WST identities. Renormalized WST
identities riafferm the basic simplicity of the one-loop structure of the
theory. These results should provide a useful guide to the two-loop
renormalization.
%--
\subsection{$\beta$ and $\Gamma$ at one loop}
\label{subsec:bt1}
%--
Two key ingredients are the parameter $\beta_t$, introduced in Sect.~2.3 
of I (tadpole renormalization), and the parameter $\Gamma$, defined in 
Sect.~3.1 of I (all-order diagonalization of the neutral sector). 
We recall that one can define different tadpole schemes, and
in I, we discussed the so-called $\beta_h$ and $\beta_t$ schemes; since here 
no ambiguities arise, we drop everywhere the subscript $t$ from $\beta$. To 
show the one-loop results, we expand $\beta$ and $\Gamma$ in the 
unrenormalized $SU(2)$ coupling constant $g$,
%--
\bq
\beta = \sum_{n=1}^{\infty}\, g^{2 n}\, \beta_{n},
\qquad
\Gamma = \sum_{n=1}^{\infty}\, g^{2 n}\, \Gamma_{n}.
\label{eq:one:ASDak}
\eq
%--
The complete answer for $\beta_{1}$ reads as
%--
\bqa
\beta_{1} &=& \frac{1}{16\, \pi^2 x_{\ssH}}
\Bigl\{  3 \sum_{i=1}^3 
  \Bigl[ x_{d,i}^2 a_0(m_{d,i}) + 
       x_{u,i}^2 a_0(m_{u,i}) \Bigr]
  + \sum_{i=1}^3 x_{l,i}^2 a_0(m_{l,i})
\nl
{}&-& \frac{1}{4 \ctw^2} \Bigl( 
\frac{n-1}{\ctw^2}+ \frac{x_{\ssH}}{2} \Bigr)
  a_0(M_0)  -  \frac{1}{2} \Bigl( n - 1 +\frac{x_{\ssH}}{2} \Bigr)
  a_0(M)
- \frac{3}{8} x_{\ssH}^2 a_0(\mh)\Bigr\},
\label{eq:one:beta1res}
\eqa
%--
where we recall that $x_i = m^2_i/M^2$, $l_i$ is the charged lepton and 
$u_i, d_i$ are the up and down quarks of the $i$th fermion doublet. The 
one-loop tadpole $a_0$ is given in \eqn{defAfun} and the explicit result for 
$\Gamma_1$ is
%--
\bq
\Gamma_1 = \frac{n-2}{16\, \pi^2} \, a_0(M).
\label{eq:one:gamma1res}
\eq
%--
Note that in \eqnsc{eq:one:beta1res}{eq:one:gamma1res} we dropped the 
subscript $R$, since bare quantities disappear after the introduction of 
renormalized parameters and fields.
%--
\subsection{One-loop counterterms \label{olCT}}
%--
In this subsection we provide the full list of one-loop counterterms in the 
$\MSB$ scheme. Counterterms are defined in \eqns{exppar}{fermRENchi}, and UV 
factors are extracted through \eqn{eq:one:pole1ctaaaDUV}. Here we introduce 
short-hand notations for sums over fermions ($l\to$ charged leptons, 
$u,d\to$ quarks), 
%--
\bq
X_{l}^{j} = \sum_{i=1}^3 x_{l,i}^{j}, \qquad \quad
X_{u}^{j} = \sum_{i=1}^3 x_{u,i}^{j}, \qquad \quad
X_{d}^{j} = \sum_{i=1}^3 x_{d,i}^{j}.
\label{eq:one:SUMMA}
\eq
%--
%--
\noindent
\underline{\emph{Gauge parameters.}}
Gauge-parameter counterterms can be expressed by means of field, mass and 
coupling-constant counterterms,
%--
\bqa
\Delta Z_{\gpA}^{(1)} &=&
\frac{1}{2}\, \Delta Z_{\ssA\ssA}^{(1)},
\qquad\qquad
\Delta Z_{\gpAZ}^{(1)} = - \,\Delta Z_{\ssA\ssZ}^{(1)},
\nl
\Delta Z_{\gpZ}^{(1)} &=& \frac{1}{2}\, \Delta Z_{\ssZ}^{(1)},
\qquad\qquad
\Delta Z_{\xi_{\varphi^0}}^{(1)} =  -\, \frac{1}{2}\,
\Bigl( \, \Delta Z_{\varphi^0}^{(1)} \,+\, \Delta Z_{\ssM}^{(1)}\, \Bigr)\,
+\, \Delta Z_{\ctw}^{(1)},
\nl 
\Delta Z_{\gpW}^{(1)} &=& \frac{1}{2} \,\Delta Z_{\ssW}^{(1)},
\qquad\qquad
\Delta Z_{\xi_{\varphi}}^{(1)} = -\, \,\frac{1}{2}\,\Bigl(\,
\Delta Z_{\varphi}^{(1)}\, +\, \Delta Z_{\ssM}^{(1)}\, \Bigr).
\label{eq:one:oneCOUNTER}
\eqa
The gauge-fixing term defined in \eqn{eq:th:gfun} is invariant under 
renormalization. We will discuss this issue in \subsect{sec:one:LABEL}.
\vspace{0.2cm}
%--

\noindent
\underline{\emph{Gauge-boson and Higghs-Kibble fields.}}
%--
\bqa
\Delta Z^{(1)}_{\ssA\ssA} &=& 
\frac{23}{3} \stw^2,\qquad\qquad\qquad\qquad\ \ \ \
\Delta Z_{\ssA\ssZ}^{(1)} = - \frac{\stw}{3}
   \Bigl( \frac{41}{2}\frac{1}{ \ctw} - 23 \, \ctw \Bigr),
\nl
\Delta Z_{\ssZ}^{(1)} &=& \frac{1}{3}
   \Bigl( \frac{41}{2}\frac{1}{ \ctw^2} - 41 + 23 \, \ctw^2 \Bigr),\qquad
\Delta Z_{\varphi^0}^{(1)} =
   - 1 - \frac{1}{2} 
   \Bigl[ \frac{1}{\ctw^2} - X_{l} - 3 \, \Bigl( X_{d} + X_{u} \Bigr) \Bigl],
\nl
\Delta Z^{(1)}_{\ssW} &=& \frac{5}{6},\qquad\qquad\qquad\qquad\qquad\ \ 
\Delta Z^{(1)}_{\varphi} = \Delta Z^{(1)}_{\varphi^0}.
\label{csymI}
\eqa
%--
\noindent
\underline{\emph{Masses and couplings.}}
Here we discover a nice feature of the neutral-sector diagonalization
introduced in Sect.~3.1 of I, since in the full-fledged Standard Model we 
derive a QED-like relation involving $\Delta Z_g^{(1)}$,
$\Delta Z_{\stw}^{(1)}$ and $\Delta Z_{\ssA\ssA}^{(1)}$,
%--
\bqa
\Delta Z^{(1)}_{\ssM} &=&
    - \frac{3}{4}\frac{1}{ \ctw^2}
    - \frac{7}{3}
    + \frac{1}{x_{\ssH}}
    \Bigl[ 
       \frac{3}{2}\frac{1}{ \ctw^4} + 3  - 2 \, X_{l}^2 - 6 \,\Bigl( X_{d}^2 
+ X_{u}^2 \Bigr)
    \Bigr] + \frac{1}{2} \Bigl[ \frac{3}{2} x_{\ssH} + X_{l}
    + 3 \, \Bigl( X_{d} + X_{u} \Bigr)\Bigr],
\nl
\Delta Z^{(1)}_{\ctw} &=& \frac{1}{2}
  \Bigl(   \frac{41}{6}\frac{1}{ \ctw^2} - \frac{29}{2} + \frac{23}{3} \,
\ctw^2 \Bigr),\qquad
\Delta Z^{(1)}_{\stw} = - \frac{\ctw^2}{\stw^2} \, \Delta Z^{(1)}_{\ctw},\qquad
\Delta Z_g^{(1)} = - \Delta Z_{\stw}^{(1)} - 
\frac{1}{2}\, \Delta Z_{\ssA\ssA}^{(1)}.
\label{eq:one:orrendo}
%--
\eqa
In other words, electric-charge renormalization depends only on the photon 
vacuum-polarization function, also in the (complete) Standard Model.
\vspace{0.2cm}
%--

\noindent
\underline{\emph{Higgs-boson field and mass.}}
%--
\bqa
\Delta Z^{(1)}_{\ssH} &=&  - 1 - \frac{1}{2}
  \Bigl[ \frac{1}{\ctw^2}  - X_{l}  - 3 \,\Bigl(  X_{d} + X_{u}  \Bigr)
  \Bigr],
\quad
\Delta Z^{(1)}_{\ssM_{\ssH}} = \frac{3}{2}
  \Bigl[ \frac{1}{2}\frac{1}{\ctw^2} + 1 - \frac{1}{2}\, x_{\ssH}
  - \frac{1}{3}\, X_{l}  - \Bigl( X_{d} + X_{u} \Bigr)  \Bigr].
\label{eq:one:oneCOUNTERstop}
\eqa
%--
\noindent
\underline{\emph{Fermion fields and masses.}}
%--
Here $\nu$ is the neutrino field associated with the charged lepton $l$,
and $u$ and $d$ are up and down quarks belonging to the same doublet. 
Furthermore, $g_s$ is the $SU(3)$ coupling constant of strong interactions.
For left-handed spinor-field counterterms we get
%--
\bqa
\Delta Z_{\nu_L}^{(1)} &=&
 \frac{1}{2}\,\Bigl( \frac{1}{2}\frac{1}{\ctws} + 
1 + \frac{1}{2}\, x_{l}  \Bigr),\qquad
%--
\Delta Z_{u_L}^{(1)} =
  \frac{1}{4} \Bigl(
  \frac{1}{9}\,\frac{1}{ \ctws} + \frac{26}{9}
  + x_u + x_d + \frac{16}{3}\,\frac{g^2_s}{g^2} \Bigr),
%--
\nl
%--
\Delta Z_{l_L}^{(1)} &=& \Delta Z_{\nu_L}^{(1)},
  \qquad\qquad\qquad\ \ \ \ \ \
%--
\Delta Z_{d_L}^{(1)} =\Delta Z_{u_L}^{(1)}.
\label{eq:one:FER1L}
\eqa
%--
Expressions for the right-handed components are given by
%--
\bqa
\Delta Z_{\nu_R}^{(1)} &=& 0, \qquad\qquad\qquad\qquad
%--
\Delta Z_{u_R}^{(1)} =
  \frac{4}{9}\frac{1}{ \ctws}
  - \frac{4}{9}
  + \frac{1}{2}\,x_u + \frac{4}{3}\,\frac{g^2_s}{g^2},
%--
\nl
%--
\Delta Z_{l_R}^{(1)} &=&
  \frac{1}{\ctws} - 1 +  \frac{1}{2}\, x_l,\qquad\ \
%--
\Delta Z_{d_R}^{(1)} =
  \frac{1}{9}\frac{1}{ \ctws}
  - \frac{1}{9}
  + \frac{1}{2}\,x_d + \frac{4}{3}\,\frac{g^2_s}{g^2}.
\label{eq:one:FER1R}
\eqa
%---
Finally, fermion-mass counterterms depend on the sum over the fermion families
because of tadpole renormalization (the explicit expression for $\beta_1$ can 
be found in \eqn{eq:one:beta1res}),
%--
\bqa
\Delta Z_{m_l}^{(1)} &=&
  3 \, \frac{\stw^2}{\ctw^2}
  + \frac{1}{x_{\ssH}}
  \Bigl[ \frac{3}{2}\frac{1}{ \ctw^4} + 3  - 2\, X_{l}^2
  - 6 \,\Bigl( X_{u}^2 + X_{d}^2 \Bigr)
  \Bigr]
  + \frac{3}{4} \Bigl( x_{\ssH} - x_{l} \Bigr),
\nl
\Delta Z_{m_u^{(1)}} &=&
  \frac{2}{3}\,\frac{\stw^2}{\ctw^2}
  + \frac{1}{x_{\ssH}} \Bigl[
  \frac{3}{2}\frac{1}{ \ctw^4} + 3 
  - 2\, X_{l}^2 - 6\, \Bigl( X_{u}^2 + X_{d}^2 \Bigr)
  \Bigr]
  + \frac{3}{4} \Bigl(
  x_{\ssH} - x_{u} + x_{d}
  \Bigr)  + 8\,\frac{g^2_s}{g^2},
\nl
\Delta Z_{m_d}^{(1)} &=&
  - \frac{1}{3}\,\frac{s^2_{\theta}}{\ctw^2}
  + \frac{1}{x_{\ssH}} \Bigl[
  \frac{3}{2}\frac{1}{ \ctw^4}+ 3    - 2\, X_{l}^2
  -6 \,\Bigl( X_{u}^2 + X_{d}^2 \Bigr)
  \Bigr]
  + \frac{3}{4} \Bigl( x_{\ssH} + x_{u} - x_{d} \Bigr)
  + 8\,\frac{g^2_s}{g^2}.
\label{eq:one:FER2}
\eqa
%--
%--
\subsection{Finite renormalization \label{FiniteR}}
%--
The last step in one-loop renormalization is the connection between 
renormalized quantities and POs. Since all quantities at this stage are
UV-free, we yerm it \emph{finite renormalization}.
Note that the absorption of UV divergencies into local counterterms is, to 
some extent, a trivial step; finite renormalization, instead, requires more 
attention. For example, beyond one loop one cannot use on-shell masses but 
only complex poles for all unstable particles.
The complete formulation of finite renormalization will be given in III.
However, let us show some examples where the concept of an on-shell can be 
employed. Suppose that we renormalize a physical (pseudo-)observable $F$,
%--
\bq
F = F_{\ssB} + g^2\,F_{1\ssL}(M^2) + g^4\,F_{2\ssL}(M^2),
\eq
%--
where $M$ is some renormalized mass which appears at one and two loops in
$F_{1\ssL}$ and $F_{2\ssL}$ but does not show up in the Born term $F_{\ssB}$.
In this case we can use the concept of an on-shell mass identifying
$M = M_{\ssO\ssS}$ for the two-loop term and performing a finite mass 
renormalization at one loop,
%--
\bq
M^2 = M^2_{\ssO\ssS} \, \Bigl\{ 1\, +\, \frac{g^2}{16\,\pi^2}\,\Bigl[\,
\Reb\,\Sigma^{(1)}_{\ssM}\bmid_{p^2=-M^2_{\ssO\ssS}} - 
\,\delta Z^{(1)}_{\ssM} \Bigr]\, \Bigr\} =
M^2_{\ssO\ssS} + g^2\,\Delta M^2,
\label{stillOMS}
\eq
%--
where $M_{\ssO\ssS}$ is the on-shell mass and $\Sigma$ is extracted
from the required one-particle irreducible Green function. \eqn{stillOMS} 
is still meaningful (no dependence on gauge parameters) and will be used 
inside the one-loop result,
%--
\bq
F = F_{\ssB} + g^2\,F_{1\ssL}(M^2_{\ssO\ssS}) + g^4\,\Bigl[
F_{2\ssL}(M^2_{\ssO\ssS}) + F'_{1\ssL}(M^2_{\ssO\ssS})\,\Delta M^2\Bigr],
\eq
%--
where
%--
\bq
F'_{1\ssL}(M^2_{\ssO\ssS})\, = \,
\frac{\partial F_{1\ssL}(M^2)}{\partial M^2}\vert_{M^2=M^2_{\ssO\ssS}}.
\eq
%--
Here we show some examples of one-loop finite renormalization.
\vspace{0.2cm}
%--

\noindent
\underline{\textbf{$W$ boson.}}
For the $W$ boson we write
%--
\bq
M^2 = M_{\ssW}^2\,\Bigl[
1 + \frac{g^2}{16 \pi^2} \Bigl( \Reb\, \Sigma_{\ssW\ssW} - 
\delta Z_{\ssM}^{(1)} \Bigr) \Bigr],
\label{OSW}
\eq
%--
where the expression for the counterterm can be read in \eqn{eq:one:orrendo}.
The quantity within square brackets in \eqn{OSW} is finite by construction,
and it is convenient to write the decomposition
%--
\bq\label{eq:alpha:Ret}
\Sigma_{\ssW\ssW} = \Sigma_{\ssW\ssW}^{f} + \Sigma_{\ssW\ssW}^{b} -
2 ( \beta_{1} + \Gamma_{1} ).
\eq
%--
$\beta_{1}$ and $\Gamma_{1}$ can be read in \eqn{eq:one:beta1res}
and \eqn{eq:one:gamma1res}, whereas the first two terms in the r.h.s. of 
\eqn{eq:alpha:Ret} refer to fermionic and bosonic contributions.
The fermionic part for one generation (one leptonic family, $x_l = 
m_l^2\slash M^2$, and one up-down quark doublet, $x_u = m_u^2\slash M^2$,
$x_d = m_d^2\slash M^2$) is
%--
\bqa
\Sigma_{\ssW\ssW}^{f(1)} &=&
\frac{1}{n-1}
\Bigl\{
\frac{1}{2}
\Bigl(n-2-x_l\Bigr) x_l a_0(m_l) \nl
{}&+& \frac{3}{2}
\Bigl( x_u - x_d + n -2 \Bigr) x_d a_0(m_d) +
\Bigl( x_d - x_u + n -2 \Bigr) x_u a_0(m_u) \nl
{}&+& \frac{3}{2}
\Bigl[
\Bigl( n-1 \Bigr)
\Bigl( 1 - x_d - x_u \Bigr)
+ 2 \Bigl( x_d + x_u \Bigr) - \Bigl( x_d - x_u \Bigr)^2
\Bigr] b_{0}(1,1\,,\,M^2\,,\,m_u,m_d)\nl
{}&+& \frac{1}{2} \Bigl[
\Bigl( n-1 \Bigr) \Bigl( 1 - x_l \Bigr)
+ 2 x_l - x_l^2 - 1
\Bigr] b_{0}(1,1\,,\,M^2\,,\,0,m_l)
\Bigr\},
\eqa
%--
and the bosonic component reads as follows,
%--
\bqa
\Sigma_{\ssW\ssW}^{b} &=&
\frac{1}{4 \ctw^2}
\Bigl[
4 \Bigl(n-2\Bigr) \stw^2 + \frac{1}{n-1}\frac{1}{\ctw^2}
- 4 n - \frac{4}{n-1} + 11 \Bigr] a_0(M_0)
\nl
{}&+& \frac{1}{4} \Bigl( 10 - \frac{1}{n-1}\frac{1}{\ctw^2}
- 4 n - \frac{1}{n-1} x_{\ssH}   \Bigr)a_0(M)
+ \frac{1}{4} \Bigl( \frac{1}{n-1}x_{\ssH} -1
\Bigr) x_{\ssH} a_0(\mh)
\nl
{}&+& \frac{1}{n-1}
\Bigl[
\frac{1}{4 \ctw^{4}} - \frac{2}{\ctw^2} + 4 + \Bigl( n-1 \Bigr)
\Bigl( - 11 + 4 \stw^2 + \frac{2}{\ctw^2} \Bigr) \Bigr]
 b_{0}(1,1\,,\,M^2\,,\,M_0,M)
\nl
{}&+& \frac{1}{n-1} \Bigl( \frac{x_{\ssH}^2}{4} - x_{\ssH} + n -1
\Bigr) b_{0}(1,1\,,\,M^2\,,\,M,\mh) - 4 \stw^2 b_{0}(1,1\,,\,M^2\,,\,0,M).
\eqa
%--
\noindent
\underline{\textbf{Fermions.}}
For a fermion $f$ we obtain
%--
\bq
\label{eq:alpha:Arid}
m_f^2 = m_{f,{\ssO\ssS}}^2\,\Bigl[ 1 + \frac{g^2}{8 \pi^2}
\Bigl( \Sigma_{f} - \frac{\delta Z_{m,f}^{(1)}}{2} \Bigr)\Bigr],
\label{OSf}
\eq
%--
where the fermion-mass counterterms can be read in \eqn{eq:one:FER2},
and the physical mass in \eqn{eq:alpha:Arid} coincides with the on-shell one.
Using
%--
\bq
Q_l = -1, \, Q_u = \frac{2}{3}, \, Q_d = -\,\frac{1}{3}, 
\; N^c_l = 1, \, N^c_q = 3,
\qquad v_f = I^{(3)}_f - 2\,Q_f\,\stws,
\eq
%--
where $I^{(3)}_f$ is the third component of isospin, we provide the needed 
expressions for arbitrary up and down partners in a given fermion doublet. 
For a generic up-type fermion we get
%--
\bqa
\Sigma_{u} &=&
- \frac{1}{8} x_{\ssH} a_0(\mh)
- \frac{1}{8} \Bigl[  
1 + \frac{1}{x_u}\Bigl( n-2+x_d\Bigr)
\Bigr] a_0(M)
- \frac{1}{8 \ctw^2} \Bigl[ 1 + \frac{1}{x_u \ctw^2}
\Bigl( n-2 \Bigr) \Bigr( v_{u}^2 + \frac{1}{4} \Bigl) 
\Bigr] a_0(M_0)
\nl
{}&+& \frac{1}{8} \Bigl\{
\Bigl( n-2 \Bigr)
\Bigl[ 4 \stw^2 Q_{u}^2 + \frac{1}{\ctw^2} \Bigl(v_{u}^2 + \frac{1}{4} \Bigr)
\Bigr] + 2 x_u \Bigr\} a_0(m_u)
+ \frac{1}{8} \Bigl[
1 + \frac{1}{x_u} \Bigl( n -2 + x_d \Bigr)
\Bigr] x_d a_0(m_d) 
\nl
{}&+& \frac{1}{8} \Bigl\{ n-3 + \frac{1}{x_u} \Bigl[
\Bigl( n-3 \Bigr) x_d - n + 2 + x_d^2 \Bigr] + x_u - 2\ x_d
\Bigr\} b_{0}(1,1\,,\,m_u^2\,,\,M,m_d)
\nl
{}&+& \frac{1}{8 \ctw^2}
\Bigl\{
1 + \Bigl( n-2 \Bigr)
\Bigl[1 - \frac{1}{\ctw^2 x_u} \Bigl(v_{u}^2 + \frac{1}{4} \Bigr)
\Bigr]
- 4 \Bigl(v_{u}^2 + \frac{1}{4} \Bigr)
\Bigr\} b_{0}(1,1\,,\,m_u^2\,,\,M_0,m_u)\nl
{}&+& \frac{1}{8} \Bigl( 4 x_u - x_{\ssH} \Bigr)
b_{0}(1,1\,,\,m_u^2\,,\,\mh,m_u)
- 2\ \stw^2 Q_{u}^2 b_{0}(1,1\,,\,m_u^2\,,\,0,m_u)
- \beta_{1}.
\eqa
%--
\noindent
For a d-type fermion, the result reads as
%--
\bqa
\Sigma_{d} &=&
- \frac{1}{8} x_{\ssH} a_0(\mh)
- \frac{1}{8} \Bigl[  
1 + \frac{1}{x_d}\Bigl( n-2+x_u\Bigr)
\Bigr] a_0(M)
-\frac{1}{8 \ctw^2} \Bigl[ 1 + \frac{1}{x_d \ctw^2}
\Bigl( n-2 \Bigr) \Bigr( v_{d}^2 + \frac{1}{4} \Bigl) 
\Bigr]a_0(M_0)
\nl
{}&+& \frac{1}{8} \Bigl\{
\Bigl( n-2 \Bigr)
\Bigl[
4 \stw^2 Q_{d}^2 + \frac{1}{\ctw^2}
\Bigl(v_{d}^2 + \frac{1}{4} \Bigr)
\Bigr]
+ 2 x_d
\Bigr\} a_0(m_d)
+ \frac{1}{8} \Bigl[
1 + \frac{1}{x_d} \Bigl( n -2 + x_u \Bigr)
\Bigr] x_u a_0(m_u) \nl
{}&+& \frac{1}{8}
\Bigl\{
n-3 + \frac{1}{x_d} \Bigl[
\Bigl( n-3 \Bigr) x_u - n + 2 + x_u^2
\Bigr]
+ x_d - 2\ x_u
\Bigr\} b_{0}(1,1\,,\,m_d^2\,,\,M,m_u)\nl
{}&+& \frac{1}{8 \ctw^2}
\Bigl\{
1 + \Bigl( n-2 \Bigr)
\Bigl[1 - \frac{1}{\ctw^2 x_d} \Bigl(v_{d}^2 + \frac{1}{4} \Bigr)
\Bigr]
- 4 \Bigl(v_{d}^2 + \frac{1}{4} \Bigr)
\Bigr\} b_{0}(1,1\,,\,m_d^2\,,\,M_0,m_d)\nl
{}&+& \frac{1}{8} \Bigl( \frac{1}{2} x_d - x_{\ssH} \Bigr)
b_{0}(1,1\,,\,m_d^2\,,\,\mh,m_d)
- 2\ \stw^2 Q_{d}^2 b_{0}(1,1\,,\,m_d^2\,,\,0,m_d)
- \beta_{1}.
\eqa
%--
\noindent
\underline{\textbf{Higgs boson.}}
%--
We also provide the result for the Higgs-boson mass,
%--
\bq
\mhs = M_{\ssH,{\ssO\ssS}}^2\,\Bigl[
1 + \frac{g^2}{16 \pi^2} \Bigl(
\Reb\, \Sigma_{\ssH\ssH} - \delta Z_{\mh}^{(1)} \Bigr) \Bigr],
\label{OSH}
\eq
%--
where the counterterm can be found in \eqn{eq:one:oneCOUNTERstop},
and we use a decomposition similar to \eqn{eq:alpha:Ret},
$\Sigma_{\ssH\ssH} = \Sigma_{\ssH\ssH}^{f} + \Sigma_{\ssH\ssH}^{b}$.
For one generation of fermions we get
%-
\bq
\Sigma_{\ssH\ssH}^{f(1)} =
3\,\sum_{f=l,u,d}\,\frac{x^2_f}{x_{\ssH}}\,a_0(m_f)
%\frac{1}{x_{\ssH}}
%\Bigl[
%3 \ x_u^2 a_0(m_u) + 3 \ x_d^2 a_0(m_d) + 
%x_l^2 a_0(m_l)
%\Bigr]
%\nl
+ \sum_{f=l,u,d}\,N^c_f\,x_f\,
\Bigl( \frac{1}{2} - 2 \frac{x_f}{x_{\ssH}} \Bigr)\,
 b_{0}(1,1\,,\,\mh^2\,,\,m_f,m_f)
%{}&+& 3\,\sum_{q=u,d}\,x_q \Bigl( \frac{1}{2} - 2 \frac{x_q}{x_{\ssH}}
%\Bigr) b_{0}(1,1\,,\,\mh^2\,,\,m_q,m_q)
%+ x_l \Bigl( \frac{1}{2} - 2 \frac{x_l}{x_{\ssH}}
%\Bigr) b_{0}(1,1\,,\,\mh^2\,,\,m_l,m_l),
\eq
%--
whereas the bosonic component reads as follows,
%--
\bqa\label{eq:alpha:Jiu}
\Sigma_{\ssH\ssH}^{b} &=&
\frac{1}{x_{\ssH}} \Bigl( n - 1 + \frac{x_{\ssH}}{2}\Bigr)
a_0(M) + \frac{1}{\ctw^2 x_{\ssH}} \Bigl(
\frac{n-1}{2 \ctw^2} + \frac{x_{\ssH}}{4} \Bigr) a_0(M_0)
+ \frac{3}{4} x_{\ssH} a_0(\mh) 
\nl
{}&-& 3\,\sum_{i=1,3}\sum_{f=l,u,d}\,N^c_{f,i}\,\frac{x^2_{f,i}}{x_{\ssH}}\,
a_0(m_{f,i})
%
%{}&-& \frac{3}{x_{\ssH}}
%\sum_{i=1}^3 \Bigl[
%3  x_{u,i}^2 a_0(m_{u,i})+ 3  x_{d,i}^2 a_0(m_{d,i})
%+ x_{l,i}^2 a_0(m_{l,i}) \Bigr]
%
- \Bigl[ 1 - \frac{1}{x_{\ssH}} \Bigl( n-1+\frac{x_{\ssH}^2}{4}
\Bigr) \Bigr]
b_{0}(1,1\,,\,\mh^2\,,\,M,M)
\nl
{}&-&\Bigl[
\frac{1}{2} \frac{1}{\ctw^2} - \frac{1}{x_{\ssH}} \Bigl(
\frac{n-1}{2 \ctw^{4}} +\frac{x_{\ssH}^2}{8} \Bigr) \Bigr]
b_{0}(1,1\,,\,\mh^2\,,\,M_0,M_0)
+ \frac{9}{8} x_{\ssH}^2 b_{0}(1,1\,,\,\mh^2\,,\,\mh,\mh).
\eqa
%--
The sum over the fermion masses in \eqn{eq:alpha:Jiu} depends on the 
$\beta_{1}$ parameter (see \eqn{eq:one:beta1res}), which we included 
explicitly in the computation of the Higgs-boson self-energy.
%--
\subsection{Ward-Slavnov-Taylor identities}\label{sec:one:LABEL}
%--
Ward-Slavnov-Taylor (WST) identities play a crucial role in showing that the 
electroweak theory is renormalizable and respects unitarity at the same 
time. In fact, we can prove with their help that transition amplitudes are 
the same in every non-singular gauge.
Having derived the one-loop counterterms, we analyze how renormalization 
affects the WST identities of the theory; note that here we consider only
a particular class of WST identities, the so-called \emph{doubly-contracted} 
ones with two external gauge-boson fields.
Let us consider the gauge-fixing functions ${\cal{C}}^{a}$ of \eqn{eq:th:gfun} 
written in terms of bare quantities,
%--
\bqa\label{eq:one:GAUGEbac}
{\cal{C}}^{\ssA} &=&
- \frac{1}{\gpA^{b}}
\partial_{\mu} A_{\mu}^{b}
- \gpAZ^{b}
\partial_{\mu} Z_{\mu}^{b},
\quad
{\cal{C}}^{\ssZ} = 
- \frac{1}{\gpZ^{b}} \partial_{\mu} Z_{\mu}^{b} 
+ \xi_{\varphi^{0}}^{b}
  M^b_0 \varphi^{b,0},
\quad
{\cal{C}}^{\pm} =
- \frac{1}{\gpW^{b}} \partial_{\mu} W^{b,\pm}_{\mu} 
+ \xi_{\varphi}^{b}
  M^b \varphi^{b, \pm}.
\eqa 
%--
Then, we introduce renormalized quantities and
we set the renormalized gauge parameters to unity,
%--
\bqa\label{eq:one:GAUGEREN}
{\cal{C}}^{\ssA} &=&
- Z_{\ssA\ssA}^{\frac{1}{2}} Z^{-1}_{\gpA}
\partial_{\mu} A_{\mu} - \Bigl(
Z_{\ssA\ssZ}^{\frac{1}{2}} Z^{-1}_{\gpA}
+ Z_{\ssZ}^{\frac{1}{2}} Z_{\gpAZ} 
\Bigr)
\partial_{\mu} Z_{\mu},
\nl
{\cal{C}}^{\ssZ} &=& 
 -Z_{\ssZ}^{\frac{1}{2}} Z_{\gpZ}^{-1}\partial_{\mu} Z_{\mu} 
+ Z_{\varphi^{0}}^{\frac{1}{2}} Z_{\ssM}^{\frac{1}{2}}
  Z_{c_\theta}^{-1}
 Z_{\xi_{\varphi^{0}}}
  M_0 \varphi^{0},
\nl
{\cal{C}}^{\pm} &=& 
 -Z_{\ssW}^{\frac{1}{2}} Z_{\gpW}^{-1}\partial_{\mu} W_{\mu}^{\pm} 
+ Z_{\varphi}^{\frac{1}{2}} Z_{\ssM}^{\frac{1}{2}} Z_{\xi_{\varphi}}
  M \varphi^{\pm}.
\eqa 
%--
Next, we associate a set of \emph{source} vertices to the Fourier transform
of a gauge-fixing function ${\cal{C}}^{a}$.

\begin{figure}[ht]
%\framebox[14.5cm]{\parbox{10cm}{
\begin{eqnarray*}
\hbox{
\begin{picture}(80,20)(0,0)
\SetScale{1}
\SetWidth{1}
\Text(30,10)[b]{\scriptsize{$A$}}
\Text(70,10)[b]{\scriptsize{$A$}}
\Photon(20,0)(40,0){3}{4}
\Photon(60,0)(80,0){3}{4}
\CCirc(50,0){10}{Black}{Gray}
\CBoxc(20,0)(4,4){Black}{Black}
\CBoxc(80,0)(4,4){Black}{Black}
\Text(50,0)[]{\scriptsize{$2$}}
\Text(90,0)[l]{\scriptsize{$=0$}}
\end{picture}
}
\qquad\qquad&&\\
\qquad&&\\
\qquad&&\\
\hspace{-1.5cm}
\hbox{
\begin{picture}(80,20)(0,0)
\SetScale{1}
\SetWidth{1}
\Text(30,10)[b]{\scriptsize{$Z$}}
\Text(70,10)[b]{\scriptsize{$Z$}}
\Photon(20,0)(40,0){3}{4}
\Photon(60,0)(80,0){3}{4}
\CCirc(50,0){10}{Black}{Gray}
\Text(50,0)[]{\scriptsize{$2$}}
\CBoxc(20,0)(4,4){Black}{Black}
\CBoxc(80,0)(4,4){Black}{Black}
\Text(90,0)[l]{\scriptsize{$+$}}
\end{picture}
\begin{picture}(80,20)(0,0)
\SetScale{1}
\SetWidth{1}
\Text(30,10)[b]{\scriptsize{$Z$}}
\Text(70,10)[b]{\scriptsize{$\varphi^0$}}
\Photon(20,0)(40,0){3}{4}
\DashLine(60,0)(80,0){3}
\CCirc(50,0){10}{Black}{Gray}
\Text(50,0)[]{\scriptsize{$2$}}
\CBoxc(20,0)(4,4){Black}{Black}
\CBoxc(80,0)(4,4){Black}{Black}
\Text(90,0)[l]{\scriptsize{$+$}}
\end{picture}
\begin{picture}(80,20)(0,0)
\SetScale{1}
\SetWidth{1}
\Text(30,10)[b]{\scriptsize{$\varphi^0$}}
\Text(70,10)[b]{\scriptsize{$Z$}}
\DashLine(20,0)(40,0){3}
\Photon(60,0)(80,0){3}{4}
\CCirc(50,0){10}{Black}{Gray}
\Text(50,0)[]{\scriptsize{$2$}}
\CBoxc(20,0)(4,4){Black}{Black}
\CBoxc(80,0)(4,4){Black}{Black}
\Text(90,0)[l]{\scriptsize{$+$}}
\end{picture}}&&
\hspace{-0.7cm}
\hbox{
\begin{picture}(80,20)(0,0)
\SetScale{1}
\SetWidth{1}
\SetScale{1}
\Text(30,10)[b]{\scriptsize{$\varphi^0$}}
\Text(70,10)[b]{\scriptsize{$\varphi^0$}}
\DashLine(20,0)(40,0){3}
\DashLine(60,0)(80,0){3}
\CCirc(50,0){10}{Black}{Gray}
\Text(50,0)[]{\scriptsize{$2$}}
\CBoxc(20,0)(4,4){Black}{Black}
\CBoxc(80,0)(4,4){Black}{Black}
\Text(90,0)[l]{\scriptsize{$=0$}}
\end{picture}
}\\
\qquad&&\\
\qquad&&\\
\hspace{-1.5cm}
\hbox{
\begin{picture}(80,20)(0,0)
\SetScale{1}
\SetWidth{1}
\Text(30,10)[b]{\scriptsize{$W$}}
\Text(70,10)[b]{\scriptsize{$W$}}
\Photon(20,0)(40,0){3}{4}
\Photon(60,0)(80,0){3}{4}
\CCirc(50,0){10}{Black}{Gray}
\Text(50,0)[]{\scriptsize{$2$}}
\CBoxc(20,0)(4,4){Black}{Black}
\CBoxc(80,0)(4,4){Black}{Black}
\Text(90,0)[l]{\scriptsize{$+$}}
\end{picture}
\begin{picture}(80,20)(0,0)
\SetScale{1}
\SetWidth{1}
\Text(30,10)[b]{\scriptsize{$W$}}
\Text(70,10)[b]{\scriptsize{$\varphi$}}
\Photon(20,0)(40,0){3}{4}
\DashLine(60,0)(80,0){3}
\CCirc(50,0){10}{Black}{Gray}
\Text(50,0)[]{\scriptsize{$2$}}
\CBoxc(20,0)(4,4){Black}{Black}
\CBoxc(80,0)(4,4){Black}{Black}
\Text(90,0)[l]{\scriptsize{$+$}}
\end{picture}
\begin{picture}(80,20)(0,0)
\SetScale{1}
\SetWidth{1}
\Text(30,10)[b]{\scriptsize{$\varphi$}}
\Text(70,10)[b]{\scriptsize{$W$}}
\DashLine(20,0)(40,0){3}
\Photon(60,0)(80,0){3}{4}
\CCirc(50,0){10}{Black}{Gray}
\Text(50,0)[]{\scriptsize{$2$}}
\CBoxc(20,0)(4,4){Black}{Black}
\CBoxc(80,0)(4,4){Black}{Black}
\Text(90,0)[l]{\scriptsize{$+$}}
\end{picture}}
&&
\hspace{-0.7cm}
\hbox{
\begin{picture}(80,20)(0,0)
\SetScale{1}
\SetWidth{1}
\Text(30,10)[b]{\scriptsize{$\varphi$}}
\Text(70,10)[b]{\scriptsize{$\varphi$}}
\DashLine(20,0)(40,0){3}
\DashLine(60,0)(80,0){3}
\CCirc(50,0){10}{Black}{Gray}
\Text(50,0)[]{\scriptsize{$2$}}
\CBoxc(20,0)(4,4){Black}{Black}
\CBoxc(80,0)(4,4){Black}{Black}
\Text(90,0)[l]{\scriptsize{$=0$}}
\end{picture}
}\\
\qquad&&\\
\qquad&&\\
\hspace{-1.5cm}
\hbox{
\begin{picture}(80,20)(0,0)
\SetScale{1}
\SetWidth{1}
\Text(30,10)[b]{\scriptsize{$A$}}
\Text(70,10)[b]{\scriptsize{$Z$}}
\Photon(20,0)(40,0){3}{4}
\Photon(60,0)(80,0){3}{4}
\CCirc(50,0){10}{Black}{Gray}
\Text(50,0)[]{\scriptsize{$2$}}
\CBoxc(20,0)(4,4){Black}{Black}
\CBoxc(80,0)(4,4){Black}{Black}
\Text(90,0)[l]{\scriptsize{$+$}}
\end{picture}
\begin{picture}(80,20)(0,0)
\SetScale{1}
\SetWidth{1}
\Text(30,10)[b]{\scriptsize{$A$}}
\Text(70,10)[b]{\scriptsize{$\varphi^0$}}
\Photon(20,0)(40,0){3}{4}
\DashLine(60,0)(80,0){3}
\CCirc(50,0){10}{Black}{Gray}
\Text(50,0)[]{\scriptsize{$2$}}
\CBoxc(20,0)(4,4){Black}{Black}
\CBoxc(80,0)(4,4){Black}{Black}
\Text(90,0)[l]{\scriptsize{$=0$}}
\end{picture}}&&
\end{eqnarray*}
\vspace{0.4cm}
%}}
\caption[]{Doubly-contracted WST identities with two external gauge bosons at 
$\ord{g^2}$. Gray circles denote the sum of the needed Feynman diagrams at 
$\ord{g^2}$. Source vertices are given in \fig{fig:one:Fsources}.}
\label{fig:one:WIbare}
\end{figure}
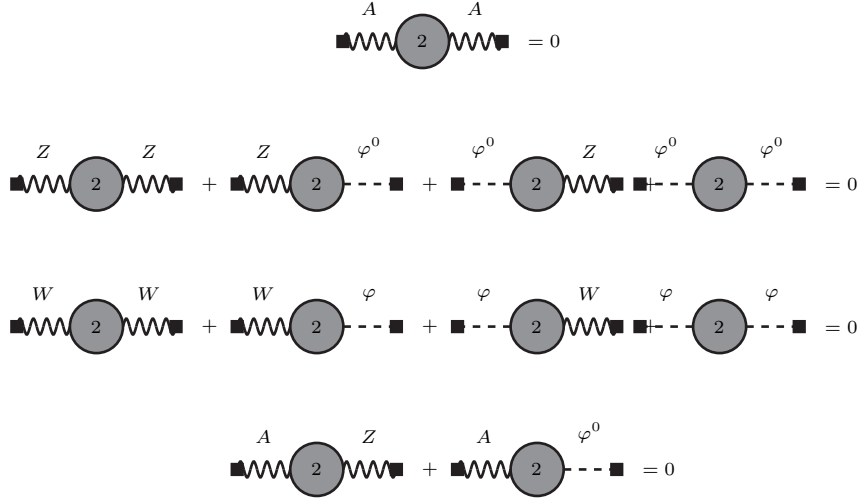

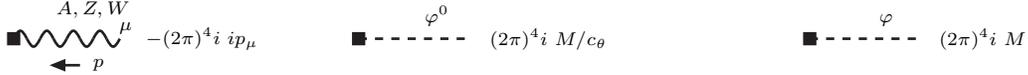
\begin{figure}[ht]
\SetWidth{1}
\begin{center}
%\framebox{
\begin{picture}(400,40)(0,-10)
\CBoxc(0,10)(4,4){Black}{Black}
\Photon(0,10)(40,10){3}{4}
\Text(30,18)[b]{\scriptsize{$A,Z,W$}}
\Text(50,10)[l]{\scriptsize{$-(2\pi)^4i \ i p_\mu $} }
\LongArrow(25,0)(15,0)
\Text(30,0)[l]{\scriptsize{$p$}}
\Text(40,12)[lb]{\scriptsize{$\mu$}}
\CBoxc(130,10)(4,4){Black}{Black}
\DashLine(130,10)(170,10){3}
\Text(160,15)[b]{\scriptsize{$\varphi^0$}}
\Text(180,10)[l]{\scriptsize{$(2\pi)^4i \ M\slash c_\theta $} }
\CBoxc(300,10)(4,4){Black}{Black}
\DashLine(300,10)(340,10){3}
\Text(330,15)[b]{\scriptsize{$\varphi$}}
\Text(350,10)[l]{\scriptsize{$(2\pi)^4i \ M$} }
\end{picture}
%}
\end{center}
\caption[]{Renormalized source vertices for the WST identities of 
\fig{fig:one:WIbare}.}
\label{fig:one:Fsources}
\end{figure}

Let us consider ${\cal{C}}^{\pm}$, where we get the two source vertices of 
\fig{fig:one:sources}. A doubly-contracted two-point WST identity is obtained 
by connecting two sources through vertices and propagators. Here we get, at
every order in perturbation theory, the identity of \fig{fig:one:WIren1}.
%--
\begin{figure}[ht]
\SetWidth{1}
\begin{center}
%\framebox{
\begin{picture}(210,70)(-10,-10)
\CCirc(0,50){2}{Black}{Black}
\Photon(0,50)(40,50){3}{4}
\LongArrow(25,45)(15,45)
\Text(20,40)[t]{\scriptsize{$p$}}
\Text(40,40)[]{\scriptsize{$\mu$}}
\Text(30,55)[b]{\scriptsize{$W$}}
\Text(50,50)[l]{\scriptsize{$-(2\pi)^4i \
i p_{\mu} Z_{W}^{1\slash 2 } Z^{-1}_{\gpW} $} }
\CCirc(0,10){2}{Black}{Black}
\DashLine(0,10)(40,10){3}
\LongArrow(25,5)(15,5)
\Text(20,0)[t]{\scriptsize{$p$}}
%\Text(40,0)[]{\scriptsize{$\mu$}}
\Text(30,15)[b]{\scriptsize{$\varphi$}}
\Text(50,10)[l]{\scriptsize{
$(2\pi)^4i \ 
M Z_{\varphi}^{1\slash 2 } Z_{\ssM}^{1\slash 2 } Z_{\xi_{\varphi}} $
}}
\end{picture}
%}
\end{center}
\caption[]{Sources related to the gauge-fixing functions ${\cal{C}}^{\pm}$ 
defined in \eqn{eq:one:GAUGEREN}. The momentum $p$ is flowing inwards.}
\label{fig:one:sources}
\end{figure}
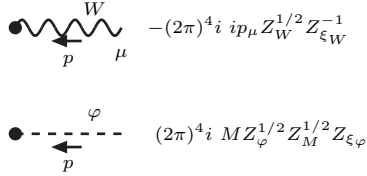
%--
\begin{figure}[ht]
%\framebox[14.5cm]{\parbox{10cm}{
\begin{eqnarray*}
\hspace{-1.5cm}
\hbox{
\begin{picture}(80,20)(0,0)
\SetScale{1}
\SetWidth{1}
\Text(30,10)[b]{\scriptsize{$W$}}
\Text(70,10)[b]{\scriptsize{$W$}}
\Photon(20,0)(40,0){3}{4}
\Photon(60,0)(80,0){3}{4}
\CCirc(50,0){10}{Black}{Gray}
\CCirc(20,0){2}{Black}{Black}
\CCirc(80,0){2}{Black}{Black}
\Text(90,0)[l]{\scriptsize{$+$}}
\end{picture}
\begin{picture}(80,20)(0,0)
\SetScale{1}
\SetWidth{1}
\Text(30,10)[b]{\scriptsize{$W$}}
\Text(70,10)[b]{\scriptsize{$\varphi$}}
\Photon(20,0)(40,0){3}{4}
\DashLine(60,0)(80,0){3}
\CCirc(50,0){10}{Black}{Gray}
\CCirc(20,0){2}{Black}{Black}
\CCirc(80,0){2}{Black}{Black}
\Text(90,0)[l]{\scriptsize{$+$}}
\end{picture}
\begin{picture}(80,20)(0,0)
\SetScale{1}
\SetWidth{1}
\Text(30,10)[b]{\scriptsize{$\varphi$}}
\Text(70,10)[b]{\scriptsize{$W$}}
\DashLine(20,0)(40,0){3}
\Photon(60,0)(80,0){3}{4}
\CCirc(50,0){10}{Black}{Gray}
\CCirc(20,0){2}{Black}{Black}
\CCirc(80,0){2}{Black}{Black}
\Text(90,0)[l]{\scriptsize{$+$}}
\end{picture}}&&
\hbox{
\begin{picture}(80,20)(0,0)
\SetScale{1}
\SetWidth{1}
\SetScale{1}
\hspace{-0.7cm}
\Text(30,10)[b]{\scriptsize{$\varphi$}}
\Text(70,10)[b]{\scriptsize{$\varphi$}}
\DashLine(20,0)(40,0){3}
\DashLine(60,0)(80,0){3}
\CCirc(50,0){10}{Black}{Gray}
\CCirc(20,0){2}{Black}{Black}
\CCirc(80,0){2}{Black}{Black}
\Text(90,0)[l]{\scriptsize{$=0$}}
\end{picture}
}\\
\end{eqnarray*}
%}}
\caption[]{Doubly-contracted WST identity with two external ${\cal{C}}^{\pm}$ 
sources. Gray circles contain all the irreducible and reducible Feynman
diagrams contributing to the needed Green functions. Black dotted circles 
represent sources. Their expression can be read in \fig{fig:one:sources}.}
\label{fig:one:WIren1}
\end{figure}
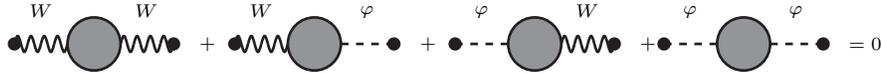
%--
We analyze the WST identity at $\ord{g^2}$. After expanding the 
renormalization constants $Z_{i}$ up to $\ord{g^2}$ we obtain
%--
\bqa
{\cal{C}}^{\pm} &=&
-\Bigl[ 1 + \frac{g^{2}}{16\pi^{2}}
\Bigl(\frac{1}{2}\delta Z_{\ssW}^{(1)} -
\delta Z_{\gpW}^{(1)}\Bigr)
\Bigr] \partial_{\mu} W_{\mu}^{\pm}
+ \Bigl[ 1 + \frac{g^{2}}{16\pi^{2}}
\Bigl(\frac{1}{2}\delta Z_{\varphi}^{(1)} +
\frac{1}{2}\delta Z_{\ssM}^{(1)} +
\delta Z_{\xi_{\varphi}}^{(1)}\Bigr)
\Bigr] M \varphi^{\pm}.
\eqa
%--
and we get the source terms of \fig{fig:one:specsources}.
As a consequence, the doubly-contracted WST identity with two external 
${\cal{C}}^{\pm}$ sources at $\ord{g^2}$ of \fig{fig:one:WIren1} receives two 
different contributions, whose graphical representations are given in 
\fig{fig:one:WIren2}. In the first line we have Green functions
containing Feynman diagrams at $\ord{g^2}$ and source vertices at $\ord{1}$.
The second line, instead, contains Green functions
with Feynman diagrams at $\ord{1}$ and source vertices at $\ord{g^2}$.
Finally we use \eqn{eq:one:oneCOUNTER},
%--
\bqa\label{eq:one:oneCOUNTER2}
\delta Z_{\gpW}^{(1)} &&=
\frac{1}{2} \delta Z_{W}^{(1)},\qquad \quad
\delta Z_{\xi_{\varphi}}^{(1)} = - \frac{1}{2} \Bigl( \delta Z_{\varphi}^{(1)}
+ \delta Z_{\ssM}^{(1)} \Bigr),
\eqa
%--
and we are immediately left with the four contributions displayed in the first 
line of \fig{fig:one:WIren2}. Note that this is exactly the doubly-contracted 
identity with two external ${\cal{C}}^{\pm}$-source vertices which appears in 
the 't Hooft-Feynman gauge.
%--
\begin{figure}[ht]
\SetWidth{1}
\begin{center}
%\framebox{
\begin{picture}(250,140)(-10,0)
\Photon(0,130)(40,130){3}{4}
\CBoxc(0,130)(4,4){Black}{Black}
\Text(30,135)[b]{\scriptsize{$W$}}
\LongArrow(25,122)(15,122)
\Text(25,120)[t]{\scriptsize{$p$}}
\Text(50,127)[rt]{\scriptsize{$\mu$}}
\Text(50,130)[l]{\scriptsize{
$-(2\pi)^4i \ i p_\mu $} }
\DashLine(0,90)(40,90){3}
\CBoxc(0,90)(4,4){Black}{Black}
\Text(30,95)[b]{\scriptsize{$\varphi$}}
\Text(50,90)[l]{\scriptsize{
$(2\pi)^4i \ M  
$} }
\Photon(0,50)(40,50){3}{4}
\CBoxc(0,50)(4,4){Black}{White}
\Text(30,55)[b]{\scriptsize{$W$}}
\LongArrow(25,42)(15,42)
\Text(25,40)[t]{\scriptsize{$p$}}
\Text(50,47)[rt]{\scriptsize{$\mu$}}
\Text(50,50)[l]{\scriptsize{
$-(2\pi)^4i \ i p_\mu \frac{g^{2}}{16 \pi^{2}} \Bigl( \frac{1}{2}\delta 
Z_{W}^{(1)}
- \delta Z_{\gpW}^{(1)} \Bigr)$} }
\DashLine(0,10)(40,10){3}
\CBoxc(0,10)(4,4){Black}{White}
\Text(30,15)[b]{\scriptsize{$\varphi$}}
\Text(50,10)[l]{\scriptsize{
$(2\pi)^4i \ M  \frac{g^2}{16 \pi^2} \Bigl[
\frac{1}{2} \Bigl(\delta Z_{\varphi}^{(1)}
 + \delta Z_{\ssM}^{(1)} \Bigr)
+ \delta Z_{\xi_{\varphi}}^{(1)}
\Bigr]$} }
\end{picture}
%}
\end{center}
\caption[]{
Source vertices related to the gauge-fixing functions ${\cal{C}}^{\pm}$.
We start from \fig{fig:one:sources} and expand the renormalization 
constants up to $\ord{g^2}$. The black squares denote source vertices at 
$\ord{1}$, whereas the white squares denote source vertices at $\ord{g^2}$.}
\label{fig:one:specsources}
\end{figure}
%--
\begin{figure}[h]
%\framebox[14.5cm]{\parbox{20cm}{
\begin{eqnarray*}
\hspace{-1.5cm}
\hbox{
\begin{picture}(80,20)(0,0)
\SetScale{1}
\SetWidth{1}
\Text(30,10)[b]{\scriptsize{$W$}}
\Text(70,10)[b]{\scriptsize{$W$}}
\Photon(20,0)(40,0){3}{4}
\Photon(60,0)(80,0){3}{4}
\CCirc(50,0){10}{Black}{Gray}
\Text(50,0)[]{\scriptsize{$2$}}
\CBoxc(20,0)(4,4){Black}{Black}
\CBoxc(80,0)(4,4){Black}{Black}
%\Text(50,-15)[t]{\scriptsize{$\ord{g^2}$}}
\Text(90,0)[l]{\scriptsize{$+$}}
\end{picture}
\begin{picture}(80,20)(0,0)
\SetScale{1}
\SetWidth{1}
\Text(30,10)[b]{\scriptsize{$W$}}
\Text(70,10)[b]{\scriptsize{$\varphi$}}
\Photon(20,0)(40,0){3}{4}
\DashLine(60,0)(80,0){3}
\CCirc(50,0){10}{Black}{Gray}
\Text(50,0)[]{\scriptsize{$2$}}
\CBoxc(20,0)(4,4){Black}{Black}
\CBoxc(80,0)(4,4){Black}{Black}
%\Text(50,-15)[t]{\scriptsize{$\ord{g^2}$}}
\Text(90,0)[l]{\scriptsize{$+$}}
\end{picture}
\begin{picture}(80,20)(0,0)
\SetScale{1}
\SetWidth{1}
\Text(30,10)[b]{\scriptsize{$\varphi$}}
\Text(70,10)[b]{\scriptsize{$W$}}
\DashLine(20,0)(40,0){3}
\Photon(60,0)(80,0){3}{4}
\CCirc(50,0){10}{Black}{Gray}
\Text(50,0)[]{\scriptsize{$2$}}
\CBoxc(20,0)(4,4){Black}{Black}
\CBoxc(80,0)(4,4){Black}{Black}
%\Text(50,-15)[t]{\scriptsize{$\ord{g^2}$}}
\Text(90,0)[l]{\scriptsize{$+$}}
\end{picture}}
&&
\hspace{-0.7cm}
\hbox{
\begin{picture}(80,20)(0,0)
\SetScale{1}
\SetWidth{1}
\Text(30,10)[b]{\scriptsize{$\varphi$}}
\Text(70,10)[b]{\scriptsize{$\varphi$}}
\DashLine(20,0)(40,0){3}
\DashLine(60,0)(80,0){3}
\CCirc(50,0){10}{Black}{Gray}
\Text(50,0)[]{\scriptsize{$2$}}
\CBoxc(20,0)(4,4){Black}{Black}
\CBoxc(80,0)(4,4){Black}{Black}
%\Text(50,-15)[t]{\scriptsize{$\ord{g^2}$}}
\Text(90,0)[l]{\scriptsize{$+$}}
\end{picture}
}\\
\qquad&&\\
\qquad&&\\
\hspace{-1.5cm}
\hbox{
\begin{picture}(80,20)(0,0)
\SetScale{1}
\SetWidth{1}
\Text(30,10)[b]{\scriptsize{$W$}}
\Text(70,10)[b]{\scriptsize{$W$}}
\Photon(20,0)(50,0){3}{4}
\Photon(50,0)(80,0){3}{4}
\CBoxc(20,0)(4,4){Black}{Black}
\CBoxc(80,0)(4,4){Black}{White}
\Text(90,0)[l]{\scriptsize{$+$}}
\end{picture}
\begin{picture}(80,20)(0,0)
\SetScale{1}
\SetWidth{1}
\Text(30,10)[b]{\scriptsize{$W$}}
\Text(70,10)[b]{\scriptsize{$W$}}
\Photon(20,0)(50,0){3}{4}
\Photon(50,0)(80,0){3}{4}
\CBoxc(20,0)(4,4){Black}{White}
\CBoxc(80,0)(4,4){Black}{Black}
\Text(90,0)[l]{\scriptsize{$+$}}
\end{picture}
\begin{picture}(80,20)(0,0)
\SetScale{1}
\SetWidth{1}
\Text(30,10)[b]{\scriptsize{$\varphi$}}
\Text(70,10)[b]{\scriptsize{$\varphi$}}
\DashLine(20,0)(50,0){3}
\DashLine(50,0)(80,0){3}
\CBoxc(20,0)(4,4){Black}{Black}
\CBoxc(80,0)(4,4){Black}{White}
\Text(90,0)[l]{\scriptsize{$+$}}
\end{picture}}&&
\hbox{
\hspace{-0.8cm}
\begin{picture}(80,20)(0,0)
\SetScale{1}
\SetWidth{1}
\SetScale{1}
\Text(30,10)[b]{\scriptsize{$\varphi$}}
\Text(70,10)[b]{\scriptsize{$\varphi$}}
\DashLine(20,0)(50,0){3}
\DashLine(50,0)(80,0){3}
\CBoxc(20,0)(4,4){Black}{White}
\CBoxc(80,0)(4,4){Black}{Black}
\Text(90,0)[l]{\scriptsize{$=0$}}
\end{picture}
}\\
\end{eqnarray*}
%}}
\caption[]{
Doubly-contracted WST identity with two external ${\cal{C}}^{\pm}$ sources 
at $\ord{g^2}$. Gray circles contain all the irreducible and reducible Feynman
diagrams contributing to the Green functions at $\ord{g^2}$.
Black squares and white squares represent source vertices. Their expression
can be read in \fig{fig:one:specsources}.}
\label{fig:one:WIren2}
\end{figure}
%--
In order to get the complete set of doubly-contracted WST identities we
insert the explicit expressions for the one-loop counterterms of 
\eqn{eq:one:oneCOUNTER} in \eqn{eq:one:GAUGEREN}. Besides
\eqn{eq:one:oneCOUNTER2} we get
%--
\bqa\label{eq:one:oneCOUNTER3}
\delta Z_{\gpA}^{(1)} &=&
\frac{1}{2} \delta Z_{\ssA\ssA}^{(1)},
\quad
\delta Z_{\gpAZ}^{(1)} = - \delta Z_{\ssA\ssZ}^{(1)},
\quad
\delta Z_{\gpZ}^{(1)} =
\frac{1}{2} \delta Z_{\ssZ}^{(1)},
\quad
\delta Z_{\xi_{\varphi^{0}}}^{(1)} = \delta Z_{\ctw}^{(1)} - \frac{1}{2}
\Bigl( \delta Z_{\varphi^0}^{(1)} + \delta Z_{\ssM}^{(1)} \Bigr).
\eqa
%--
Using \eqn{eq:one:oneCOUNTER2} and \eqn{eq:one:oneCOUNTER3} in 
\eqn{eq:one:GAUGEREN} we finally obtain
%--
\bq\label{eq:one:GAUGEREN5}
{{\cal{C}}^{\ssA}} =
- \partial_{\mu} A_{\mu} + \ord{g^4},
\qquad
{\cal{C}}^{\ssZ} = -\partial_{\mu} Z_{\mu} + M_0 \varphi^{0} + \ord{g^4},
\qquad
{\cal{C}^{\pm}} =  -\partial_{\mu} W^{\pm}_{\mu} + M \varphi^{\pm} + \ord{g^4}.
\eq 
%--
The whole set of one-loop WST identities is given in \fig{fig:one:WIbare},
whereas source terms can be read in \fig{fig:one:Fsources}. The results are 
exactly the doubly-contracted identities with two external source vertices 
in 't Hooft-Feynman gauge.
%--
%--
%--
\section{$\beta$ and $\Gamma$ at two loops}\label{sec:one:beta2}
%--
The first step in extending our renormalization procedure beyond one loop
has to do, once again, with tadpole renormalization ($\beta$, 
Subsections~\ref{sus:b1} and ~\ref{sus:b2}) and neutral-sector diagonalization
($\Gamma$, Subsections~\ref{sus:g1} and ~\ref{sus:g2}).
%--
\subsection{$\beta_2$: reducible contributions}
\label{sus:b1}
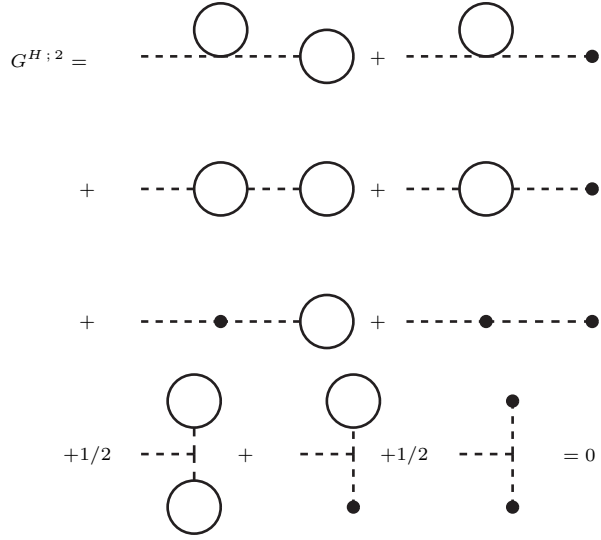
\begin{figure}[ht]
\SetWidth{1}
\begin{center}
\begin{picture}(400,250)(0,0)
\DashLine(120,175)(140,175){3}     %225 175
\CCirc(150,175){10}{Black}{White}
\DashLine(160,175)(180,175){3}
\CCirc(190,175){10}{Black}{White}
\DashLine(220,175)(240,175){3}
\CCirc(250,175){10}{Black}{White}
\DashLine(260,175)(290,175){3}
\CCirc(290,175){2}{Black}{Black}
\DashLine(120,125)(150,125){3}     %175 125
\DashLine(150,125)(180,125){3}
\CCirc(190,125){10}{Black}{White}
\CCirc(150,125){2}{Black}{Black}
\DashLine(220,125)(250,125){3}
\DashLine(250,125)(290,125){3}
\CCirc(290,125){2}{Black}{Black}
\CCirc(250,125){2}{Black}{Black}
\DashLine(120,225)(150,225){3}     %125 225
\CCirc(150,235){10}{Black}{White}
\DashLine(150,225)(180,225){3}
\CCirc(190,225){10}{Black}{White}
\DashLine(220,225)(250,225){3}
\CCirc(250,235){10}{Black}{White}
\DashLine(250,225)(290,225){3}
\CCirc(290,225){2}{Black}{Black}
%--
\DashLine(120,75)(140,75){3}
\DashLine(140,75)(140,85){3}
\DashLine(140,75)(140,65){3}
\CCirc(140,95){10}{Black}{White}
\CCirc(140,55){10}{Black}{White}
\DashLine(180,75)(200,75){3}
\DashLine(200,75)(200,95){3}
\DashLine(200,75)(200,55){3}
\CCirc(200,95){10}{Black}{White}
\CCirc(200,55){2}{Black}{Black}
\DashLine(240,75)(260,75){3}
\DashLine(260,75)(260,95){3}
\DashLine(260,75)(260,55){3}
\CCirc(260,95){2}{Black}{Black}
%\CCirc(260,55){10}{Black}{White}
\CCirc(260,55){2}{Black}{Black}
%
%\DashLine(300,75)(320,75){3}
%\DashLine(320,75)(320,95){3}
%\DashLine(320,75)(320,55){3}
%\CCirc(320,95){2}{Black}{Black}
%\CCirc(320,55){2}{Black}{Black}
%
\Text(100,225)[r]{\scriptsize{$G^{\ssH\,;\,2} = $}}
\Text(210,225)[]{\scriptsize{$+$}}
\Text(210,175)[]{\scriptsize{$+$}}
\Text(210,125)[]{\scriptsize{$+$}}
\Text(160,75)[]{\scriptsize{$+$}}
\Text(220,75)[]{\scriptsize{$+1 \slash 2$}}
%\Text(280,75)[]{\scriptsize{$+$}}
%
\Text(100,175)[]{\scriptsize{$+$}}
\Text(100,125)[]{\scriptsize{$+$}}
\Text(100,75)[]{\scriptsize{$+1 \slash 2$}}
\Text(280,75)[l]{\scriptsize{$= 0$}}
\end{picture}
\end{center}
%--
\caption[]{
The reducible contributions to the renormalization of the Higgs-boson 
vacuum expectation value at $\ord{g^3}$. Nine families of Feynman diagrams 
give a contribution. Dashed lines represent a Higgs boson. Dotted one-leg 
vertices are $\beta_{1}$-dependent vertices at $\ord{g}$. Dotted two-leg 
vertices, instead, depend on $\beta_{1}$ and on one-loop counterterms.
Combinatorial factors for two-loop reducible diagrams are given by the 
products of those for one-loop diagrams, except for two cases, where we 
indicated explicitly the additional multiplicative factor.}
\label{fig:one:tad2redsa}
\end{figure}

We consider the contributions of \fig{fig:one:tad2redsa} to the reducible Green
function with one external Higgs boson at $\ord{g^3}$,
%--
\bq
\Gred^{\ssH\,;\,2} = \sum_{i=1}^{9} F_{i\rre}^{\ssH\,;\,2},
\label{eq:one:beta1swah}
\eq
%--
where the sum runs over the nine families of Feynman diagrams of 
\fig{fig:one:tad2redsa}. The sum over the first six families is given by
%--
\bq
\sum_{i=1}^{6}\,F_{i\rre}^{\ssH\,;\,2} =
\Bigl[ \sum_{j=1}^3\,F^{\ssH\ssH\,;\,1}_{j\irr}\Bigr]\,
\frac{1}{(2 \pi)^4 i}\,\frac{1}{p^2+\mhs}
\Bigl[ \sum_{k=1}^2\,F^{\ssH\,;\,1}_{k\irr}\Bigr],
\label{eq:one:beta1swah1}
\eq
%--
where $F^{\ssH\ssH\,;\,1}_{j\irr}$ denotes one of the three 
families of Feynman diagrams which contribute to the Green function with 
two external Higgs bosons at $\ord{g^2}$ and $F^{\ssH\,;\,1}_{k\irr}$ is one 
of the two families of Feynman diagrams which contribute to the Green function
with one external Higgs boson at $\ord{g}$.
%--
%--
As a consequence of our choice for $\beta_{1}$ we get
%--
\bq
F^{\ssH\,;\,1}_{1\irr} + F^{\ssH\,;\,1}_{2\irr} = 0
\qquad \to \qquad
\sum_{i=1}^{6} F_{i\rre}^{\ssH\,;\,2} = 0.
\label{eq:one:beta1swas}
\eq
%--
For the last three diagrams of \fig{fig:one:tad2redsa} we obtain
another vanishing component,
%--
\bq
\sum_{i=7}^{9}\,F_{i\rre}^{\ssH\,;\,2} =
-\frac{1}{(2 \pi)^4 i} 
\frac{3}{4} \frac{g \mhs}{M}
\frac{1}{( p^2+\mhs )^2}
\Bigl[ F^{\ssH\,;1}_{1\irr} +
F^{\ssH\,;\,1}_{2\,;\,{\rm irr}}\Bigr]^2 = 0.
\label{eq:one:beta1swah1we}
\eq
%--
As a result, $\beta_1$ receives contributions only from irreducible 
diagrams.
%--
\subsection{$\beta_2$: irreducible contributions}
\label{sus:b2}
%--
In order to derive $\beta_2$ we take in account only irreducible diagrams and
we write
%--
\bq\label{eq:one:beta2fam}
\Girr^{\ssH\,;\,2} =
\sum_{j=1}^{6} F_{j\irr}^{\ssH\,;\,2} = 0,
\eq
%--
where the sum is over the six families of irreducible diagrams 
of \fig{fig:one:tad22}.
%--
\begin{figure}[ht]
\SetWidth{1}
\begin{center}
\begin{picture}(400,150)(0,-10)
\DashLine(120,125)(140,125){3}
\CCirc(150,125){10}{Black}{White}
\CCirc(170,125){10}{Black}{White}
\DashLine(220,125)(240,125){3}
\CCirc(250,125){10}{Black}{White}
\Line(250,115)(250,135)
\DashLine(120,75)(150,75){3}
\CCirc(150,75){10}{Black}{White}
\Line(140,75)(160,75)
\DashLine(220,75)(250,75){3}
\CCirc(250,75){10}{Black}{White}
\CCirc(260,75){2}{Black}{Black}
\DashLine(120,25)(150,25){3}
\CCirc(150,25){10}{Black}{White}
%\Line(140,25)(160,25)
\CCirc(140,25){2}{Black}{Black}
\DashLine(220,25)(250,25){3}
\CBoxc(250,25)(4,4){Black}{Black}
\Text(100,125)[r]{\scriptsize{$\Girr^{\ssH\,;\,2} = $}}
\Text(210,125)[]{\scriptsize{$+$}}
\Text(210,75)[]{\scriptsize{$+$}}
\Text(210,25)[]{\scriptsize{$+$}}
\Text(100,75)[]{\scriptsize{$+$}}
\Text(100,25)[]{\scriptsize{$+$}}
\Text(340,25)[l]{\scriptsize{$= 0$}}
\end{picture}
\end{center}
\caption[]{
The irreducible contributions to the renormalization of the Higgs-boson 
vacuum expectation value at $\ord{g^3}$. Six families of Feynman diagrams give 
a contribution. Dotted vertices depend on $\beta_{1}$, $\Gamma_{1}$ or the 
one-loop counterterms. The black-square vertex, instead, represents a 
$\beta_{2}$-dependent vertex.}
\label{fig:one:tad22}
\end{figure}
%--
Note that here we are dealing with three kinds of diagrams. The first three 
families contain two-loop diagrams at $\ord{g^3}$. The fourth and the fifth 
family contain one-loop diagrams obtained with the insertion of a dotted 
vertex which represent a $\beta_{1}$, $\Gamma_{1}$ or one-loop counterterm 
insertion. Explicit expressions for $\beta_{1}$ and $\Gamma_{1}$ can be read 
respectively in \eqnsc{eq:one:beta1res}{eq:one:gamma1res}.
One-loop counterterms are given in 
\eqns{eq:one:oneCOUNTER}{eq:one:oneCOUNTERstop}, 
\eqns{eq:one:FER1L}{eq:one:FER1R} and \eqn{eq:one:FER2}.
The last family contains one diagram with a $\beta_{2}$-dependent vertex.
%--
Here we present the UV divergent residue for $\beta_{2}$ neglecting the 
fermion masses, except for the top-quark mass $m_{t}$. Although not reported 
in this paper, the full result is available. As usual, we write the 
decomposition
%--
\bq
\beta_2 =
\lpar \frac{1}{\ep} - \DUV \rpar\,\lpar
\frac{\beta_{2\uv,1}}{\ep} + \beta_{2\uv,2}\rpar
+ \DUV^2\,\beta_{2\uv,3} + \beta_{2\Fin}.
\label{eq:one:beta2dec}
\eq
%--
The expression for $\beta_{2\uv,1}$ is given by
%--
\bqa
\beta_{2\uv,1} &=& \frac{1}{128 \pi^4} \Bigl[
\frac{9}{x_{\ssH}^2}
\Bigl(
 - \frac{1}{16 \ctw^{8}}
 + \frac{1}{2}\frac{x_{t}^2}{\ctw^4}
 - \frac{1}{4 \ctw^4}
 + x_{t}^2
 - x_{t}^4
 - \frac{1}{4}
\Bigr)
\nl
{}&+&
\frac{1}{2 x_{\ssH}}
\Bigl(
  \frac{185}{8}\frac{1}{ \ctw^{6}}
 - \frac{9}{2}\frac{x_{t}}{\ctw^4}
 - \frac{153}{4} \frac{1}{\ctw^4}
 + \frac{13}{2}\frac{ x_{t}^2}{\ctw^2}
 + \frac{65}{4}\frac{1}{ \ctw^2}
 - 9 x_{t}
 - x_{t}^2
 - 9 x_{t}^3
 - \frac{27}{2}
\Bigr)
\nl
{}&+&
\frac{1}{4}
\Bigl(
  - \frac{53}{12}                            
  - \frac{11}{16}\frac{1}{ \ctw^4}
  - \frac{43}{12}\frac{1}{ \ctw^2}
  + \frac{3}{2}\frac{ x_{t}}{\ctw^2}
  + 3 x_{t}
\Bigr) 
+ \frac{x_{\ssH}}{8}\Bigl( -\frac{15}{4}\frac{1}{ \ctw^2}
 + 9 x_{t} - \frac{25}{2}\Bigr)
\nl
{}&+& \frac{45}{64} x_{\ssH}^2 
+ 48 \frac{g_\ssS^2}{g^2} \frac{x_t^2}{x_{\ssH}}
  \Bigr].
\eqa
%--
The expression for $\beta_{2\uv,2}$ reads as
%--
\bqa
\beta_{2\uv,2} &=&
\frac{1}{128 \pi^4} \Bigl\{
\frac{3}{x_{\ssH}^2}
\Bigl(
           \frac{1}{16 \ctw^{8}}
          - \frac{x_{t}^2}{\ctw^4}
          + \frac{1}{4 \ctw^4}
          - 2 x_{t}^2
          + 3 x_{t}^4
          + \frac{1}{4}\Bigr)
\nl
{}&+&
\frac{1}{x_{\ssH}}\Bigl(
          - \frac{47}{6}\frac{1}{ \ctw^{6}}
          - \frac{19}{8}\frac{x_{t}}{\ctw^4}
          + \frac{287}{24}\frac{1}{ \ctw^4}
          + 10 \frac{x_{t}}{\ctw^2}
          - \frac{2}{3}\frac{ x_{t}^2}{\ctw^2}
          - \frac{151}{24}\frac{1}{ \ctw^2}
          - \frac{35}{4} x_{t}
         + \frac{19}{6} x_{t}^2
\nl
{}&+& \frac{15}{4} x_{t}^3
          + \frac{127}{6}
\Bigr) + \frac{1}{2}\Bigl(
    \frac{9}{4}\frac{1}{ \ctw^4}
        - \frac{3}{8}\frac{ x_{t}}{\ctw^2}
        - \frac{23}{24}\frac{1}{ \ctw^2}
        - \frac{3}{4} x_{t}
        - 3 x_{t}^2
        + \frac{143}{24}\Bigr)
+ \frac{x_{\ssH}}{16}\Bigl( \frac{3}{\ctw^2} - 9 x_{t} + 1 \Bigr)
- \frac{3}{8} x_{\ssH}^2
\nl
{}&+&
 \frac{1}{4}\frac{\ln \ctw^2}{\ctw^2}
\Bigl[
9 \frac{1}{x_{\ssH}^2 \ctw^2}
\Bigl(
   \frac{1}{4 \ctw^4}
 - x_{t}^2
 + \frac{1}{2}
\Bigr)
+ \frac{1}{x_{\ssH}}
\Bigl(
 - \frac{13}{4}\frac{1}{ \ctw^2}
 - \frac{3}{2} x_{t}^2
 + \frac{3}{4}
\Bigr)
\nl
{}&+&
\frac{1}{8} \Bigl(
  \frac{17}{2}\frac{1}{ \ctw^2} -\frac{13}{3} \Bigr)
+ \frac{3}{16} x_{\ssH}\Bigr]
+ \frac{1}{8} \ln x_{\ssH}\Bigl[
 - \frac{9}{2}
 - \frac{9}{4}\frac{1}{ \ctw^4}
 + \frac{3}{8}\frac{ x_{\ssH}}{\ctw^2}
 + \frac{13}{4} x_{\ssH}
 - \frac{9}{8} x_{\ssH}^2
 + 9 x_{t}^2\Bigr]
\nl
{}&+&
x_{t}^2 \ln x_{t} \Bigl[
  \frac{9}{4}\frac{1}{ \ctw^4 x_{\ssH}^2} 
 - \frac{3}{8}\frac{1}{ \ctw^2 x_{\ssH}}
 + \frac{9}{2}\frac{1}{ x_{\ssH}^2}
 - 9 \frac{x_{t}^2}{x_{\ssH}^2}
 - \frac{13}{4}\frac{1}{x_{\ssH}}
 + \frac{9}{8}\Bigr]
- 8 \frac{g_\ssS^2}{g^2} \frac{x_t^2}{x_{\ssH}}
\Bigr\}.
\eqa
%--
Now, the expression for $\beta_{2\uv,3}$ is given by
%--
\bqa
\beta_{2\uv,3} &=&
\frac{1}{128 \pi^4} \Bigl[ \frac{1}{x_{\ssH}}
\Bigl(
  \frac{185}{64}\frac{1}{ \ctw^{6}}
 - \frac{9}{16}\frac{ x_{t}}{\ctw^4}
 + \frac{13}{16}\frac{ x_{t}^2}{\ctw^2}
 + \frac{17}{32}\frac{1}{ \ctw^2}
 - \frac{79}{16}\frac{1}{ \ctw^4}
 - \frac{9}{8} x_{t}
+ \frac{x_{t}^2}{2} - \frac{9}{8} x_{t}^3 - 5\Bigr)
\nl
{}&+& \frac{1}{16}\Bigl( - 1              
  + \frac{3}{2}\frac{1}{ \ctw^4}
  - \frac{17}{4}\frac{1}{ \ctw^2} + 3 x_{t} - 9 x_{t}^2
  + \frac{3}{2}\frac{ x_{t}}{\ctw^2}\Bigr)
+ \frac{3}{32} x_{\ssH}\Bigl(
           - \frac{5}{4}\frac{1}{ \ctw^2} + 3 x_{t} - 5 \Bigr)
\nl
{}&+& \frac{27}{128} x_{\ssH}^{2 }
+ 12 \frac{g_\ssS^2}{g^2} \frac{x_t^2}{x_{\ssH}}
\Bigr].         
\eqa
%--
\begin{figure}[ht]
\SetWidth{1}
\begin{center}
\begin{picture}(350,200)(-30,0)
\Line(0,130)(30,130)
\CCirc(40,130){10}{Black}{Gray}
\Line(50,130)(70,130)
\Text(45,147.5)[l]{\scriptsize{$H$}}
\DashLine(40,140)(40,155){2}
\CCirc(40,160){5}{Black}{White}
\CCirc(40,170){5}{Black}{White}
\Line(100,130)(130,130)
\CCirc(140,130){10}{Black}{Gray}
\Line(150,130)(180,130)
\Text(145,147.5)[l]{\scriptsize{$H$}}
\DashLine(140,140)(140,160){2}
\CCirc(140,160){5}{Black}{White}
\Line(135,160)(145,160)
\Line(200,130)(230,130)
\CCirc(240,130){10}{Black}{Gray}
\Line(250,130)(280,130)
\Text(245,147.5)[l]{\scriptsize{$H$}}
\DashLine(240,140)(240,160){2}
\CCirc(240,160){5}{Black}{White}
\Line(240,155)(240,165)
\Text(90,130)[]{\scriptsize{$+$}}
\Text(190,130)[]{\scriptsize{$+$}}
\Text(-10,50)[]{\scriptsize{$+$}}
\Line(0,50)(30,50)
\CCirc(40,50){10}{Black}{Gray}
\Line(50,50)(70,50)
\Text(45,67.5)[l]{\scriptsize{$H$}}
\DashLine(40,60)(40,75){2}
\CCirc(40,80){5}{Black}{White}
\CCirc(40,85){2}{Black}{Black}
\Line(100,50)(130,50)
\CCirc(140,50){10}{Black}{Gray}
\Line(150,50)(180,50)
\Text(145,67.5)[l]{\scriptsize{$H$}}
\DashLine(140,60)(140,75){2}
\CCirc(140,80){5}{Black}{White}
\CCirc(140,75){2}{Black}{Black}
\Line(200,50)(230,50)
\CCirc(240,50){10}{Black}{Gray}
\Line(250,50)(280,50)
\Text(245,67.5)[l]{\scriptsize{$H$}}
\DashLine(240,60)(240,75){2}
\CBoxc(240,75)(4,4){Black}{Black}
\Text(90,50)[]{\scriptsize{$+$}}
\Text(190,50)[]{\scriptsize{$+$}}
\Text(290,50)[l]{\scriptsize{$=0$}}
\end{picture}
\end{center}
%--
\caption[]{
$\ord{g^3}$ tadpole insertions contributing to the reducible two-point Green 
function at $\ord{g^k}$. The gray circle denotes the three-point Green 
function, with an additional Higgs boson line, at $\ord{g^{k-3}}$.
These diagrams cancel as a consequence of our choice for $\beta_{2}$.
Dotted and squared vertices are the same introduced in \fig{fig:one:tad22}.}
\label{fig:one:LALALA2}
\end{figure}
%--
It is worth nothing that, as a consequence of our choice for $\beta_2$, 
reducible Feynman diagrams containing a tadpole subdiagram at $\ord g^3$ should
not be included in any computation (this is displayed in 
\fig{fig:one:LALALA2}).
%--
\subsection{$\Gamma_2$: reducible contributions}\label{sus:g1}
%--
Let us start with the reducible contributions of \fig{fig:one:REDaz}, whose
%--
\begin{figure}[ht]
\SetWidth{1}
\begin{center}
\begin{picture}(360,80)(-60,0)
\Text(0,40)[r]{\scriptsize{$G^{AZ\,;\,2}_{\mu\nu\rre} = $}}
\Text(12,50)[b]{\scriptsize{$A$}}
\Text(50,50)[b]{\scriptsize{$A$}}
\Text(88,50)[b]{\scriptsize{$Z$}}
\Text(112,50)[b]{\scriptsize{$A$}}
\Text(150,50)[b]{\scriptsize{$Z$}}
\Text(188,50)[b]{\scriptsize{$Z$}}
\Text(212,50)[b]{\scriptsize{$A$}}
\Text(250,50)[b]{\scriptsize{$\varphi^{0}$}}
\Text(288,50)[b]{\scriptsize{$Z$}}
\Text(100,40)[]{\scriptsize{$+$}}
\Text(200,40)[]{\scriptsize{$+$}}
\Text(10,35)[t]{\scriptsize{$\mu$}}
\Text(90,35)[t]{\scriptsize{$\nu$}}
\Text(20,30)[t]{\scriptsize{$p$}}
\LongArrow(15,20)(25,20)
\Photon(10,40)(30,40){3}{3}
\Photon(40,40)(60,40){3}{3}
\Photon(70,40)(90,40){3}{3}
\CCirc(35,40){8}{Black}{Gray}
\CCirc(65,40){8}{Black}{Gray}
\Photon(110,40)(130,40){3}{3}
\Photon(140,40)(160,40){3}{3}
\Photon(170,40)(190,40){3}{3}
\CCirc(135,40){8}{Black}{Gray}
\CCirc(165,40){8}{Black}{Gray}
\Photon(210,40)(230,40){3}{3}
\DashLine(240,40)(260,40){3}
\Photon(270,40)(290,40){3}{3}
\CCirc(265,40){8}{Black}{Gray}
\CCirc(235,40){8}{Black}{Gray}
\Text(35,40)[]{\scriptsize{$2$}}
\Text(135,40)[]{\scriptsize{$2$}}
\Text(235,40)[]{\scriptsize{$2$}}
\Text(65,40)[]{\scriptsize{$2$}}
\Text(165,40)[]{\scriptsize{$2$}}
\Text(265,40)[]{\scriptsize{$2$}}
%--
\end{picture}
\end{center}
\caption[]{
Reducible contributions to the Green function with one external photon and 
one external $Z$ boson at $\ord{g^4}$.
Gray circles denote the sum over the irreducible Feynman diagrams at 
$\ord{g^2}$.}
\label{fig:one:REDaz}
\end{figure}
%--
contribution to the Green function with one external photon and one external 
$Z$ boson at $\ord{g^4}$ can be written through the two-point Green functions 
at one loop involving neutral gauge bosons,
%--
\bqa
G^{\ssA\ssZ\,;\,2}_{\mu\nu\rre} &=&
\frac{1}{(2\pi)^4i}
\Bigl\{
G^{\ssA\ssA\,;\,1}_{\mu\alpha\irr}
\frac{1}{p^2}
G^{\ssA\ssZ\,;\,1}_{\alpha\nu\irr}
+
G^{\ssA\ssZ\,;\,1}_{\mu\alpha\irr}
\frac{1}{p^2+M^2_0}
G^{\ssZ\ssZ\,;\,1}_{\alpha\nu\irr}
+
G^{\ssA\varphi^0\,;\,1}_{\mu\irr}
\frac{1}{p^2+M^2_0}
G^{\varphi^{0}\ssZ\,;\,1}_{\nu\irr}
\Bigr\}.
\eqa
%--
We extract form factors as
%--
\bqa
\label{formFactors}
G^{ij;2}_{\mu\nu}\, &=&\, 
G^{ij;2}_{d}(p^2)\, \delta_{\mu\nu}\, +\,
G^{ij;2}_{pp}(p^2)\, p_{\mu}p_{\nu} \qquad (i, j \quad \texttt{vectors}),
\nl
G^{ij;2}_{\mu}\, &=&\, 
G^{ij;2}_{p}(p^2)\, i\, p_{\mu}
\qquad (i \quad \texttt{vector}, \quad j \quad \texttt{scalar})
\eqa
%--
and obtain
%--
\bq
G^{\ssA\ssZ\,;\,2}_{d\rre}(p^2) = \frac{1}{(2\pi)^4i}
G^{\ssA\ssZ\,;\,1}_{d\irr}(p^2)\Bigl\{
G^{\ssA\ssA\,;\,1}_{d\irr}(p^2)\frac{1}{p^2} +
G^{\ssZ\ssZ\,;\,1}_{d\irr}(p^2)\,\frac{1}{p^2+M^2_0}\Bigr\}.
\label{eq:one:ix}
\eq
%--
The photon self-energy shows factorization,
%--
\bq
G^{\ssA\ssA\,;\,1}_{d\irr}(p^2) = p^2 \Pi^{\ssA\ssA\,;1}_{d}(p^2).
\eq
%--
It follows that \eqn{eq:one:ix}, at $p^2=0$, can be written as
%--
\bq
G^{\ssA\ssZ\,;\,2}_{d\rre}(0) = \frac{1}{(2\pi)^4i}
G^{\ssA\ssZ\,;\,1}_{d\irr}(0)\Bigl\{
\Pi^{\ssA\ssA\,;\,1}_{d}(0) + G^{\ssZ\ssZ\,;\,1}_{d\irr}(0)
\frac{1}{M^2_0} \Bigr\}.
\label{eq:one:ix2}
\eq
%--
Following our choice for $\Gamma_{1}$ we have $G^{\ssA\ssZ\,;\,1}_{d\irr}(0) 
= 0$. As a consequence, the reducible contributions of \eqn{eq:one:ix} vanish.
Therefore, $\Gamma_2$ is fixed through irreducible Feynman diagrams,
%--
\bq
G^{\ssA\ssZ\,;\,2}_{d}(0) = G^{\ssA\ssZ\,;\,2}_{d\irr}(0) = 0.
\label{eq:one:gundamwing}
\eq
%--
\subsection{$\Gamma_2$: irreducible contributions}\label{sus:g2}
%--
To fix $\Gamma_{2}$ we use 
%--
\bq
G^{\ssA\ssZ\,;\,2}_{d\irr}(0) = \sum_{i=1}^{13}
F_{d\,;\,i\irr}^{\ssA\ssZ\,;\,2}(0) = 0,
\label{eq:one:gamma1fam2}
\eq
%--
where the sum is over the form factors of the thirteen families of irreducible 
diagrams of \fig{fig:one:selfone2}. Here we get three kinds of diagrams.
The first eight families contain two-loop diagrams
at $\ord{g^4}$. Next, we have four families of one-loop diagrams obtained by 
inserting a vertex which depends on $\beta_{1}$, $\Gamma_{1}$, or on a
one-loop counterterm. Note that these diagrams depend also on the counterterms
for the gauge parameters $\delta Z_{\xi_{i}}$.
Obviously, there will be no track of the gauge parameters in the final
results since we made the identification $\xi_{i} = 1$. 
Nevertheless, gauge-parameter counterterms play an essential role in making 
every Green function UV-finite.
Finally, the last family contains a single Feynman diagram with a 
$\Gamma_{2}$-dependent vertex.

The contribution of the thirteenth family of \fig{fig:one:selfone2} is given by
%--
\bqa
F_{d\,;\,13\irr}^{\ssA\ssZ\,;\,2}(0) &=&
- (2 \pi)^4 i  g^4 M^2 \stw \slash
\ctw \Bigl\{\Gamma_{2} + \Bigl( \ctw^2 \Gamma_{1}^2
+ 2 \Gamma_{1} \beta_{1} \Bigr)
\nl
{}&+&
\frac{\Gamma_{1}}{16\pi^2} \Bigl[
\delta Z_{\stw}^{(1)} - \delta Z_{\ctw}^{(1)}
+ 2 \delta Z_{g}^{(1)} + \delta Z_{\ssM}^{(1)}
+ \frac{1}{2} \Bigl( \delta Z_{\ssA\ssA}^{(1)}
+ \delta Z_{\ssZ}^{(1)}\Bigr)\Bigr]\Bigr\}.
\label{eq:one:LASg2}
\eqa
%--
Using \eqn{eq:one:LASg2} in \eqn{eq:one:gamma1fam2} we get
%--
\bqa
\Gamma_{2} &=& \frac{1}{(2\pi)^4i}
\frac{\ctw \sum_{i=1}^{12} F_{d \vert 12\irr}^{\ssA\ssZ\,;\,2}(0)}
{g^4 \stw M^2}
- \Bigl( \ctw^2 \Gamma_{1}^2 + 2 \Gamma_{1} \beta_{1} \Bigr)
\nl
{}&-&
\frac{\Gamma_{1}}{16\pi^2}
\Bigl[ \delta Z_{\stw}^{(1)}
- \delta Z_{\ctw}^{(1)}
+ 2 \delta Z_{g}^{(1)}
+ \delta Z_{\ssM}^{(1)}
+ \frac{1}{2} \Bigl( \delta Z_{\ssA\ssA}^{(1)}
+ \delta Z_{\ssZ}^{(1)} \Bigr)\Bigr].
\label{eq:one:solveg12}
\eqa
%--
$\Gamma_{2}$ can be expressed through a linear combination of products of 
one- and two-loop scalar integrals, evaluated at zero-momentum transfer.
As usual, we write the decomposition
%--
\bq
\Gamma_{2} =
\lpar \frac{1}{\ep^2} - \DUV \rpar\,
\lpar \frac{\Gamma_{2\uv,1}}{\ep} + \Gamma_{2\uv,2} \rpar
+ \DUV^2\,\Gamma_{2\uv,3} + \Gamma_{2\Fin},
\label{eq:one:gamma2dec}
\eq
%--
where the UV residues are
%--
\bq\label{eq:one:gamma1resp}
\Gamma_{2\uv,1} = \frac{169}{1536 \pi^4},
\quad
\Gamma_{2\uv,2} = - \frac{137}{6144 \pi^4},
\quad
\Gamma_{2\uv,3} = \frac{111}{2048 \pi^4}.
\eq
%--
\section{Two-loop self-energies}
\label{sec:one:twoloop}
%--
In this section we will show that two-loop self-energies develop 
UV poles whose residues are polynomials in the external momentum 
(\emph{local} residues).
Therefore, a suitable choice for the counterterms in the $\MSB$ scheme
removes UV divergencies also at two loops.
We generate the needed Feynman diagrams with 
${\cal G}raph{\cal S}hot$~\cite{GraphShot} and express every two-loop 
two-point Green function through scalar integrals using standard 
tensor-reduction techniques. Finally, we extract UV residues by means of the 
results collected in \sect{conve}.

In \subsect{subsec:not} we introduce our notations, and in 
\subsect{susec:one:WER} and \subsect{zg} we review the transverse form 
factors for the photon self-energy and the transition between the photon and 
the $Z$ boson.
Explicit results for the transverse components of the $Z$- and $W$-boson 
self-energies and for the Higgs-boson self-energy are shown
in \subsect{z}, \subsect{w} and \subsect{h}.
%--
\subsection{Notation}
\label{subsec:not}
\begin{figure}[ht]
\SetWidth{1}
\begin{center}
\begin{eqnarray*}
\begin{picture}(360,80)(-60,0)
\Text(0,40)[r]{\scriptsize{$G^{ij\,;\,2}_{\mu\nu\rre}=$}}
\Text(12,50)[b]{\scriptsize{$i$}}
\Text(88,50)[b]{\scriptsize{$j$}}
\Text(100,40)[]{\scriptsize{$+$}}
\Text(200,40)[]{\scriptsize{$+$}}
\Text(10,35)[t]{\scriptsize{$\mu$}}
\Text(90,35)[t]{\scriptsize{$\nu$}}
\Text(20,30)[t]{\scriptsize{$p$}}
\LongArrow(15,20)(25,20)
\Line(10,40)(35,40)
\Line(35,40)(65,40)
\Line(65,40)(90,40)
\CCirc(35,48){8}{Black}{White}
\CCirc(65,48){8}{Black}{White}
\Line(110,40)(135,40)
\Line(135,40)(165,40)
\Line(165,40)(190,40)
\CCirc(135,48){8}{Black}{White}
\CCirc(165,40){8}{Black}{White}
\Line(210,40)(235,40)
\Line(235,40)(265,40)
\Line(265,40)(290,40)
\CCirc(265,40){2}{Black}{Black}
\CCirc(235,48){8}{Black}{White}
\end{picture}\\
\begin{picture}(360,80)(-60,0)
\Text(0,40)[r]{\scriptsize{$+$}}
\Text(100,40)[]{\scriptsize{$+$}}
\Text(200,40)[]{\scriptsize{$+$}}
\Line(10,40)(30,40)
\Line(40,40)(65,40)
\Line(65,40)(90,40)
\CCirc(35,40){8}{Black}{White}
\CCirc(65,48){8}{Black}{White}
\Line(110,40)(130,40)
\Line(140,40)(160,40)
\Line(170,40)(190,40)
\CCirc(135,40){8}{Black}{White}
\CCirc(165,40){8}{Black}{White}
\Line(210,40)(230,40)
\Line(240,40)(265,40)
\Line(265,40)(290,40)
\CCirc(265,40){2}{Black}{Black}
\CCirc(235,40){8}{Black}{White}
\end{picture}\\
\begin{picture}(360,80)(-60,0)
\Text(0,40)[r]{\scriptsize{$+$}}
\Line(10,40)(35,40)
\Line(35,40)(65,40)
\Line(65,40)(90,40)
\CCirc(35,40){2}{Black}{Black}
\CCirc(65,48){8}{Black}{White}
\Line(110,40)(135,40)
\Line(135,40)(160,40)
\Line(170,40)(190,40)
\CCirc(135,40){2}{Black}{Black}
\CCirc(165,40){8}{Black}{White}
\Line(210,40)(235,40)
\Line(235,40)(265,40)
\Line(265,40)(290,40)
\CCirc(265,40){2}{Black}{Black}
\CCirc(235,40){2}{Black}{Black}
\end{picture}
\end{eqnarray*}
\end{center}
\caption[]{
Reducible contributions to the two-point Green function
with external boson fields $i$ and $j$ at $\ord{g^4}$.
The dotted two-leg vertex represents a $\beta_1$-,
$\Gamma_1$- or one-loop-counterterm-dependent vertex.}
\label{fig:one:reddiag}
\end{figure}
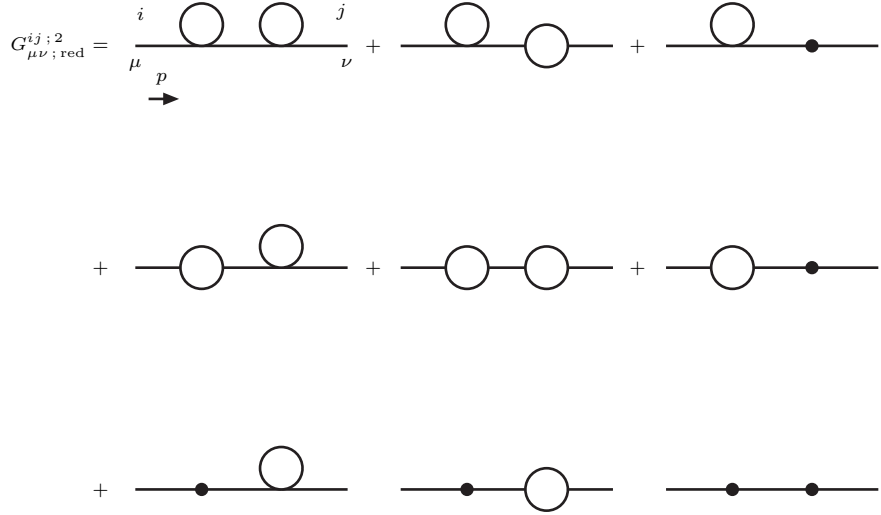

%--
Reducible two-loop two-point Green functions with two external boson lines 
are displayed in \fig{fig:one:reddiag}. They are UV finite by construction, 
since one-loop counterterms have already been fixed.
%--
%--
\noindent
Therefore, we can limit our discussion to irreducible two-loop two-point 
Green functions, which can be written through the form factors introduced 
in \eqn{formFactors},
%--
\bq\label{eq:one:IrreTwo}
G^{ij\,;\,2}_{a\irr}(s) = \sum_{k=1}^{13}
F_{a\irr\,;\,k}^{ij\,;\,2}(s).
\eq
%--
Here $i$ and $j$ are the external fields, the subscript $a$ labels the form 
factor and the sum runs over the thirteen families of Feynman diagrams of 
\fig{fig:one:selfone2}. We shall organize the list of results according
to the following decomposition:
%--
\bq\label{eq:one:IrreTwo2}
G^{ij\,;\,2}_{a\irr}(s) = G^{ij\,;\,2\bos}_{a\irr}(s)
+ G^{ij\,;\,2\lep}_{a\irr}(s)
+ G^{ij\,;\,2\,;\,lq}_{a\irr}(s)
+ G^{ij\,;\,2\,;\,tb}_{a\irr}(s),
\label{fterms}
\eq
%--
where we split the complete result into purely-bosonic diagrams (bos),
diagrams with one lepton sub-loop (lep), diagrams with one light-quark sub-loop
(lq; $u$, $d$, $c$ and $s$) and diagrams with one top-bottom sub-loop (tb).
Furthermore, we extract UV poles using \eqn{eq:one:polect12} and we always 
leave out an overall factor $(2\pi)^4\,i$. 
In the following we show results neglecting the masses of the leptons
and the five light quarks. Note, however, that the full result is available.
%--
\subsection{The photon self-energy}
\label{susec:one:WER}
%--
When computing the photon self-energy at a given order in perturbation 
theory we adopt the convention of excluding diagrams which contain light 
quarks coupled to photons or gluons with no other hard scale.
In this case the correct treatment requires the use of non-perturbative 
dispersion relations~\cite{Eidelman:1995ny}. 
So there are two preliminary steps: first, we exclude from the 
computation two-loop diagrams which contain only light quarks and photons 
(or gluons). Next, we derive again one-loop counterterms for light-quark 
fields and masses neglecting one-loop diagrams which contain only light 
quarks and photons (or gluons). The results for up- and down-quark fields 
reads as
%--
\bqa
\Delta Z_{u_L}^{(1)} &=&
  \frac{1}{4} \Bigl(
  \frac{1}{9}\frac{1}{\ctws} +
  \frac{10}{9}
  + \frac{16}{9}\,\ctws
  + x_u + x_d   \Bigl),
%--
\quad
%--
\Delta Z_{d_L}^{(1)} =
  \frac{1}{4} \Bigl(
\frac{1}{9}\frac{1}{\ctws} +
  \frac{22}{9} +
   \frac{4}{9}\,\ctws
  + x_u + x_d  \Bigr),
\nl
\Delta Z_{u_R}^{(1)} &=&
  \frac{1}{4} \Bigl(
  \frac{16}{9}\frac{1}{ \ctws}
  - \frac{32}{9}
  + \frac{16}{9}\,\ctws
  + 2\,x_u \Bigr),
%--
\qquad
%--
\Delta Z_{d_R}^{(1)} =
  \frac{1}{4} \Bigl(
  \frac{4}{9}\frac{1}{\ctws}  - \frac{8}{9} + \frac{4}{9}\,\ctws
  + 2\,x_d \Bigr).
\label{eq:one:FER1QEDR}
\eqa
%--
%
%\bqa\label{eq:one:FER1QED}
%\delta Z_u^{(1)} &=&
%  \frac{1}{9} \Bigl(
%  \frac{17}{8 \ctw^2}
%  - \frac{11}{4}
%  + 4 \ctw^2 \Bigl)
%  + \frac{1}{8} \Bigl(x_d
%  + 3 x_u \Bigr),
%\qquad
%\delta Z_d^{(1)} =
%  \frac{1}{9} \Bigl(
%  \frac{5}{8 \ctw^2}
%  + \frac{7}{4}
%  + \ctw^2 \Bigl)
%  + \frac{1}{8} \Bigl(3 x_d
%  + x_u \Bigr).
%\eqa
%--
Light-quark mass counterterms are given by
%--
\bqa\label{eq:one:FER2QED}
\Delta Z_{m_u}^{(1)} &=&
 - \frac{1}{3} \Bigl( - 2\, \frac{1}{\ctw^2} + 10  - 8 \ctw^2 \Bigr)
+ \frac{3}{x_{\ssH}} \Bigl[ \frac{1}{2}\frac{1}{\ctw^4} + 1
- 2 \Bigl( X_u^2 + X_d^2 \Bigr) - \frac{2}{3} X_{l}^2 \Bigr]
 + \frac{3}{4} \Bigl( x_{\ssH} - x_u + x_d \Bigr)
,
\nl
\Delta Z_{m_d}^{(1)} &=&
 - \frac{1}{3} \Bigl(  \frac{1}{\ctw^2} + 1  - 2 \ctw^2 \Bigr)
+ \frac{3}{x_{\ssH}} \Bigl[ \frac{1}{2}\frac{1}{\ctw^4} + 1
- 2 \Bigl( X_u^2 + X_d^2 \Bigr) - \frac{2}{3} X_{l}^2 \Bigr]
 + \frac{3}{4} \Bigl( x_{\ssH} + x_u - x_d \Bigr).
\eqa
%--

\noindent
\textbf{Two-loop result}
\vspace{0.2cm}

\noindent
%--
Before showing explicit expressions for the individual components of 
\eqn{eq:one:IrreTwo2} let us recall that scaled masses are defined by 
$x_{i} = M_{i}^2 \slash M^2$, where $M$ is the renormalized $W$-boson mass, 
and where $y_{i} = M_{i}^2\slash s$, with $s= -p^2$; $p$ is the external 
momentum. Each term in the r.h.s. of \eqn{eq:one:IrreTwo2} develops a 
non-local residue containing the (logarithmic) function
%--
\bq
L_\beta(x)\, =\, \ln\, \frac{\beta(x)+1}{\beta(x)-1},
\eq
%--
where $\beta$ can be found in \eqn{deflambda}.
We will show explicitly that non-local residues cancel when summing up all 
the contributions, and UV poles will be subtracted by polynomial counterterms.
\vspace{0.2cm}

\noindent
\underline{\emph{Two light-quark doublets}}
\bq
G^{\ssA\ssA\,;\,2\,;\,lq}_{d\irr\uv;1}(s) \,=\,
2\,G^{\ssA\ssA\,;\,2\,;\,lq}_{d\irr\uv;3}(s) \,=\,
 -  \, \frac{g^4\stw^2}{\pi^4}\, \frac{3}{64} \, s,
\eq
%--
\bqa
\label{lqa}
G^{\ssA\ssA\,;\,2\,;\,lq}_{d\irr\uv;2}(s) &=&
 \ \frac{g^4\stw^2}{\pi^4}
\frac{1}{8} \Bigl\{ \frac{s}{108} \Bigl(
\frac{115}{8} + \frac{73}{8}\frac{1}{\ctw^2} + 17 \ctw^2
\Bigr) + \frac{3}{2} M^2
+ \frac{L_{\beta}(y_{\ssW})}{\beta(y_{\ssW})}\Bigl[
-\frac{s}{8} + M^2 \Bigl( \frac{1}{4} + 3 y_{\ssW} \Bigr)\Bigr]\Bigr\}.
\eqa
%--
\noindent
\underline{\emph{Top-bottom doublet}}
\bq
G^{\ssA\ssA\,;\,2\,;\,tb}_{d\irr\uv;1}(s) =
2\,G^{\ssA\ssA\,;\,2\,;\,tb}_{d\irr\uv;3}(s) =
-  \frac{g^4\stw^2}{\pi^4} \frac{3}{128} s,
\eq
%--
\bqa
G^{\ssA\ssA\,;\,2\,;\,tb}_{d\irr\uv;2}(s) &=&
 \ \frac{g^4\stw^2}{\pi^4}\Bigl\{
\frac{s}{64}\Bigl[ \frac{1}{27} \Bigl(
\frac{73}{8} \frac{1}{\ctw^2} + \ctw^2 \Bigr)
+ \frac{1}{8} \Bigl( 9 -\frac{13}{3} x_t \Bigr) \Bigr]
\nl
{}&+& \frac{3}{32} M^2 \Bigl( 1 - \frac{3}{2}x_t - x_t^2 \Bigr)
 + \frac{1}{16}\frac{L_{\beta}(y_{\ssW})}{\beta(y_{\ssW})}
\Bigl[ -\frac{s}{8} + \frac{M^2}{4} \Bigl( 1 -\frac{9}{2} x_t + 3 x_t^2 \Bigr)
\nl
{}&+& 3  M^2 y_{\ssW}\Bigl(
1 -\frac{3}{2} x_t - \frac{3}{4} x_t^2 \Bigr) \Bigr] 
+\frac{s}{36}\frac{g_\ssS^2}{g^2}
\Bigr\}.
\eqa
%--
Note the presence of mixed electroweak-QCD corrections proportional to the 
strong coupling constant $g_{\ssS}$, which do not show up in \eqn{lqa} because
they are subtracted from the light-quark components.
\vspace{0.2cm}

%--
\noindent
\underline{\emph{Three lepton doublets}}
\bq
G^{\ssA\ssA\,;\,2\lep}_{d\irr\uv;1}(s) =
2\,G^{\ssA\ssA\,;\,2\lep}_{d\irr\uv;3}(s) =
-   \frac{g^4\stw^2}{\pi^4} \frac{3}{128} s,
\eq
%--
\bqa
G^{\ssA\ssA\,;\,2\lep}_{d\irr\uv;2}(s) &=&
 \frac{g^4\stw^2}{\pi^4}\Bigl\{\frac{s}{512}
\Bigl(1 + 15 \frac{1}{\ctw^2}\Bigr) + \frac{3}{32} M^2
+ \frac{1}{16}\frac{L_{\beta}(y_{\ssW})}{\beta(y_{\ssW})}
\Bigl[ -\frac{s}{8} + M^2 \Bigl( \frac{1}{4} + 3 y_{\ssW} \Bigr) \Bigr]
\Bigr\}.
\eqa
%--
\noindent
\underline{\emph{Bosonic contributions}}
\bq
G^{\ssA\ssA\,;\,2\bos}_{d\irr\uv;1}(s) =
 \frac{g^4\stw^2}{\pi^4}
\frac{47}{768}\, s,
%\frac{s}{384} 
%\Bigl(-241 + \frac{529}{2} \ctw^2 \Bigr),
\eq
%--
\bqa
G^{\ssA\ssA\,;\,2\bos}_{d\irr\uv;2}(s) &=&
 \frac{g^4\stw^2}{\pi^4}
\Bigl\{ \frac{s}{256} \Bigl( - \frac{21}{4} + \frac{1}{\ctw^2}
\Bigr) + \frac{3}{8} M^2 \Bigl(
- 1 + \frac{3}{8} x_t + \frac{1}{4} x_t^2 \Bigr)
\nl
{}&+& \frac{1}{4}\frac{L_{\beta}(y_{\ssW})}{ \beta(y_{\ssW})} 
\Bigl[ \frac{s}{8} + \frac{M^2}{2} \Bigl(
-\frac{1}{2} + \frac{9}{16} x_t - \frac{3}{8} x_t^2 \Bigr) +
M^2 y_{\ssW} \Bigl( -3 +\frac{9}{8} x_t + \frac{9}{16} x_t^2 \Bigr)
\Bigr]\Bigr\},
\eqa
%--
\bq
G^{\ssA\ssA\,;\,2\bos}_{d\irr\uv;3}(s) =
 \frac{g^4\stw^2}{\pi^4}
\frac{33}{1024}\, s.
%\frac{s}{1536} \Bigl( -215 + \frac{529}{2}\ctw^2 \Bigr).
\eq
%--
\noindent
\underline{\emph{Total contributions}}
\vspace{0.2cm}

\noindent
The sum of the four components gives
%--
\bq
G^{\ssA\ssA\,;\,2}_{d\irr\uv;1}(s) =
-\, \frac{g^4\stw^2}{\pi^4}
\frac{25}{768}\, s,
%\frac{s}{384} \Bigl(-277 + \frac{529}{2} \ctw^2 \Bigr),
\eq
%--
\bq
G^{\ssA\ssA\,;\,2}_{d\irr\uv;2}(s) =
 \frac{g^4\stw^2}{\pi^4}
\frac{s}{192} \Bigl(\frac{433}{144} 
+ \frac{113}{12}\frac{1}{ \ctw^2} + \frac{35}{9} \ctw^2 - \frac{13}{8} x_t
\, + \frac{16}{3} \frac{g_{\ssS}^2}{g^2}
 \Bigr),
\eq
%--
\bq
G^{\ssA\ssA\,;\,2}_{d\irr\uv;3}(s) =
- \frac{g^4\stw^2}{\pi^4}
\frac{15}{1024} s.
%\frac{s}{1536} \Bigl( -287 + \frac{529}{2}\ctw^2 \Bigr),
\eq
%--
Non-local residues cancel and the UV residue
can be reabsorbed by the counterterm contributions,
%--
\bq
G^{\ssA\ssA\,;\,2\ct}_{d\irr\uv;a}(s) =
 \frac{g^4 s }{256\pi^4} \Delta Z_{\ssA\ssA\,;\,a}^{(2)}, \qquad a=1,2,3.
\eq
%--
\vspace{0.2cm}

\noindent
\textbf{An alternative method}
\vspace{0.2cm}

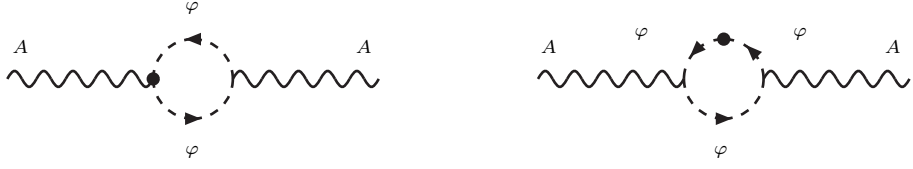
\begin{figure}[ht]
\SetWidth{1}
\begin{center}
%\framebox{
\begin{picture}(400,100)(0,0)
\Photon(30,50)(85,50){3}{5}
\Photon(115,50)(170,50){3}{5}
\Photon(230,50)(285,50){3}{5}
\Photon(315,50)(370,50){3}{5}
\DashArrowArc(100,50)(15,0,180){3}
\DashArrowArc(300,50)(15,0,90){3}
\DashArrowArc(300,50)(15,90,180){3}
\DashArrowArc(100,50)(15,180,360){3}
\DashArrowArc(300,50)(15,180,360){3}
\Text(35,60)[b]{\scriptsize{$A$}}
\Text(165,60)[b]{\scriptsize{$A$}}
\Text(235,60)[b]{\scriptsize{$A$}}
\Text(365,60)[b]{\scriptsize{$A$}}
\Text(100,75)[b]{\scriptsize{$\varphi$}}
\Text(100,25)[t]{\scriptsize{$\varphi$}}
\Text(270,65)[b]{\scriptsize{$\varphi$}}
\Text(330,65)[b]{\scriptsize{$\varphi$}}
\Text(300,25)[t]{\scriptsize{$\varphi$}}
\CCirc(85,50){2}{Black}{Black}
\CCirc(300,65){2}{Black}{Black}
\end{picture}
%}
\end{center}
\caption[]{
Counterterm-dependent Feynman diagrams with a $\varphi^{\pm}$ loop. Dotted 
vertices represent counterterm insertions.}
\label{fig:one:Xc}
\end{figure}

\noindent
Here we outline another strategy to verify the cancellation of logarithmic 
residues. Let us consider \fig{fig:one:Xc}, corresponding to two Feynman 
diagrams with a Higgs-Kibble $\varphi^{\pm}$ loop and a one-loop counterterm 
insertion.
%--
\noindent
The contributions of these two diagrams is given by
%--
\bqa\label{eq:one:ST1}
H_{\mu\nu\,;\,1}^{\ssA\ssA\,;\,2} &=&
\frac{g^4 \stw^2}{16 \pi^2} \mu^{4-n}
\int d^{n}q
\frac{(2q+p)_{\mu} (2q+p)_{\nu}}{(q^2+M^2)[(q+p)^2+M^2]}
\,\lpar \delta Z_{g}^{(1)} + \delta Z_{\stw}^{(1)}
+ \delta Z_{\varphi}^{(1)} + \frac{ \delta Z_{\ssA\ssA}^{(1)}}{2}\rpar,
\eqa
%--
\bqa\label{eq:one:ST2}
H_{\mu\nu\,;\,2}^{\ssA\ssA\,;\,2} &=&
-\frac{g^4 \stw^2}{16 \pi^2} \mu^{4-n}
\int d^{n}q \frac{(2q+p)_{\mu} (2q+p)_{\nu}}{(q^2+M^2)^2
[(q+p)^2+M^2]}
\Bigl[  \lpar  q^2+M^2 \rpar  \delta Z_{\varphi}^{(1)} + M^2
\lpar \delta Z_{\ssM}^{(1)} + 2 \delta Z_{\xi_{\varphi}}^{(1)}
\rpar \Bigr].
\eqa
%--
After summing \eqns{eq:one:ST1}{eq:one:ST2} the terms which depend on 
$\delta Z_{\varphi}^{(1)}$ cancel out,
%--
\bqa
H_{\mu\nu\,;\,1}^{\ssA\ssA\,;\,2} &+&
H_{\mu\nu\,;\,2}^{\ssA\ssA\,;\,2} =
\frac{g^4 \stw^2}{16 \pi^2} \mu^{4-n} \int d^{n}q
\frac{(2q+p)_{\mu} (2q+p)_{\nu}}{(q^2+M^2) [(q+p)^2+M^2]}
\nl
{}&\times& \Bigl[ \delta Z_{g}^{(1)} + \delta Z_{\stw}^{(1)}
+ \frac{ \delta Z_{\ssA\ssA}^{(1)}}{2} - \frac{M^2}{q^2+M^2}
\lpar \delta Z_{\ssM}^{(1)} + 2 \delta Z_{\xi_{\varphi}}^{(1)}\rpar 
\Bigr].
\eqa
%--
Moreover, after using the relation
%--
\bq
\delta Z_{g}^{(1)} = - \delta Z_{\stw}^{(1)} - \frac{1}{2}\,
\delta Z_{\ssA\ssA}^{(1)},
\eq
%--
the answer contains only the $W$-mass- and gauge-parameter-dependent
counterterms,
%--
\bqa
H_{\mu\nu\,;\,1}^{\ssA\ssA\,;\,2} &+&
H_{\mu\nu\,;\,2}^{\ssA\ssA\,;\,2} =
-\frac{g^4 \stw^2 M^2}{16 \pi^2}  \lpar \delta Z_{\ssM}^{(1)}
+ 2 \delta Z_{\xi_{\varphi}}^{(1)} \rpar 
\,\mu^{4-n} \int d^{n}q
\frac{(2q+p)_{\mu} (2q+p)_{\nu}}{(q^2+M^2)^2 [(q+p)^2+M^2]}.
\eqa
%--
Furthermore, all Feynman diagrams with the insertion of counterterms can be 
paired like in \fig{fig:one:Xc} and we can repeat the argument for any of 
the pairs.
Consequently, counterterms for masses and gauge parameters suffice in
removing logarithmic residues. Here we outline the procedure: we take in 
account the irreducible Green function at $\ord{g^2}$ which are written as
follows:
%--
\bq\label{eq:one:ASDE}
G^{\ssA\ssA\,;\,1}_{d\irr}(s) =
G^{\ssA\ssA\,;\,1\fer}_{d\irr}(s) +
G^{\ssA\ssA\,;\,1\bos}_{d\irr}(s),
\eq
%--
where the first term receives contributions from the three lepton families
and the top quark (light-quark components are subtracted), whereas the second 
one gets contributions from bosonic loops. The essential point is that only 
bare quantities appear in \eqn{eq:one:ASDE}; in other words, the gauge-fixing 
terms are
%--
\bqa\label{eq:th:gfunSAS}
{\cal{C}}^{\ssA} &=& 
           -\frac{1}{\gpA^b} \partial_{\mu} A_{\mu}^b 
           - \gpAZ^b \partial_{\mu} Z^b_{\mu},
\quad
{\cal{C}}^{\ssZ} =
           -\frac{1}{\gpZ^b} \partial_{\mu} Z^b_{\mu} 
           + \xi^b_{\varphi^{0}} M^b_0 \varphi^{0,b},
\quad
{\cal{C}}^{\pm} = 
           -\frac{1}{\gpW^b} \partial_\mu W^{\pm,b}_{\mu} 
           + \xi_{\varphi}^b M^b \varphi^{\pm,b},
\eqa
%--
where we used the superscript $b$ to denote bare quantities.
Next, we use fermion mass renormalization, $m^b = Z_{m}^{1\slash 2} m$, where 
the renormalization constant is expanded as
%--
\bq
Z_{m} = 1 + \frac{g^2}{16\pi^2} \delta Z_{m}^{(1)} + \ord{g^4},
\eq
%--
and the counterterms for lepton and top masses can be read in 
\eqn{eq:one:FER2}. We rewrite the first term in the r.h.s. of 
\eqn{eq:one:ASDE} as
%--
\bqa\label{eq:one:ASDEsaa}
G^{\ssA\ssA\,;\,1\fer}_{d\irr}(s) &=&
G^{\ssA\ssA\,;\,1\fer}_{d\irr\ren}(s) +
\Delta G^{\ssA\ssA\,;\,2\fer}_{d\irr\ren}(s),
\eqa
%--
where functions in in the r.h.s. of \eqn{eq:one:ASDEsaa} depend on 
renormalized quantities. The first term in r.h.s. of \eqn{eq:one:ASDEsaa} 
represents the one-loop fermionic contribution to the renormalized photon 
self-energy, whereas the second term gives an additional contribution to the 
two-loop result.

We verified that, after adding the second term in the r.h.s. of 
\eqn{eq:one:ASDEsaa} to two-loop diagrams, the answer is free from 
logarithmic residues. Therefore, fermion-mass renormalization removes all 
logarithms in the residue of the simple UV pole for the fermionic part.

\begin{figure}[ht]
\SetWidth{1}
\begin{center}
%\framebox[13cm]{\parbox{13cm}{
\begin{eqnarray*}
\begin{picture}(320,60)(-60,0)
%
%\Text(0,40)[r]{\scriptsize{$\widetilde{\ca{G}}^{ij\,;\,2}_{\mu\nu}=$}}
%
\Text(90,40)[]{\scriptsize{$+$}}
\Text(20,50)[b]{\scriptsize{$W$}}
\ArrowLine(10,40)(30,40)
\ArrowLine(40,40)(65,40)
\CCirc(35,40){8}{Black}{Gray}
\ArrowLine(110,40)(135,40)
\ArrowLine(135,40)(165,40)
\CCirc(135,40){2}{Black}{Black}
\Text(135,50)[b]{\scriptsize{$\delta Z_{\ssM}^{(1)}$}}
\end{picture}
\\
\begin{picture}(320,60)(-60,0)
\Text(90,40)[]{\scriptsize{$+$}}
\Text(20,50)[b]{\scriptsize{$\varphi$}}
\DashArrowLine(10,40)(30,40){2}
\DashArrowLine(40,40)(65,40){2}
\CCirc(35,40){8}{Black}{Gray}
\DashArrowLine(110,40)(135,40){2}
\DashArrowLine(135,40)(165,40){2}
\CCirc(135,40){2}{Black}{Black}
\Text(135,50)[b]{\scriptsize{$\delta Z_{\ssM}^{(1)}$}}
\end{picture}
\\
\begin{picture}(320,60)(-60,0)
\Text(90,40)[]{\scriptsize{$+$}}
\Text(20,50)[b]{\scriptsize{$X$}}
\DashArrowLine(10,40)(30,40){1}
\DashArrowLine(40,40)(65,40){1}
\CCirc(35,40){8}{Black}{Gray}
\DashArrowLine(110,40)(135,40){1}
\DashArrowLine(135,40)(165,40){1}
\CCirc(135,40){2}{Black}{Black}
\Text(135,50)[b]{\scriptsize{$\delta Z_{\ssM}^{(1)}$}}
\end{picture}
\end{eqnarray*}
%}}
\end{center}
\caption[]{
$W$-mass counterterm insertion in the charged one-loop transitions.
The $W-W$ one is UV finite, whereas the same is not true
for $\varphi -\varphi$ and ghost-ghost transitions.}
\label{fig:one:1234}
\end{figure}
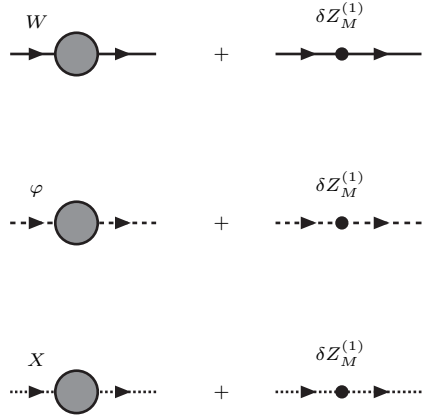

\begin{figure}[ht]
\SetWidth{1}
\begin{center}
%\framebox[13cm]{\parbox{13cm}{
\begin{eqnarray*}
\begin{picture}(360,80)(-60,0)
\Text(90,40)[]{\scriptsize{$+$}}
\Text(20,50)[b]{\scriptsize{$W$}}
\ArrowLine(10,40)(30,40)
\ArrowLine(40,40)(65,40)
\CCirc(35,40){8}{Black}{Gray}
\ArrowLine(110,40)(135,40)
\ArrowLine(135,40)(165,40)
\CCirc(135,40){2}{Black}{Black}
\Text(135,50)[b]{\scriptsize{$\delta Z_{\ssM}^{(1)}$}}
\end{picture}\\
\begin{picture}(360,80)(-60,0)
\Text(90,40)[]{\scriptsize{$+$}}
\Text(20,50)[b]{\scriptsize{$\varphi$}}
\DashArrowLine(10,40)(30,40){2}
\DashArrowLine(40,40)(65,40){2}
\CCirc(35,40){8}{Black}{Gray}
\DashArrowLine(110,40)(135,40){2}
\DashArrowLine(135,40)(165,40){2}
\CCirc(135,40){2}{Black}{Black}
\Text(135,50)[b]{\scriptsize{$\delta Z_{\ssM}^{(1)}
+ 2 \delta Z^{(1)}_{\xi_{\varphi}}$}}
\end{picture}\\
\begin{picture}(360,80)(-60,0)
\Text(20,50)[b]{\scriptsize{$W$}}
\Text(55,50)[b]{\scriptsize{$\varphi$}}
\ArrowLine(10,40)(30,40)
\DashArrowLine(40,40)(65,40){1}
%\Line(65,40)(90,40)
\CCirc(35,40){8}{Black}{Gray}
\ArrowLine(110,40)(135,40)
\DashArrowLine(135,40)(165,40){1}
%\Line(265,40)(290,40)
%\CCirc(265,40){2}{Black}{Black}
\CCirc(135,40){2}{Black}{Black}
\Text(135,50)[b]{\scriptsize{$
 \delta Z_{\gpW}^{(1)}
- \delta Z_{\xi_{\varphi}}^{(1)}
$}}
\end{picture}\\
\begin{picture}(360,80)(-60,0)
\Text(20,50)[b]{\scriptsize{$X$}}
\DashArrowLine(10,40)(30,40){1}
\DashArrowLine(40,40)(65,40){1}
\CCirc(35,40){8}{Black}{Gray}
\DashArrowLine(110,40)(135,40){1}
\DashArrowLine(135,40)(165,40){1}
\CCirc(135,40){2}{Black}{Black}
\Text(135,50)[b]{\scriptsize{$\delta Z_{\ssM}^{(1)}
+ \delta Z_{\gpW}^{(1)}
+ \delta Z_{\xi_{\varphi}}^{(1)}
$}}
\end{picture}
\end{eqnarray*}
%}}
\end{center}
\caption[]{
The correct recipe for renormalization of mass-dependent UV poles in the 
charged sector.}
\label{fig:one:12345}
\end{figure}
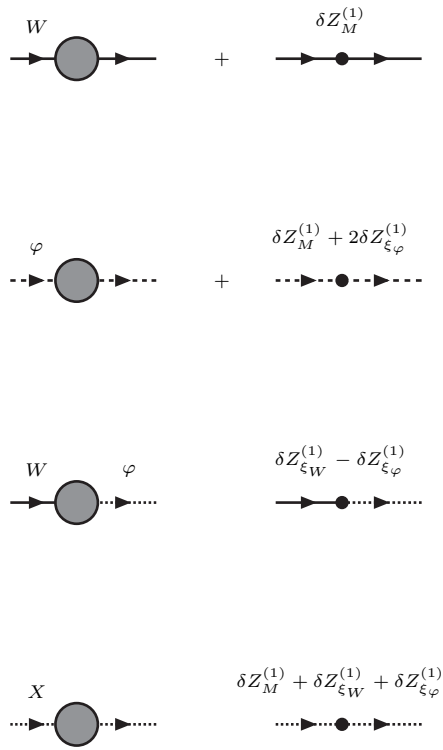

However, a non-local residue remains in the bosonic part.
Unfortunately, a simple procedure of $W$-mass renormalization is not enough
to get rid of logarithmic residues in the bosonic component and
the reason is that in a bosonic loop we may have three different fields, the 
$W$, the $\varphi$ and the charged ghost fields, and only one mass is 
available. The situation is illustrated in \fig{fig:one:1234} 
where the black dot denotes insertion of a counterterm $\delta Z_{\ssM}$.
The latter is fixed in order to remove the UV pole in the $W$ 
self-energy and one verifies that the total in the second and third line 
of \fig{fig:one:1234} ($\varphi$ and $X$ self-energies, respectively)
is not UV finite.
%--
%--
The procedure has to be changed if we want to make the result in the bosonic 
sector as similar as possible to the one in the fermionic sector.
A diagrammatic interpretation of the method is displayed 
in \fig{fig:one:12345}. For the bosonic part we have to take into account 
Dyson-resummed propagators. Let us consider the following integral, 
corresponding to a $\varphi$ loop in the photon self-energy,
%--
\bqa
H_{\mu\nu}^{\ssA\ssA\,;\,1} &=& (g^b)^2(s^b_{\theta})^2
\mu^{4-n}\int d^{n} q
\,\frac{[q^2+(\gpW^b)^2 M^2_b] [(q+p)^2 
+(\gpW^b)^2 M^2_b]} {[ q^2 + \gpW^b \xi^b_{\varphi} M^2_b]
[(q+p)^2+\gpW^b\xi^b_{\varphi}M^2_b]}
(2 q_{\mu} + p_{\mu}) (2 q_{\nu} + p_{\nu}).
\eqa
%--
We introduce renormalized quantities for the gauge parameters and the 
$W$-boson mass,
%--
\bq
M^b = Z_{\ssM}^{1\slash 2} M,\qquad
\gpW = Z_{\gpW} \gpW,\qquad
\xi_{\varphi} = Z_{\xi_{\varphi}} \xi_{\varphi},
\eq
%--
and expand the renormalization constants,
%--
\bq
Z_{i} = 1 + \frac{g^2}{16 \pi^2} \delta Z_{i}^{(1)} + \ord{g^2}.
\eq
%--
For the propagators we get
%--
\bqa\label{eq:one:HJHJ}
[q^2 + \xi_{\varphi}^b \gpW^b M^2_b]^{-1} &=& (q^2+M^2)^{-1}
- \frac{g^2}{16 \pi^2} (\delta Z_{\gpW} + \delta Z_{\xi_{\varphi}}
+ \delta Z_{\ssM} )  M^2 (q^2+M^2)^{-2} + \ldots.
\eqa
%--
Therefore, each one-loop bosonic diagram can be written as
%--
\bqa\label{eq:one:ASDEsaaa}
H^{\ssA\ssA\,;\,1}_{d\irr}(s) &=&
H^{\ssA\ssA\,;\,1}_{d\irr\ren}(s) +
\Delta H^{\ssA\ssA\,;\,2}_{d\irr\,;\,\ren}(s),
\eqa
%--
where the second term in the r.h.s. of \eqn{eq:one:ASDEsaaa} gives an 
additional contribution to the two-loop photon self-energy which depends on 
the $W$-mass and gauge-parameter counterterms. Collecting all the diagrams, 
we get 
%--
\bqa\label{eq:one:ASDEsaawe}
G^{\ssA\ssA\,;\,1\bos}_{d\irr}(s) &=&
G^{\ssA\ssA\,;\,1\bos}_{d\irr\ren}(s)
+ \Delta G^{\ssA\ssA\,;\,2\bos}_{d\irr\ren}(s).
\eqa
%--
We verified that $W$-mass and gauge-parameter renormalization removes all 
the non-local residues also for the bosonic component of the photon two-loop 
self-energy.

We emphasize the importance of using the $R_{\xi\xi}$ gauge of 
\eqn{eq:th:gfun}: renormalization of gauge parameters is essential in removing 
UV divergencies at one loop; furthermore, unitarity requires that any 
subtraction term must be local.

Summarizing, we have been able to verify that the electroweak theory can be 
made (two-loop) UV finite by adding local counterterms with two different 
methods, each leading to the same result.
In other words, the well-known one-loop result that self-energies suffice
in performing renormalization can be extended up to two loops. Although
counterterms have been explicitly included it remains true that:
%--
\bei
\item[--] In extracting $\alpha$ from Thomson scattering at zero momentum 
transfer we find four classes of two-loop diagrams:

\bei
\item[I)] irreducible two-loop vertices and wave-function factors, product
of one-loop corrected vertices with one-loop wave-function factors;

\item[II)] one-loop vacuum polarization $\,\otimes\,$ one-loop vertices or
one-loop wave-function factors;

\item[III)] irreducible two-loop $AA, AZ, A\phi^0$ transitions;

\item[IV)] reducible two-loop $AA, AZ, A\phi^0$ transitions.

\eei

We have verified that the non-vanishing contribution originates from III and 
IV only and, within these terms, only the reducible and irreducible $AA$ 
transition survives. For this result the role of $\Gamma$ is vital.
%--
\item[--] In extracting the Fermi coupling constant from the muon lifetime
all corrections to
%--
\bq
\frac{\gf}{\sqrt{2}} = \frac{g^2}{8\,M^2}\,(1 + \Delta g)
\eq
%--
which do not originate from the $W$ self-energy and that are UV 
(and IR) finite at one loop remain finite at two loops after
one-loop renormalization (i.e. two-loop counterterms are not needed), 
\eei
%--
The proof has been obtained by using $\GS$~\cite{GraphShot}, generating the 
whole set of corrections. The result follows by algebraic methods (see 
\appendx{limits}), i.e. full reduction of tensor structures without using the 
explicit expressions for the scalar integrals.
%--
\subsection{The $\gamma$-$Z$ transition}
\label{zg}
%--
For the transition between the photon and the $Z$ boson we show again all the 
various components of \eqn{eq:one:IrreTwo2}.
\vspace{0.2cm}

\noindent
\underline{\emph{Two light-quark doublets}}
%--
\bq
G^{\ssA\ssZ\,;\,2\,;\,lq}_{d\irr\uv;1}(s) =
2\,G^{\ssA\ssZ\,;\,2\,;\,lq}_{d\irr\uv;3}(s) =
 \frac{g^4}{\pi^4}\frac{\stw}{\ctw} \frac{3}{64}
\Bigl( - \ctw^2 s + \frac{M^2}{2} \Bigr),
\eq
%--
\bqa
G^{\ssA\ssZ\,;\,2\,;\,lq}_{d\irr\uv;2}(s) &=&
 \frac{g^4}{\pi^4}\frac{\stw}{\ctw}
\Bigl\{  \frac{s}{54} \Bigl( 1 - \frac{137}{256} \frac{1}{\ctw^2}
+ \frac{251}{128} \ctw^2 \Bigr) + \frac{M^2}{16}
\Bigl( \frac{29}{32} + 3 \ctw^2 \Bigr)
\nl
{}&+& \frac{1}{8}\frac{L_{\beta}(y_{\ssW})}{ \beta(y_{\ssW})} 
\Bigl[ - s \frac{\ctw^2}{8} + M^2
\Bigl( - \frac{3}{16} + \frac{\ctw^2}{4} \Bigr)
+ M^2 y_{\ssW} \Bigl( 1 + 3 \ctw^2 \Bigr)\Bigr]
+ \frac{g_\ssS^2}{g^2} s \Bigl( \frac{5}{72} c_\theta^2 - \frac{11}{288}\Bigr)
\Bigr\}.
\eqa
%--
\noindent
\underline{\emph{Top-bottom doublet}}
%--
\bq
G^{\ssA\ssZ\,;\,2\,;\,tb}_{d\irr\uv;1}(s) =
2\,G^{\ssA\ssZ\,;\,2\,;\,tb}_{d\irr\uv;3}(s) =
 \frac{g^4}{\pi^4}\frac{\stw}{\ctw} \frac{3}{128}
\Bigl( - \ctw^2 s + \frac{M^2}{2} \Bigr),
\eq
%--
\bqa
G^{\ssA\ssZ\,;\,2\,;\,tb}_{d\irr\uv;2}(s) &=&
 \frac{g^4}{\pi^4}\frac{\stw}{\ctw}
\Bigl\{ \frac{s}{12} \Bigl( \frac{1}{9} - \frac{137}{2304} \frac{1}{\ctw^2}
- \frac{13}{128} \ctw^2 x_t + \frac{251}{1152} \ctw^2 + \frac{17}{256} x_t
\Bigr)
\nl
{}&+& 
\frac{M^2}{32} \Bigl( \frac{29}{32} - \frac{9}{2} \ctw^2 x_t -
3 \ctw^2 x_t^2 + 3 \ctw^2 - \frac{3}{4} x_t +
\frac{3}{2} x_t^2 \Bigr)
\nl
{}&+& \frac{1}{16}\frac{L_{\beta}(y_{\ssW})}{ \beta(y_{\ssW})} 
\Bigl[ -\frac{s}{8} \ctw^2 + \frac{M^2}{4} \Bigl(
- \frac{3}{4} -\frac{9}{2}\ctw^2 x_t + 3 \ctw^2 x_t^2
+ \ctw^2 + \frac{9}{4} x_t - \frac{3}{2} x_t^2 \Bigr)
\nl
{}&+& M^2 y_{\ssW} \Bigl( 1 - \frac{9}{2} \ctw^2 x_t
- \frac{9}{4} \ctw^2 x_t^2 + 3 \ctw^2 - \frac{9}{8} x_t + \frac{3}{4} x_t^2
\Bigr) \Bigr]
+ \frac{g_\ssS^2}{g^2} s \Bigl( \frac{5}{144} c_\theta^2 - \frac{11}{576}\Bigr)
 \Bigr\}.
\eqa
%--
\noindent
\underline{\emph{Three lepton doublets}}
%--
\bq
G^{\ssA\ssZ\,;\,2\lep}_{d\irr\uv;1}(s) =
2\,G^{\ssA\ssZ\,;\,2\lep}_{d\irr\uv;3}(s) =
 \frac{g^4}{\pi^4}\frac{\stw}{\ctw} \frac{3}{128}
\Bigl( - \ctw^2 s + \frac{M^2}{2} \Bigr),
\eq
%--
\bqa
G^{\ssA\ssZ\,;\,2\lep}_{d\irr\uv;2}(s) &=&
 \frac{g^4}{\pi^4}\frac{\stw}{\ctw}
\Bigl\{ \frac{s}{64} \Bigl( 3 -\frac{27}{16} \frac{1}{\ctw^2}
+ \frac{1}{8} \ctw^2 \Bigr) + \frac{M^2}{32} \Bigl(
\frac{29}{32} + 3 \ctw^2 \Bigr)
\nl
{}&+& \frac{1}{16} \frac{L_{\beta}(y_{\ssW})} {\beta(y_{\ssW})}
\Bigl[ - \frac{s}{8}\ctw^2 + \frac{M^2}{4} \Bigl(
- \frac{3}{4} + \ctw^2 \Bigr) + M^2 y_{\ssW} \Bigl( 1 + 3 \ctw^2
\Bigr)\Bigr]\Bigr\}.
\eqa
%--
\noindent
\underline{\emph{Bosonic contributions}}
%--
\bq
G^{\ssA\ssZ\,;\,2\bos}_{d\irr\uv;1}(s) =
 \frac{g^4}{\pi^4}\frac{\stw}{\ctw}
\frac{1}{64} \Bigl[ \frac{s}{2} \Bigl( - \frac{943}{18} + 119 \ctw^2
- \frac{529}{9} \ctw^4 \Bigr) - 3 M^2 \Bigr],
\eq
%--
\bqa
G^{\ssA\ssZ\,;\,2\bos}_{d\irr\uv;2}(s) &=&
 \frac{g^4}{\pi^4}\frac{\stw}{\ctw}
\Bigl\{ \frac{s}{512} \Bigl( - \frac{1}{\ctw^2} - \frac{21}{2} \ctw^2
\Bigr) + \frac{M^2}{8} \Bigl(
-\frac{29}{32} + \frac{9}{8} \ctw^2 x_t
+ \frac{3}{4} \ctw^2 x_t^2
- 3 \ctw^2 + \frac{3}{16} x_t - \frac{3}{8} x_t^2 \Bigr)
\nl
{}&+& \frac{1}{4}\frac{L_{\beta}(y_{\ssW})}{ \beta(y_{\ssW})}
\Bigl[ \frac{s}{8} \ctw^2 + \frac{M^2}{4}
\Bigl(  \frac{3}{4} + \frac{9}{8}\ctw^2 x_t - \frac{3}{4}\ctw^2 x_t^2
- \ctw^2 - \frac{9}{16} x_t + \frac{3}{8} x_t^2 \Bigr)
\nl
{}&+& M^2 y_{\ssW} \Bigl( - 1 + \frac{9}{8} \ctw^2 x_t 
+ \frac{9}{16} \ctw^2 x_t^2 - 3 \ctw^2 + \frac{9}{32} x_t
- \frac{3}{16} x_t^2 \Bigr) \Bigr] \Bigr\},
\eqa
%--
\bq
G^{\ssA\ssZ\,;\,2\bos}_{d\irr\uv;3}(s) =
 \frac{g^4}{\pi^4}\frac{\stw}{\ctw} \frac{1}{128}
\Bigl[ \frac{s}{12} \Bigl( - \frac{1886}{3} + 383 \ctw^2
- \frac{529}{3} \ctw^4 \Bigr) - 3 M^2 \Bigr].
\eq
%--
\noindent
\underline{\emph{Total contributions}}
\vspace{0.2cm}

\noindent
The sum of the four components gives
\bq
G^{\ssA\ssZ\,;\,2}_{d\irr\uv;1}(s) =
 \frac{g^4}{\pi^4}\frac{\stw}{\ctw} \frac{s}{128}
\Bigl( - \frac{943}{18} + 107 \ctw^2 - \frac{529}{9} \ctw^4 \Bigr),
\eq
\bq
G^{\ssA\ssZ\,;\,2}_{d\irr\uv;2}(s) =
 \frac{g^4}{\pi^4}\frac{\stw}{\ctw} \frac{s}{64}
\Bigl[ \frac{43}{9} -\frac{199}{72} \frac{1}{\ctw^2} - 
\frac{13}{24} \ctw^2 x_t
+ \frac{331}{144} \ctw^2 + \frac{17}{48} x_t + 
\frac{g_\ssS^2}{g^2} \Bigl( \frac{20}{3} c_\theta^2 - \frac{11}{3}\Bigr)\Bigr],
\eq
%--
\bq
G^{\ssA\ssZ\,;\,2}_{d\irr\uv;3}(s) =
 \frac{g^4}{\pi^4}\frac{\stw}{\ctw} \frac{s}{1536}
\Bigl( - \frac{1886}{3} + 311 \ctw^2 - \frac{529}{3} \ctw^4 \Bigr),
\eq
%--
\noindent
and can be absorbed by the contribution of the counterterms,
%--
\bq
G^{\ssA\ssZ\,;\,2\,;\,ct}_{d\irr\uv;a}(s) =
 \frac{g^4 s }{256\pi^4} \Delta Z_{\ssA\ssZ\,;\,a}^{(2)},\qquad a=1,2,3.
\eq
%--
\subsection{The $Z$-boson self-energy}
\label{z}
%--
\noindent
The explicit expressions for the individual components of the $Z$-boson 
self-energy are rather lengthy. Here we provide the total result for the 
coefficients of the UV factors and we show that non-local residues cancel,
%--
\bqa
G^{\ssZ\ssZ\,;\,2}_{d\irr\uv;1}(s) &=&
\frac{g^4}{\pi^4 \ctw^2} 
\Bigl[ 
\frac{s}{576}
\Bigl( 
\frac{1681}{4} 
- \frac{5453}{4} \ctw^2 
+ \frac{5813}{4} \ctw^4
- 529 \ctw^{6} 
\Bigr)
\nl
{}&+& \frac{M^2}{32}
\Bigl( 
\frac{10385}{288}
+ \frac{9}{8}\frac{1}{\ctw^{8} x_{\ssH}^2}
- \frac{25}{4}\frac{1}{\ctw^{6} x_{\ssH}}
- 9 \frac{x_t^2}{\ctw^4x_{\ssH}^2}
+ \frac{9}{2}\frac{1}{\ctw^4x_{\ssH}^2}
+ \frac{9}{4}\frac{x_t}{\ctw^4x_{\ssH}}
\nl
{}&+& \frac{59}{8} \frac{1}{\ctw^4x_{\ssH}}
+ \frac{4667}{576} \frac{1}{\ctw^4}
+ \frac{1}{4} \frac{x_t^2}{\ctw^2 x_{\ssH}}
- 5 \frac{1}{\ctw^2 x_{\ssH}}
- \frac{1}{32} \frac{x_t}{\ctw^2}
- \frac{425}{18} \frac{1}{\ctw^2}
\nl
{}&-& \frac{1771}{72} \ctw^2
+ \frac{529}{72} \ctw^4
- 18 \frac{x_t^2}{x_{\ssH}^2}
+ 18  \frac{x_t^4}{x_{\ssH}^2}
+ \frac{9}{2} \frac{1}{x_{\ssH}^2}
+ \frac{9}{2} \frac{x_t}{x_{\ssH}}
+ 9 \frac{x_t^2}{x_{\ssH}}
- \frac{9}{4} \frac{x_t^3}{x_{\ssH}}
\nl
{}&-&  \frac{1}{x_{\ssH}}
- \frac{5}{16} x_{\ssH}
- \frac{9}{64} x_{\ssH}^2
- \frac{23}{16} x_t
- \frac{117}{32} x_t^2
\Bigr)
+ \frac{g_\ssS^2}{g^2}M^2 \Bigl(
-\frac{3}{4} \frac{x_t^2}{x_\ssH}
+ \frac{3}{32}x_t
\Bigr)
\Bigr],
\eqa
%--
\bqa
G^{\ssZ\ssZ\,;\,2}_{d\irr\uv;2}(s) &=&
 \frac{g^4}{\pi^4 \ctw^2} 
\Bigl[ \frac{s}{384}
\Bigl(- 43
+ \frac{199}{12}\frac{1}{\ctw^2}
+ \frac{17}{4} \ctw^2 x_t
+ \frac{77}{2} \ctw^2
- \frac{13}{4} \ctw^4 x_t
+\frac{331}{24} \ctw^4- \frac{17}{8} x_t \Bigr)
\nl
{}&+&\frac{M^2}{32}
\Bigl(
- \frac{5}{12}
+ \frac{379}{96}\frac{1}{\ctw^{6} x_{\ssH}}
+ \frac{19}{16}\frac{x_t}{\ctw^4 x_{\ssH}}
- \frac{289}{48}\frac{1}{\ctw^4 x_{\ssH}}
- \frac{3383}{2304}\frac{1}{\ctw^4}
- 5 \frac{x_t}{\ctw^2 x_{\ssH}}
\nl
{}&+& \frac{1}{3} \frac{x_t^2}{\ctw^2 x_{\ssH}}
+ \frac{77}{24}\frac{1}{\ctw^2 x_{\ssH}}
- \frac{3}{32}\frac{x_{\ssH}}{\ctw^2}
- \frac{85}{384}\frac{x_t}{\ctw^2}
+ \frac{4}{3}\frac{1}{\ctw^2}
+ \frac{35}{8}\frac{x_t}{x_{\ssH}}
- \frac{1}{3}\frac{x_t^2}{x_{\ssH}}
\nl
{}&-& \frac{15}{8}\frac{x_t^3}{x_{\ssH}}
- \frac{32}{3}\frac{1}{x_{\ssH}}
+ \frac{9}{32} x_tx_{\ssH}
- \frac{3}{16} x_{\ssH}
+ \frac{63}{256} x_{\ssH}^2
- \frac{25}{192} x_t
- \frac{21}{128} x_t^2
\Bigr)
\nl
{}&+& \frac{g_\ssS^2}{g^2}
s \Bigl(
\frac{11}{192}
- \frac{11}{96} c_\theta^2
+ \frac{5}{48} c_\theta^4
\Bigr)
+ \frac{g_\ssS^2}{g^2}
M^2
\Bigl(
\frac{1}{8} \frac{x_t^2}{x_\ssH}
- \frac{5}{128} x_t
\Bigr)
\Bigr],
\eqa
%--
\bqa
G^{\ssZ\ssZ\,;\,2}_{d\irr\uv;3}(s) &=&
 \frac{g^4}{\pi^4 \ctw^2} 
\Bigl[\frac{s}{2304}
\Bigl(
\frac{1681}{4}
- \frac{5453}{4} \ctw^2
+ \frac{5753}{4} \ctw^4
- 529 \ctw^{6}
\Bigr)
\nl
{}&+& \frac{M^2}{128}
\Bigl(
 \frac{9737}{288}
+\frac{9}{8}\frac{1}{\ctw^{8} x_{\ssH}^2}
-\frac{25}{4}\frac{1}{\ctw^{6} x_{\ssH}}
- 9 \frac{x_t^2}{\ctw^4x_{\ssH}^2} 
+\frac{9}{2}\frac{1}{\ctw^4x_{\ssH}^2}
+\frac{9}{4}\frac{x_t}{\ctw^4x_{\ssH}}
\nl 
{}&+& \frac{59}{8} \frac{1}{\ctw^4x_{\ssH}}
+\frac{4667}{576} \frac{1}{\ctw^4}
+\frac{1}{4} \frac{x_t^2}{\ctw^2 x_{\ssH}}
-2  \frac{1}{\ctw^2 x_{\ssH}}
-\frac{1}{32} \frac{x_t}{\ctw^2}
-\frac{425}{18} \frac{1}{\ctw^2}
\nl
{}&-& \frac{1771}{72} \ctw^2
+\frac{529}{72} \ctw^4
- 18 \frac{x_t^2}{x_{\ssH}^2}
+ 18 \frac{x_t^4}{x_{\ssH}^2}
+\frac{9}{2} \frac{1}{x_{\ssH}^2}
+\frac{9}{2}\frac{x_t}{x_{\ssH}}
+ 9 \frac{x_t^2}{x_{\ssH}}
-\frac{9}{4} \frac{x_t^3}{x_{\ssH}}
\nl
{}&+& 5 \frac{1}{x_{\ssH}}
-\frac{5}{16}  x_{\ssH}
-\frac{9}{64} x_{\ssH}^2
-\frac{23}{16} x_t
-\frac{117}{32} x_t^2
\Bigr)
+ \frac{g_\ssS^2}{g^2}
M^2
\Bigl(
- \frac{3}{16} \frac{x_t^2}{x_\ssH}
+ \frac{3}{128} x_t
\Bigr)
\Bigr].
\eqa
As evident, the two-loop counterterms contain only local residues,
%--
\bq
G_{d\irr\uv;a}^{\ssZ\ssZ\,;\,2\ct}(s) =
\frac{g^4}{256 \pi^4}\
\Bigl[ s \ \Delta Z^{(2)}_{\ssZ;a} - \frac{M^2}{\ctw^2} \Bigl(
\Delta Z^{(2)}_{\ssZ;a} + \Delta Z^{(2)}_{\ssM;a}
- 2 \Delta Z^{(2)}_{\ctw;a} \Bigr) \Bigr].
\eq
%--
\subsection{The $W$-boson self-energy}
\label{w}
%--
\noindent
For the $W$-boson self-energy we provide the total result which shows
once again absence of non-local residues.
%--
\bqa
G^{\ssW\ssW\,;\, 2}_{d\irr\uv;1}(s) &=&
 \frac{g^4}{\pi^4} 
\Bigl[
-\frac{25}{768}s
+\frac{M^2}{32}
\Bigl(
\frac{49}{18}
+ \frac{9}{8} \frac{1}{\ctw^{8} x_{\ssH}^2}
- \frac{25}{4}\frac{1}{\ctw^{6} x_{\ssH}}
\nl
{}&-& 9\frac{x_t^2}{\ctw^4 x_{\ssH}^2}
+ \frac{9}{2} \frac{1}{\ctw^4 x_{\ssH}^2}
+ \frac{9}{4} \frac{x_t}{\ctw^4 x_{\ssH}}
+ \frac{59}{8} \frac{1}{\ctw^4 x_{\ssH}}
+ \frac{145}{64} \frac{1}{\ctw^4}
+ \frac{1}{4} \frac{x_t^2}{\ctw^2 x_{\ssH}}
\nl
{}&-& 5 \frac{1}{\ctw^2 x_{\ssH}}
- \frac{1}{32} \frac{x_t}{\ctw^2}
- \frac{27}{16} \frac{1}{\ctw^2}
- 18 \frac{x_t^2}{x_{\ssH}^2}
+ 18  \frac{x_t^4}{x_{\ssH}^2}
+ \frac{9}{2} \frac{1}{x_{\ssH}^2}
+ \frac{9}{2} \frac{x_t}{x_{\ssH}}
\nl
{}&+& 9 \frac{x_t^2}{x_{\ssH}}
- \frac{9}{4} \frac{x_t^3}{x_{\ssH}}
-  \frac{1}{x_{\ssH}}
- \frac{5}{16} x_{\ssH}
- \frac{9}{64} x_{\ssH}^2
-\frac{23}{16} x_t
-\frac{117}{32} x_t^2
\Bigr)
+ \frac{g_\ssS^2}{g^2}
M^2
\Bigl(
- \frac{3}{4} \frac{x_t^2}{x_\ssH}
+ \frac{3}{32} x_t
\Bigr)
\Bigr],
\eqa
%--
\bqa
G^{\ssW\ssW\,;\,2}_{d\irr\uv;2}(s) &=&
 \frac{g^4}{\pi^4} 
\Bigl[
\frac{s}{512}
\Bigl(
\frac{63}{2}
+ 3 \frac{1}{\ctw^2}
- \frac{3}{2} x_t
\Bigr)
\nl
{}&+& \frac{M^2}{32}
\Bigl(
-\frac{5}{12}
+ \frac{379}{96} \frac{1}{\ctw^{6} x_{\ssH}}
+ \frac{19}{16} \frac{x_t}{\ctw^4 x_{\ssH}}
- \frac{289}{48} \frac{1}{\ctw^4 x_{\ssH}}
- \frac{701}{768} \frac{1}{\ctw^4}
\nl
{}&-& 5 \frac{x_t}{\ctw^2 x_{\ssH}}
+ \frac{1}{3} \frac{x_t^2}{\ctw^2 x_{\ssH}}
+ \frac{77}{24} \frac{1}{\ctw^2 x_{\ssH}}
- \frac{3}{32} \frac{x_{\ssH}}{\ctw^2}
- \frac{85}{384} \frac{x_t}{\ctw^2}
+ \frac{4}{3} \frac{1}{\ctw^2}
+ \frac{35}{8} \frac{x_t}{x_{\ssH}}
\nl
{}&-& \frac{1}{3} \frac{x_t^2}{x_{\ssH}}
- \frac{15}{8} \frac{x_t^3}{x_{\ssH}}
- \frac{32}{3} \frac{1}{x_{\ssH}}
+ \frac{9}{32} x_t x_{\ssH}
- \frac{3}{16} x_{\ssH}
+ \frac{63}{256} x_{\ssH}^2
- \frac{25}{192} x_t
- \frac{21}{128} x_t^2
\Bigr)
\nl
{}
&+& \frac{g_\ssS^2}{g^2}
M^2
\Bigl(
\frac{1}{8} \frac{x_t^2}{x_\ssH}
- \frac{5}{128} x_t
\Bigr)
+ \frac{g_\ssS^2}{g^2}
s
\frac{3}{64}
\Bigr],
\eqa
%--
\bqa
G^{\ssW\ssW\,;\,2}_{d\irr\uv;3}(s) &= &
\frac{g^4}{\pi^4}
\Bigl[
 - \frac{15}{1024}\,s
+ 
\frac{M^2}{32}
\Bigl(
\frac{17}{144}
+ \frac{9}{32} \frac{1}{\ctw^{8} x_{\ssH}^2}
- \frac{25}{16}\frac{1}{\ctw^{6} x_{\ssH}}
\nl
{}&-& \frac{9}{4}\frac{x_t^2}{\ctw^4 x_{\ssH}^2}
+ \frac{9}{8} \frac{1}{\ctw^4 x_{\ssH}^2}
+ \frac{9}{16} \frac{x_t}{\ctw^4 x_{\ssH}}
+ \frac{59}{32} \frac{1}{\ctw^4 x_{\ssH}}
+ \frac{145}{256} \frac{1}{\ctw^4}
+ \frac{1}{16} \frac{x_t^2}{\ctw^2 x_{\ssH}}
\nl
{}&-& \frac{1}{2} \frac{1}{\ctw^2 x_{\ssH}}
- \frac{1}{128} \frac{x_t}{\ctw^2}
- \frac{27}{64} \frac{1}{\ctw^2}
- \frac{9}{2} \frac{x_t^2}{x_{\ssH}^2}
+ \frac{9}{2}  \frac{x_t^4}{x_{\ssH}^2}
+ \frac{9}{8} \frac{1}{x_{\ssH}^2}
+ \frac{9}{8} \frac{x_t}{x_{\ssH}}
\nl
{}&+& \frac{9}{4} \frac{x_t^2}{x_{\ssH}}
- \frac{9}{16} \frac{x_t^3}{x_{\ssH}}
+  \frac{5}{4} \frac{1}{x_{\ssH}}
- \frac{5}{64} x_{\ssH}
- \frac{9}{256} x_{\ssH}^2
-\frac{23}{64} x_t
-\frac{117}{128} x_t^2
\Bigr)
\nl
{}&+& \frac{g_\ssS^2}{g^2}
M^2
\Bigl(
- \frac{3}{16} \frac{x_t^2}{x_\ssH}
+ \frac{3}{128} x_t
\Bigr)
\Bigr]
.
\eqa
%--
\bq
G_{d\irr\uv;a}^{\ssW\ssW\,;\,2\ct}(s) =
 \ \frac{g^4}{256 \pi^4}\
\Bigl[ s \ \Delta Z^{(2)}_{\ssW;a} - M^2 \Bigl( \Delta Z^{(2)}_{\ssW;a} +
\Delta Z^{(2)}_{\ssM;a} \Bigr) \Bigr].
\eq
%--
\subsection{The Higgs-boson self-energy}
\label{h}
%--
\noindent
Here we provide the result for the Higgs-boson self-energy, showing the 
absence of non-local residues,
%--
\bqa
G^{\ssH\ssH\, ; \,2}_{\ssU\ssV\irr;1}(s) &=&
 \frac{g^4}{\pi^4} 
\Bigl[
\frac{s}{64}
\Bigl(
- 1
- \frac{43}{16}\frac{1}{\ctw^4}
- \frac{5}{16} \frac{x_t}{\ctw^2}
+ \frac{37}{8}\frac{1}{\ctw^2}
+ \frac{7}{8} x_t
+ \frac{9}{16} x_t^2
\Bigr)
\nl
{}&+& 
\frac{M^2}{256}
\Bigl(
\frac{31}{8} \frac{x_{\ssH}}{\ctw^4}
- \frac{15}{2} \frac{x_{\ssH} x_t}{\ctw^2}
+ \frac{13}{2} \frac{x_{\ssH}}{\ctw^2}
- \frac{9}{2} \frac{x_{\ssH}^2}{\ctw^2}
- 15 x_{\ssH} x_t
\nl
{}&+& 17 x_{\ssH}
+ 9 x_{\ssH}^2 x_t
- 9 x_{\ssH}^2
+ \frac{27}{8} x_{\ssH}^3
\Bigr)
- \frac{3}{32} \frac{g_\ssS^2}{g^2}s x_t
\Bigr],
\eqa
%--
\bqa
G^{\ssH\ssH\,;\,2}_{\ssU\ssV \irr;2}(s) &=&
 \frac{g^4}{\pi^4} 
\Bigl[
\frac{s}{512}
\Bigl(
- \frac{7}{3}
+ \frac{431}{48} \frac{1}{\ctw^4}
+ \frac{85}{24} \frac{x_t}{\ctw^2}
- \frac{101}{6} \frac{1}{\ctw^2}
+ \frac{3}{16} x_{\ssH}^2
+ \frac{25}{12} x_t
- \frac{27}{8} x_t^2
\Bigr)
\nl
{}&+& 
\frac{M^2}{256}
\Bigl(
\frac{21}{16} \frac{x_{\ssH}}{\ctw^4}
- \frac{9}{4} \frac{x_{\ssH}}{\ctw^2}
+ \frac{3}{2} \frac{x_{\ssH}^2}{\ctw^2}
+ \frac{3}{2} x_{\ssH}
+ \frac{9}{4} x_{\ssH}^2 x_t
+ 3 x_{\ssH}^2
- \frac{9}{16} x_{\ssH}^3
\Bigr)
+ \frac{5}{128} \frac{g_\ssS^2}{g^2}s x_t
\Bigr],
\eqa
%--
\bqa
G^{\ssH\ssH\,;\,2}_{\ssU\ssV\irr;3}(s) &=&
 \frac{g^4}{\pi^4} 
\Bigl[
\frac{s}{128}
\Bigl(
1 - \frac{43}{32}\frac{1}{\ctw^4}
- \frac{5}{32} \frac{x_t}{\ctw^2}
+ \frac{37}{16} \frac{1}{\ctw^2}
+\frac{7}{16} x_t
+ \frac{9}{32} x_t^2
\Bigr)
\nl
{}&+& 
\frac{M^2}{1024}
\Bigl(
\frac{31}{8}\frac{x_{\ssH}}{\ctw^4}
- \frac{15}{2} \frac{x_{\ssH} x_t}{\ctw^2}
+ \frac{13}{2} \frac{x_{\ssH}}{\ctw^2}
- \frac{9}{2} \frac{x_{\ssH}^2}{\ctw^2}
- 15 x_{\ssH} x_t
\nl
{}&+&
 23 x_{\ssH}
+ 9 x_{\ssH}^2 x_t
-9 x_{\ssH}^2
+ \frac{27}{8} x_{\ssH}^3
\Bigr)
- \frac{3}{128} \frac{g_\ssS^2}{g^2}s x_t
\Bigr].
\eqa
Therefore, we can employ a non-local subtraction term,
%--
\bq
G_{\ssU\ssV\irr;a}^{\ssH\ssH\,;\,2\ct}(s) \, =\, \frac{g^4}{256 \, \pi^4}\,
\Bigl[\, s\,  \Delta Z^{(2)}_{\ssH;a} \,-\, M_{\ssH}^2\, \Bigl(\,
\Delta Z^{(2)}_{\ssH;a}\, +\, \Delta Z^{(2)}_{\mh;a}\, \Bigr)\, \Bigr].
\eq
%--
\section{Two-Loop Counterterms}\label{sec:one:ct2}
%--
In this Section we provide the full list of two-loop counterterms (for 
reason of space fermion masses other than the top-quark one are not shown, 
except for the first result).

Let us start with field counterterms, where the sum over the fermion masses 
was defined in \eqn{eq:one:SUMMA},
%--
\bq\label{eq:one:unoI}
\Delta Z_{\ssA\ssA\,;\,1}^{(2)} =
\frac{25}{3} s_\theta^2
,
\eq
%--
\bq
\Delta Z_{\ssA\ssA\,;\,2}^{(2)} =
\frac{923}{108}
- \frac{113}{9} \frac{1}{c_\theta^2}
+ \frac{13}{6}s_\theta^2  X_u
+ \frac{7}{6} s_\theta^2  X_d
+ \frac{3}{2} s_\theta^2  X_l
- \frac{127}{108} c_\theta^2
+ \frac{140}{27} c_\theta^4
- \frac{64}{9}s_\theta^2 \frac{g_{\ssS}^2}{g^2}
,
\eq
%--
\bq
\Delta Z_{\ssA\ssA\,;\,3}^{(2)} =
\frac{15}{4} s_\theta^2
.
\eq
%--
\bq
\Delta Z_{\ssA\ssZ\,;\,1}^{(2)} =
s_\theta \Bigl(
\frac{943}{9} \frac{1}{c_\theta}
- 214 c_\theta
+ \frac{1058}{9} c_\theta^3
\Bigr)
,
\eq
%--
\bq
\Delta Z_{\ssA\ssZ\,;\,2}^{(2)} =
s_\theta
\Bigl(
\frac{199}{18} \frac{1}{c_\theta^3}
- \frac{17}{12} \frac{x_t}{c_\theta}
- \frac{172}{9} \frac{1}{c_\theta}
+ \frac{13}{6} c_\theta x_t
- \frac{331}{36} c_\theta
+ \frac{44}{3} \frac{1}{c_\theta} \frac{g_{\ssS}^2}{g^2}
- \frac{80}{3} c_\theta \frac{g_{\ssS}^2}{g^2}
\Bigr)
,
\eq
%--
\bq
\Delta Z_{\ssA\ssZ\,;\,3}^{(2)} =
s_\theta \Bigl(
\frac{943}{9} \frac{1}{c_\theta}
- \frac{311}{6} c_\theta
+ \frac{529}{18} c_\theta^3
\Bigr)
.
\eq
%--
\bqa
\Delta Z_{\ssZ\,;\,1}^{(2)} &=&
\frac{1}{9}
\Bigl(
5453 
- 1681 \frac{1}{c_\theta^2}
- 5813 c_\theta^2
+ 2116 c_\theta^4
\Bigr)
,
\eqa
%--
\bqa
\Delta Z_{\ssZ\,;\,2}^{(2)} &=&
- \frac{77}{3}
- \frac{199}{18}\frac{1}{c_\theta^4}
+ \frac{17}{12} \frac{x_t}{c_\theta^2}
+ \frac{86}{3} \frac{1}{c_\theta^2}
+ \frac{13}{6} c_\theta^2 x_t
- \frac{331}{36} c_\theta^2 
- \frac{17}{6} x_t
\nonumber \\
{}\nl{}
&-&
\frac{44}{3} \frac{1}{c_\theta^2}\frac{g_{\ssS}^2}{g^2}
- \frac{80}{3} c_\theta^2 \frac{g_{\ssS}^2}{g^2}
+ \frac{88}{3} \frac{g_{\ssS}^2}{g^2}
,
\eqa
%--
\bqa
\Delta Z_{\ssZ\,;\,3}^{(2)} &=&
\frac{5453}{36}
- \frac{1681}{36} \frac{1}{c_\theta^2}
- \frac{5753}{36} c_\theta^2
+ \frac{529}{9} c_\theta^4
.
\eqa
%--
\bqa
\Delta Z_{\ssW\,;\,1}^{(2)} &=&
\frac{25}{3}
,\qquad
\Delta Z_{\ssW\,;\,2}^{(2)} =
-\frac{63}{4} -\frac{3}{2}\frac{1}{c_\theta^2}+ \frac{3}{4} x_t
- 12 \frac{g_{\ssS}^2}{g^2} 
,\qquad
\Delta Z_{\ssW\,;\,3}^{(2)} =
\frac{15}{4}
.
\eqa
%--
\bqa
\Delta Z_{\ssH\,;\,1}^{(2)} &=&
4
+ \frac{43}{4} \frac{1}{c_\theta^4}
+ \frac{5}{4} \frac{x_t}{c_\theta^2}
- \frac{37}{2} \frac{1}{c_\theta^2}
- \frac{7}{2} x_t
- \frac{9}{4} x_t^2
+ 24 x_t \frac{g_{\ssS}^2}{g^2}
,
\eqa
%--
\bqa
\Delta Z_{\ssH\,;\,2}^{(2)} &=&
\frac{7}{6}
- \frac{431}{96} \frac{1}{c_\theta^4}
- \frac{85}{48} \frac{x_t}{c_\theta^2}
+ \frac{101}{12} \frac{1}{c_\theta^2}
- \frac{25}{24} x_t
+ \frac{27}{16} x_t^2
- \frac{3}{32} x_\ssH^2
- 10 x_t \frac{g_{\ssS}^2}{g^2}
,
\eqa
%--
\bqa
\Delta Z_{\ssH\,;\,3}^{(2)} &=&
-2
+ \frac{43}{16} \frac{1}{c_\theta^4}
+ \frac{5}{16} \frac{x_t}{c_\theta^2}
- \frac{37}{8} \frac{1}{c_\theta^2}
- \frac{7}{8} x_t
- \frac{9}{16} x_t^2
+ 6 x_t \frac{g_{\ssS}^2}{g^2}
.
\eqa
%--
\noindent
For the $W$-mass counterterm we get
%--
\bqa
\Delta Z_{\ssM\,;\,1}^{(2)} &=&
\frac{121}{9}
+9 \frac{1}{c_\theta^8 x_\ssH^2}
-50  \frac{1}{c_\theta^6 x_\ssH}
+18\frac{x_t}{c_\theta^4 x_\ssH}
-72\frac{x_t^2}{c_\theta^4 x_\ssH^2}
+36\frac{1}{c_\theta^4 x_\ssH^2}
+59\frac{1}{c_\theta^4 x_\ssH}
+\frac{145}{8}\frac{1}{c_\theta^4} \nonumber \\
{}\nl{}
&-& \frac{1}{4}\frac{x_t}{c_\theta^2}
+2\frac{x_t^2}{c_\theta^2 x_\ssH}
-40\frac{1}{c_\theta^2 x_\ssH}
-\frac{27}{2}\frac{1}{c_\theta^2}
+36\frac{x_t}{x_\ssH} 
-\frac{23}{2} x_t
-144\frac{x_t^2}{x_\ssH^2}
+72\frac{x_t^2}{x_\ssH}\nonumber \\
{}\nl{}
&-&\frac{117}{4} x_t^2
-18\frac{x_t^3}{x_\ssH}
+144\frac{x_t^4}{x_\ssH^2}
+36\frac{1}{x_\ssH^2}
-8 \frac{1}{x_\ssH}
-\frac{5}{2} x_\ssH
-\frac{9}{8} x_\ssH^2
+24 x_t \frac{g_{\ssS}^2}{g^2}
-192 \frac{x_t^2}{x_\ssH} \frac{g_{\ssS}^2}{g^2},
\eqa
%--
\bqa
\Delta Z_{\ssM\,;\,2}^{(2)} &=&
\frac{149}{12}
+\frac{379}{12}  \frac{1}{c_\theta^6 x_\ssH}
+\frac{19}{2}\frac{x_t}{c_\theta^4 x_\ssH}
- \frac{289}{6}\frac{1}{c_\theta^4 x_\ssH}
- \frac{701}{96}\frac{1}{c_\theta^4} \nonumber \\
{}\nl{}
&-& 40 \frac{x_t}{c_\theta^2 x_\ssH}
- \frac{85}{48}\frac{x_t}{c_\theta^2}
+ \frac{8}{3} \frac{x_t^2}{c_\theta^2 x_\ssH}
+ \frac{77}{3} \frac{1}{c_\theta^2 x_\ssH}
- \frac{3}{4} \frac{x_\ssH}{c_\theta^2}
+ \frac{73}{6} \frac{1}{c_\theta^2} 
+ 35 \frac{x_t}{x_\ssH}
+\frac{9}{4} x_t x_\ssH
-\frac{43}{24} x_t
- \frac{8}{3} \frac{x_t^2}{x_\ssH}\nonumber \\
{}\nl{}
&-& \frac{21}{16} x_t^2 
- 15 \frac{x_t^3}{x_\ssH}
- \frac{256}{3} \frac{1}{x_\ssH}
- \frac{3}{2} x_\ssH
+ \frac{63}{32} x_\ssH^2
 - 10 x_t\frac{g_{\ssS}^2}{g^2}
+32 \frac{x_t^2}{x_\ssH}\frac{g_{\ssS}^2}{g^2}
+12\frac{g_{\ssS}^2}{g^2}
,
\eqa
%--
\bqa
\Delta Z_{\ssM\,;\,3}^{(2)} &=&
- \frac{101}{36}
+\frac{9}{4} \frac{1}{c_\theta^8 x_\ssH^2}
-\frac{25}{2}  \frac{1}{c_\theta^6 x_\ssH}
+\frac{9}{2}\frac{x_t}{c_\theta^4 x_\ssH}
-18\frac{x_t^2}{c_\theta^4 x_\ssH^2}
+9\frac{1}{c_\theta^4 x_\ssH^2}
+\frac{59}{4}\frac{1}{c_\theta^4 x_\ssH}
+\frac{145}{32}\frac{1}{c_\theta^4} \nonumber \\
{}\nl{}
&-& \frac{1}{16}\frac{x_t}{c_\theta^2}
+\frac{1}{2}\frac{x_t^2}{c_\theta^2 x_\ssH}
-4\frac{1}{c_\theta^2 x_\ssH}
-\frac{27}{8}\frac{1}{c_\theta^2}
+9\frac{x_t}{x_\ssH} 
-\frac{23}{8} x_t
-36 \frac{x_t^2}{x_\ssH^2}
+18\frac{x_t^2}{x_\ssH}
-\frac{117}{16} x_t^2\nonumber \\
{}\nl{}
&-&\frac{9}{2}\frac{x_t^3}{x_\ssH}
+36\frac{x_t^4}{x_\ssH^2}
+9\frac{1}{x_\ssH^2}
+10 \frac{1}{x_\ssH}
-\frac{5}{8} x_\ssH
-\frac{9}{32} x_\ssH^2
+6 x_t \frac{g_{\ssS}^2}{g^2}
-48 \frac{x_t^2}{x_\ssH} \frac{g_{\ssS}^2}{g^2}
.
\eqa
%--
For the cosine of the weak-mixing angle we obtain
%--
\bq
\Delta Z_{\ctw\,;\,1}^{(2)} =
\frac{11911}{72}
- \frac{1681}{72} \frac{1}{c_\theta^4}
- \frac{205}{36} \frac{1}{c_\theta^2}
- \frac{2021}{9} c_\theta^2
+ \frac{529}{6} c_\theta^4
,
\eq
%--
\bqa
\Delta Z_{\ctw\,;\,2}^{(2)} &=&
- \frac{119}{24}
- \frac{199}{36} \frac{1}{c_\theta^4}
+ \frac{17}{24} \frac{x_t}{c_\theta^2}
+ \frac{181}{12} \frac{1}{c_\theta^2}
{}\nl{}
&+& \frac{13}{12} c_\theta^2 x_t
- \frac{331}{72} c_\theta^2
- \frac{43}{24} x_t
- \frac{22}{3} \frac{1}{c_\theta^2}\frac{g_{\ssS}^2}{g^2}
- \frac{40}{3} c_\theta^2 \frac{g_{\ssS}^2}{g^2}
+ \frac{62}{3}\frac{g_{\ssS}^2}{g^2}
,
\eqa
%--
\bq
\Delta Z_{\ctw\,;\,3}^{(2)} =
\frac{11671}{288}
- \frac{1681}{288} \frac{1}{c_\theta^4}
- \frac{205}{144} \frac{1}{c_\theta^2}
- \frac{1991}{36} c_\theta^2
+ \frac{529}{24} c_\theta^4
.
\eq
%--
Counterterms for the Higgs-boson mass read as
%--
\bqa
\Delta Z_{\mh\,;\,1}^{(2)} &=&
13
- \frac{55}{8} \frac{1}{c_\theta^4}
- \frac{35}{4}\frac{x_t}{c_\theta^2}
- \frac{9}{2}\frac{x_\ssH}{c_\theta^2}
+ 25\frac{1}{c_\theta^2}
+ 9 x_t x_\ssH \nonumber \\
{}\nl{}
&-& \frac{23}{2} x_t
+ \frac{9}{4} x_t^2
- 9 x_\ssH
+ \frac{27}{8} x_\ssH^2
- 24 x_t \frac{g_{\ssS}^2}{g^2}
,
\eqa
%--
\bqa
\Delta Z_{\mh\,;\,2}^{(2)} &=&
\frac{1}{3}
+ \frac{557}{96} \frac{1}{c_\theta^4}
+ \frac{85}{48}\frac{x_t}{c_\theta^2}
+ \frac{3}{2}\frac{x_\ssH}{c_\theta^2}
- \frac{32}{3}\frac{1}{c_\theta^2}
+ \frac{9}{4} x_t x_\ssH \nonumber \\
{}\nl{}
&+& \frac{25}{24} x_t
- \frac{27}{16} x_t^2
+ 3 x_\ssH
- \frac{15}{32} x_\ssH^2
+ 10 x_t \frac{g_{\ssS}^2}{g^2}
,
\eqa
%--
\bqa
\Delta Z_{\mh\,;\,3}^{(2)} &=&
\frac{31}{4}
- \frac{55}{32} \frac{1}{c_\theta^4}
- \frac{35}{16}\frac{x_t}{c_\theta^2}
- \frac{9}{8}\frac{x_\ssH}{c_\theta^2}
+ \frac{25}{4}\frac{1}{c_\theta^2}
+ \frac{9}{4} x_t x_\ssH \nonumber \\
{}\nl{}
&-& \frac{23}{8} x_t
+ \frac{9}{16} x_t^2
- \frac{9}{4} x_\ssH
+ \frac{27}{32} x_\ssH^2
- 6 x_t \frac{g_{\ssS}^2}{g^2}
.
\eqa
%--
\noindent
Here we present also the two-loop counterterms for the Higgs-Kibble 
$\varphi^{0}$ and $\varphi$ scalar bosons,
%--
\bq
\Delta Z_{\varphi^{0}\,;\,1}^{(2)} =\Delta Z_{\varphi\,;\,1}^{(2)}=
4
+ \frac{43}{4} \frac{1}{c_\theta^4}
+ \frac{5}{4} \frac{x_t}{c_\theta^2}
- \frac{37}{2} \frac{1}{c_\theta^2}
- \frac{7}{2} x_t
- \frac{9}{4} x_t^2
+ 24 x_t \frac{g_{\ssS}^2}{g^2},
\label{csymII}
\eq
%--
\bq
\Delta Z_{\varphi^{0}\,;\,2}^{(2)} =\Delta Z_{\varphi\,;\,2}^{(2)} =
\frac{7}{6}
- \frac{431}{96} \frac{1}{c_\theta^4}
- \frac{85}{48} \frac{x_t}{c_\theta^2}
+ \frac{101}{12} \frac{1}{c_\theta^2}
- \frac{25}{24} x_t
+ \frac{27}{16} x_t^2
- \frac{3}{32} x_\ssH^2
- 10 x_t \frac{g_{\ssS}^2}{g^2},
\label{csymIII}
\eq
%--
\bq\label{eq:one:ultI}
\Delta Z_{\varphi^{0}\,;\,3}^{(2)} =\Delta Z_{\varphi\,;\,3}^{(2)} =
- 2
+ \frac{43}{16} \frac{1}{c_\theta^4}
+ \frac{5}{16} \frac{x_t}{c_\theta^2}
- \frac{37}{8} \frac{1}{c_\theta^2}
- \frac{7}{8} x_t
- \frac{9}{16} x_t^2
+ 6 x_t \frac{g_{\ssS}^2}{g^2}.
\label{csymIV}
\eq
%--
%--
\noindent
The equalities in \eqn{csymI} and in \eqns{csymII}{csymIV} are related
to the custodial symmetry $SU(2)_{\ssL}\,\otimes\,SU(2)_{\ssR} \to
SU(2)_{\ssV}$. Finally, gauge-parameter counterterms can be expressed by 
means of 
\eqns{eq:one:unoI}{eq:one:ultI}, similarly to the one-loop case,
%--
\bqa\label{eq:one:KGA}
\delta Z_{\gpA}^{(2)} &=&
\frac{1}{2}
\Bigl\{
\delta Z_{\ssA\ssA}^{(2)} - \frac{1}{4}
\Bigl[ 
\delta Z_{\ssA\ssA}^{(1)}
\Bigr] ^2
\Bigr\},
\qquad
\delta Z_{\gpZ}^{(2)} =
\frac{1}{2}
\Bigl\{
\delta Z_{\ssZ}^{(2)} -\frac{1}{4}
\Bigl[ 
\delta Z_{\ssZ}^{(1)}
\Bigr] ^2
\Bigr\},
\nl
\delta Z_{\gpW}^{(2)} &=&
\frac{1}{2}
\Bigl\{
\delta Z_{\ssW}^{(2)} -\frac{1}{4}
\Bigl[ 
\delta Z_{\ssW}^{(1)}
\Bigr] ^2
\Bigr\},
\qquad
\delta Z_{\gpAZ}^{(2)} =
- \delta Z_{\ssA\ssZ}^{(2)}
+ \frac{1}{2}
\delta Z_{\ssA\ssZ}^{(1)}
\Bigl[
\delta Z_{\ssA\ssA}^{(1)} +
\delta Z_{\ssZ}^{(1)}
\Bigr],
\eqa
%--
\bqa\label{eq:one:KGAu}
\delta Z_{\xi_{\varphi}}^{(2)} &=&
- \frac{1}{2} \Bigl[
\delta Z_{\varphi}^{(2)} +
\delta Z_{\ssM}^{(2)}
\Bigr] + \frac{3}{8}
\Bigl\{
\Bigl[ 
\delta Z_{\varphi}^{(1)} \Bigr] ^2 +
\Bigl[ 
\delta Z_{\ssM}^{(1)}
\Bigr] ^2
\Bigr\}
+ \frac{1}{4}
\delta Z_{\varphi}^{(1)}
\delta Z_{\ssM}^{(1)},
\nl
\delta Z_{\xi_{\varphi^{0}}}^{(2)} &=&
- \frac{1}{2}
\Bigl[
\delta Z_{\varphi^{0}}^{(2)} +
\delta Z_{\ssM}^{(2)} - 2
\delta Z_{\ctw}^{(2)}
\Bigr]
+ \frac{3}{8}
\Bigl\{ \Bigl[ \delta Z_{\varphi^{0}}^{(1)}\Bigr]^2
+
\Bigl[ \delta Z_{\ssM}^{(1)}\Bigr]^2\Bigr\}
\nl
{}&+& \frac{1}{4}
\delta Z_{\varphi^{0}}^{(1)}
\delta Z_{\ssM}^{(1)}
- \frac{1}{2} \delta Z_{\ctw}^{(1)}
\Bigl[ 
\delta Z_{\varphi^{0}}^{(1)} + \delta Z_{\ssM}^{(1)} \Bigr].
\eqa
%--
After deriving the relations for the two-loop gauge-parameter counterterms,
we can verify immediately an important result: the doubly-contracted 
WST identities with two external gauge-boson fields have the 
same form derived in 't Hooft-Feynman gauge.
The gauge-fixing functions given in \eqn{eq:th:gfun} are expressed by means
of bare quantities. After introducing renormalized quantities, the 
gauge-fixing functions change; their expression, order-by-order in 
perturbation theory, can be read in \eqn{eq:one:GAUGEREN}.
By expanding \eqn{eq:one:GAUGEREN} up to two loops, and using the expression 
of \eqns{eq:one:KGA}{eq:one:KGAu} for the gauge-parameter counterterms,
we derive
%--
\bqa\label{eq:one:GAUGEREN5dou}
{\cal C}^{\ssA} &=&
- \partial_{\mu} A_{\mu} + \ord{g^6},
\quad
{\cal C}^{\ssZ} =
 -\partial_{\mu} Z_{\mu} 
+ M_{0} \Phi^{0} + \ord{g^6},
\quad
{\cal C}^{\pm} = 
 -\partial_{\mu} W_{\mu} + M \Phi
+ \ord{g^6}
\eqa 
%--
and the doubly-contracted WST identities are formally  equivalent to those 
of \fig{fig:one:WIbare}, at $\ord{g^4}$.
%--
\begin{figure}
\SetWidth{1}
\begin{center}
%\framebox[15cm]{\parbox{15cm}{
\begin{eqnarray*}
\begin{picture}(430,350)(-20,0)
\Text(200,325)[]{$+$}
\Text(200,225)[]{$+$}
\Text(200,125)[]{$+$}
\Text(200,25)[]{$+$}
\Text(0,225)[]{$+$}
\Text(0,125)[]{$+$}
\Text(0,25)[]{$+$}
\Text(24,325)[r]{$G_{\mu\nu\irr}^{\ssA\ssZ\,;\,2}=$}
\Line(25,325)(75,325)
\CCirc(100,325){25}{Black}{White}
\Line(125,325)(175,325)
\Line(75,325)(125,325)
\Text(75,295)[lt]{$F_{\mu\nu\irr\,;\,1}^{\ssA\ssZ\,;\,2}$}
\Line(225,325)(275,325)
\CCirc(300,325){25}{Black}{White}
\Line(325,325)(375,325)
\CArc(275,350)(25,-90,0)
\Text(300,295)[t]{$F_{\mu\nu\irr\,;\,2}^{\ssA\ssZ\,;\,2}$}
\Line(25,225)(75,225)
\CCirc(100,225){25}{Black}{White}
\Line(125,225)(175,225)
\CCirc(100,250){10}{Black}{White}
\Text(75,195)[lt]{$F_{\mu\nu\irr\,;\,3}^{\ssA\ssZ\,;\,2}$}
\Line(225,225)(275,225)
\CCirc(300,225){25}{Black}{White}
\Line(325,225)(375,225)
\Line(300,250)(300,200)
\Text(300,195)[t]{$F_{\mu\nu\irr\,;\,4}^{\ssA\ssZ\,;\,2}$}
\Line(25,125)(80,125)
\CCirc(100,125){20}{Black}{White}
\Line(120,125)(175,125)
\CCirc(100,155){10}{Black}{White}
\Text(75,95)[l]{$F_{\mu\nu\irr\,;\,5}^{\ssA\ssZ\,;\,2}$}
\Line(225,125)(275,125)
\CCirc(287.5,125){12.5}{Black}{White}
\CCirc(312.5,125){12.5}{Black}{White}
\Line(325,125)(375,125)
\Text(300,95)[]{$F_{\mu\nu\irr\,;\,6}^{\ssA\ssZ\,;\,2}$}
\Line(25,25)(175,25)
\CCirc(100,45){20}{Black}{White}
\Line(80,45)(120,45)
\Text(75,12)[l]{$F_{\mu\nu\irr\,;\,7}^{\ssA\ssZ\,;\,2}$}
\Line(225,25)(350,25)
\CCirc(300,37.5){12.5}{Black}{White}
\CCirc(300,62.5){12.5}{Black}{White}
\Line(250,25)(375,25)
\Text(300,12)[]{$F_{\mu\nu\irr\,;\,8}^{\ssA\ssZ\,;\,2}$}
\end{picture}\\
\begin{picture}(430,150)(-20,0)
\Text(200,125)[]{$+$}
\Text(200,25)[]{$+$}
\Text(0,125)[]{$+$}
\Text(0,25)[]{$+$}
\Line(25,125)(80,125)
\CCirc(100,125){20}{Black}{White}
\Line(120,125)(175,125)
\CCirc(100,145){2}{Black}{Black}
\Text(75,93)[l]{$F_{\mu\nu\irr\,;\,9}^{\ssA\ssZ\,;\,2}$}
\Line(225,125)(280,125)
\CCirc(300,125){20}{Black}{White}
\CCirc(280,125){2}{Black}{Black}
\Line(320,125)(375,125)
\Text(300,93)[]{$F_{\mu\nu\irr\,;\,10}^{\ssA\ssZ\,;\,2}$}
\Line(25,25)(175,25)
\CCirc(100,45){20}{Black}{White}
\CCirc(100,65){2}{Black}{Black}
\Text(75,13)[l]{$F_{\mu\nu\irr\,;\,11}^{\ssA\ssZ\,;\,2}$}
\Line(225,25)(375,25)
\CCirc(300,45){20}{Black}{White}
\CCirc(300,25){2}{Black}{Black}
%\Line(250,25)(375,25)
\Text(300,13)[]{$F_{\mu\nu\irr\,;\,12}^{\ssA\ssZ\,;\,2}$}
\end{picture}\\
\begin{picture}(430,32)(-20,-12)
\Text(0,10)[]{$+$}
\Line(25,10)(175,10)
\CBoxc(100,10)(4,4){Black}{Black}
\Text(75,-8)[]{$F_{\mu\nu\irr\,;\,13}^{\ssA\ssZ\,;\,2}$}
\end{picture}
\end{eqnarray*}
%}}
\end{center}
\caption[]{
\small{The irreducible contributions to the Green function with one external 
photon and one external $Z$ boson at $\ord{g^4}$ (denoted by the superscript 
$2$). Thirteen families of Feynman diagrams give a contribution
(each gets a subscript-label). Dotted vertices depend on $\beta_1$, 
$\Gamma_1$ or the counterterms at $\ord{g^2}$. The black-square vertex 
represents a vertex which depends on $\Gamma_2$ or the counterterms at 
$\ord{g^4}$.}}
\label{fig:one:selfone2}
\end{figure}
%--
%--
\section{Conclusions}
\label{conclu}
In this paper, we have collected all the ingredients that are needed to
construct a two-loop renormalization procedure for the minimal Standard Model
and to work out the details for constructing theoretical predictions for
physical (pseudo-)observables.

The procedure covers broadly (and not relying on other sources) all aspects
of the problem, from diagram generation, to definition of the renormalization
scheme, to evaluation (mostly numerically) of Green functions, including 
all customary approximations that are usually made as the small fermion-mass
approximation (with particular attention to the analytical extraction of 
collinear logarithms). Our scheme can deal with arbitrary scales and goes
beyond the introduction of $\MSB$ couplings in the electroweak sector.

Among the technical aspects which we consider to be innovative and that we
have introduced in this paper we point out a universal ultraviolet 
decomposition of one- and two-loop diagrams which is closed with respect
to multiplication and a non-negligible simplification in the structure
of renormalized Green functions due to the (possible) introduction of a
non-minimal $\MSB$ prescription. We also present residues of ultraviolet
poles, up to two-loop vertices, in terms of familiar one-loop functions.

Techniques for expanding one- and two-loop diagrams in different regions,
small fermion masses or small momentum transfer or both, have been
introduced and discussed as well as their relevance to renormalization.

Applications, as well as the introduction of unstable particles in the
renormalization procedure, will be presented in a third, forthcoming, paper.

In several appendices we have presented examples of our computational
techniques. Although there is no substitute for the original FORM codes in
modern technology, we are confident enough that the examples shown here 
suffice in illustrating our methods. The parallel development of $\GS$
and of $\LB$ is aimed to give to most complete documentation for obtaining
all results which are too long to be presented here; if we ask the
question `might the theory in general be a lot simpler than we thought?'
we may conclude that we have a `nearly nifty' formulation
\footnote{see Z. Bern talk, High precision for hard processes at the LHC,
Zurich 2006} showing that the Standard Model is canonically as simple as 
QED, even at the two-loop level.
%--
\Acknowledgments
%--
The contribution of Andrea Ferroglia and Massimo Passera to an early stage of 
this paper is gratefully acknowledged. We recognize the role played by 
Sandro Uccirati in several steps of our global project. Finally, we gratefully
acknowledge several discussions with Ansgar Denner, Stefan Dittmaier and 
Mikhail Kalmykov. S.A. is indebted to N.~Glover for hospitality at the 
Institute for Particle Physics Phenomenology of the University of Durham where
part of the manuscript was written.
%--
\clearpage
%--
\appendix
%--
\section{Reduction of vacuum bubbles\label{ibpvb}}
%--
Following the definitions of \sect{vacB} we give few examples of relations
between two-loop vacuum bubbles and the master bubble, which we select to be
$T_{121}$. For instance, we have
%--
\bqa
T_{211}(m_3,m_1,m_2) &=& -\,\frac{1}{2\,m^2_3}\,\Bigl[
     A_0(1,m_1)\,A_0(2,m_2) + A_0(1,m_3)\,A_0(2,m_2) 
\nl
{}&+&                 (n-3)\,T_{111}(m_3,m_1,m_2) 
       +(m^2_3+m^2_2-m^2_1)\,T_{112}(m_3,m_1,m_2)\Bigr],
\nl
T_{112}(m_3,m_1,m_2) &=&  - \frac{1}{m^2_1-m^2_2-m^2_3}\,\Bigl[
         A_0(1,m_2)\,A_0(2,m_1) 
\nl
{}&-& A_0(1,m_1)\,A_0(2,m_2) - A_0(1,m_3)\,A_0(2,m_1) 
    + A_0(1,m_3)\,A_0(2,m_2) 
\nl
{}&+& (m^2_3-m^2_2+m^2_1)\,T_{121}(m_3,m_1,m_2)\Bigr],
\nl
(2-n)\,T_{111}(m_3,m_1,m_2) &=& \frac{1}{m^2_1-m^2_2-m^2_3}
\nl
{}&\times& \Bigl\{
    (m^2_1+m^2_2-m^2_3)\,
\Bigl[A_0(1,m_1)\,A_0(2,m_2)+A_0(2,m_1)\,A_0(1,m_3)\Bigr]
\nl
{}&-& (m^2_1+m^2_2-m^2_3)\,
\Bigl[A_0(2,m_2)\,A_0(1,m_3)+A_0(2,m_1)\,A_0(1,m_2)\Bigr]
\nl
{}&+& \Bigl[ (m^2_2-m^2_3)^2-m^4_1\Bigr]\,T_{121}(m_3,m_1,m_2) \Bigr\}
       + A_0(1,m_3)\,A_0(2,m_2) 
\nl
{}&-& A_0(1,m_1)\,A_0(2,m_2) + 2\,m^2_1\,T_{121}(m_3,m_1,m_2). 
\eqa
%--
Vacuum bubbles with arbitray powers can always be reduced to a combination
of products of $A_0$ functions and of $T_{121}$. This results are relevant for
processes where all external scales can be put to zero.

An efficient algorithmic solution of the IBP equations for vacuum bubbles must 
be able to isolate special cases; an example is given by the following two 
cases:
%--
\bqa
T_{112}(0,m,m) &=& \frac{1}{2\,m^2}\,A_0(1,m)\,A_0(2,m),
\nl
T_{112}(0,m_1,m_2) &=& 
 2\,\frac{m^2_2}{m^2_1}\,A_0(1,m_1)\,A_0(3,m_2)
      -\,A_0(2,m_1)\,A_0(2,m_2)
      -\,T_{121}(0,m_1,m_2).
\eqa
%--
We can now compute the master vacuum bubble of \eqn{masterB}, which is 
representable as
%--
\bqa
T^{\rm fin}_{121}(m_1,m_2,m_3) \equiv T_{121}(m_1,m_2,m_3\,;\,0) &=&
 \frac{7}{2} - \frac{1}{2}\,\zeta(2) - \ln\frac{m^2_2}{M^2}\,
\Bigl( 1 + \frac{1}{2}\,\ln \frac{m^2_2}{M^2} \Bigr) + \intfx{x}\,T(x),
\eqa
%--
\bqa
T(x) &=& -\,\frac{Q(x)}{P(x)}\,\Bigl[ \ln m^2_2 - \ln \frac{Q(x)}{X}
\Bigr] - \ln |P(x)|\,\Bigl[ \ln m^2_2 + \ln X\Bigr] +
\frac{1}{2}\,\ln^2 |P(x)| + \li{2}{\frac{m^2_2}{M^2}\,\frac{X}{P(x)}},
\label{asusual}
\eqa
%--
\bq
Q(x) = \frac{m^2_1-m^2_3}{M^2}\,x - \frac{m^2_1}{M^2},
\qquad
P(x) = Q(x) + \frac{m^2_2}{M^2}\,X,
\eq
%--
with $X = x\,(1-x)$; special cases are given in \appendx{msll}.
%--
\section{Singularities whose origin is infrared}
%--
As mentioned in \sect{conve} after reduction we may have a migration of poles,
from the IR to the UV; in this appendix we collect two examples. 
%--
\subsection{Evaluation of $s^{\ssC}(s=0\,;\,m_1,m_2,0,0)$}
%--
In this appendix we consider the following integral, belonging to the 
$C$-family (Sect.~5 of Ref.~\cite{Ferroglia:2003yj}),
%--
\bqa
s^{\ssC}(s=0\,;\,m_1,m_2,0,0) &=& -2\,(\pi M^2)^{-\ep}\,\egam{\ep}\,
\int_0^1 dx \int_0^1 dy \int_0^y dz\,\Bigl[ x\,(1-x)\Bigr]^{-\ep/2}
(1 - y)^{\ep/2}
\nl
{}&\times& \Bigl[ m^2_x\,V^{-1-\ep}_{\ssC} + (\frac{2}{\ep} - \frac{1}{2} )\,
V^{-\ep}_{\ssC}\Bigr].
\label{UVIR}
\eqa
%--
When $s = 0$ and $m_3 = m_4 = 0$ we have $V_{\ssC} = (1-y)\,m^2_x$ and the 
first term in square brackets in \eqn{UVIR} is singular at $y = 1$. 
Furthermore,
%--
\bq
x\,(1 - x)\,m^2_x = \frac{m^2_1}{M^2} + \frac{m^2_2 - m^2_1}{M^2}\,x.
\label{defmx}
\eq
%--
We now perform the (trivial) $z$ integration and the $y$ integration using, 
e.g.,
%--
\bq
\int_0^1\,(1 - y)^{-1-\ep/2} = -\,\frac{2}{\ep},
\eq
%--
where we have naively assumed the relation $\ep_{\ssU\ssV} = - \ep_{\ssI\ssR}$.
The last $x$ integration is finally performed by using the following relation,
valid for any monomial $V(x)$,
%--
\bq
V^{-1-\ep/2}(x) = \frac{M^2}{m^2_1}\,\Bigl[ 1 + \frac{x}{\ep}\,\frac{d}{dx}
\Bigr]\,V^{-\ep}(x).
\eq
%--
Now, after integration by parts we obtain
%--
\bq
\frac{(\pi M^2)^{\ep}}{\egam{1+\ep}}\,
s^{\ssC}(0,m_1,m_2,0,0) = 
\frac{2}{\ep^2} + \Bigl[ 1 +
2\,\frac{m^2_1}{m^2_2-m^2_1}\,\ln\frac{m^2_1}{M^2} -
2\,\frac{m^2_2}{m^2_2-m^2_1}\,\ln\frac{m^2_2}{M^2}\Bigr]\,\frac{1}{\ep} +
\ord{1}.
\eq
%--
Note that the residue of the double pole has changed sign with respect to the 
non-IR configuration. 
%--
\subsection{Evaluation of $s^{\ssE}(s,m_1,m_2,0,m_4,0)$}
%--
Another example is represented by $s^{\ssE}(s,m_1,m_2,0,m_4,0)$ where we have
to compute the following three-fold integral,
%--
\bq
\int_0^1 dx \int_0^1 dy \int_0^y dz \Bigl[ x\,(1-x)\Bigr]^{-\ep/2}\,
( 1- y)^{\ep/2}\,(y - z)\,V^{-1-\ep}_{\ssE},
\label{eEfam}
\eq
%--
where the quadratic form $V_{\ssE}$ (see Sect.~6 of 
Ref.~\cite{Ferroglia:2003yj}) is defined as,
%--
\bq
V_{\ssE} = z^2 -a\,z + b\,(1-y),
\quad
a = 1- \frac{m^2_4}{s},
\qquad
b = m^2_x\,\frac{M^2}{s},
\eq
%--
with $m^2_x$ given in \eqn{defmx}.
After changing variable, $y' = 1-y$, in \eqn{eEfam} we use a functional
relation valid for quadratic forms,
%--
\bq
V^{-1-\ep} = -\,\frac{4}{a^2}\,\Bigl[ 1 + \frac{y}{\ep}\,\partial_y +
\frac{1}{2}\,(z - \frac{a}{2} )\,\partial_z\Bigr]\,V^{-\ep}.
\eq
%--
After inserting this result in \eqn{eEfam} and after integration by parts we 
obtain the result,
%--
\bqa
\frac{(\pi\,s)^{\ep}}{\egam{1+\ep}}\,s^{\ssE}(s,m_1,m_2,0,m_4,0) &=&
-\,\frac{2}{a}\,\frac{1}{\ep^2} + \Bigl[
-\,\frac{1}{a} +
(1 - \frac{1}{a})\,\ln (1 - a) -
(1 -\frac{2}{a})\,\ln (-\,a) 
\nl
{}&-&
\frac{m^2_1}{m^2_2 - m^2_1}\,\ln\frac{m^2_1}{s} +
\frac{m^2_2}{m^2_2 - m^2_1}\,\ln\frac{m^2_2}{s}\Bigr]\,\frac{1}{\ep} +
\ord{1},
\eqa
%--
where, once more, $a = 1 - m^2_4/s$; this function, for a non-IR
configuration ($m_3 \not= 0$), shows only a single pole at $n = 4$; in the case
$m_3 = 0$ (coming from reduction of integrals with irreducible scalar products 
in the numerator) there is, instead, a double pole.

Note that all these features arise from the reduction of 
$S_{pp}^{\ssE}(1,1,\cdots)$ and of $S^{\ssE}_d(1,1,\cdots)$; in the case of 
pure QED all these terms are absent and, as a matter of fact, they are also 
absent in the doubly contracted Ward identity for the $AA$ transition, even 
in the full standard model. It is only in the evaluation of $\Pi(s)$ -- 
vacuum polarization in the full standard model -- that they are needed; here, 
for the first time, the explicit form of the two-loop, two-point, form factors
is actually needed and represents a severe test for the final result.

There are virtues and vices in any reduction algorithm; note, however, that
a check on any calculation cannot do without reduction, i.e. without the
possibility of expressing the result in terms of a basis. 
%--
\section{The limit $m_f \to 0$ \label{msll}}
%--
All diagrams contributing to a certain Green function are generated by
our code $\GS$; the output of $\GS$ is fully general and fermion masses are 
kept in the final answer. However, in most cases, we are interested in 
taking the limit $m_f \to 0, \; f\not= t$, extracting at the same time all 
the collinear logarithms. 

After renormalization, we aim to compute physical observables with an 
algebraic-numerical approach; all contributions that are potentially large, 
e.g. collinear logarithms, are extracted with the use of an analytical 
calculation and only the collinear-free reminder is subject to a numerical 
evaluation. In few cases even this last step can be avoided given that also the
finite part (in the collinear limit) is expressed in terms of polylogarithms. 
In this way we are always able to produce and explicit proof of the expected 
cancellations: no power behavior in a gauge theory etc.

Collinear logarithms has two different origins; for instance
they are present in the vacuum polarization function due to QED effects; for 
two loop irreducible contributions, however, they are sub-leading; e.g. the 
correction factor is proportional to $g^4\,\ln m^2_l$ and not to 
$g^4\,\ln^2 m^2_l$. Thus, cancellation of the leading part is an important 
test of the calculation. 

Otherwise, collinear logarithms (sometimes even power behavior) may be 
induced by the reduction procedure in different pieces of the final answer,
although the total is collinear free. In this cases, where there is no 
enhancement, one could as well set $m_f$ to zero from the very beginning and
control cancellation of collinear poles in $n-4$. The equivalence of the two
formulations has not yet received a satisfactory answer.

To discuss the collinear limit we introduce scaled variables,
%--
\bq
z_i = \frac{m^2_i}{s}, \qquad L_i = \ln z_i.
\eq
%--
and collect our results.
%--
\bei
\item[--] {\underline{Miscellanea}}
\eei
%--
The following expansions (see \eqn{deflambda} for the corresponding 
definitions), which have been implemented in $\GS$, are very useful in 
deriving the limit of Green functions for small values of the light fermion 
masses:
%--
\bqa
\beta^{-1}(z_f)\,\ln\frac{\beta(z_f)+1}{\beta(z_f)-1} &=& 
       - L_f - i\,\pi
       - 2\,z_f\,(1 + L_f + i\,\pi)
       - z^2_f\,\Bigl[7 + 6\,(L_f + i\,\pi)\Bigr] + \ord{z^3_f},
\eqa
%--
\bqa
\lambda^{1/2}(1,z_t,z_b) &=& 1 - z_t - z_b + \ord{z^3_b},
\nl
\lambda^{-1/2}(1,z_t,z_b) &=&
-\,\frac{1}{z_t-1}\,\Bigl[ 1 - \frac{z_b}{z_t-1} + 
\lpar \frac{z_b}{z_t-1}\rpar^2\Bigr] + \ord{z^3_b},
\eqa
%--
while, for a massive top quark, we obtain
%--
\bqa
L(1,z_t,z_b) &=&
  - \frac{1}{2}\,L_t - \frac{1}{2}\,L_b + \ln (z_t - 1)
  - \frac{z_b}{(z_t-1)^2} - \frac{z^2_b}{(z_t-1)^3}\,
\lpar 1 + \frac{3}{2}\,\frac{1}{z_t-1}\rpar + \ord{z^3_b}.
\eqa
%--
\bei
\item[--] {\underline{$\ord{\ep}$ one-loop functions}}
\eei 
%--
$\ord{\ep}$ one-loop functions, \eqns{B0oo}{B0to}, are expanded as follows:
%--
\bqa
b_0^{\ep}(2,1,s,m_f,m_f) &=& 4\,\zeta(2) + L^2_f -
   2\,z_f\,\Bigl[ 2 - 4\,\zeta(2) + L_f - L^2_f\Bigr],
\nl
b_0^{\ep}(1,1,s,m_f,m_f) &=& - 8 + 8\,\zeta(2) -
   2\,z_f\,\Bigl[ 2 + 8\,\zeta(2) - 2\,L_f + L^2_f\Bigr] - 4\,i\,\pi,
\nl
b_0^{\ep}(1,1,s,0,m_f) &=& - 8 + 8\,\zeta(2) -
   z_f\,\Bigl[ 2 + 8\,\zeta(2) - 2\,L_f + L^2_f\Bigr] - 4\,i\,\pi,
\nl
b_0^{\rm fin}(1,1,s,\mb,\mt) &=& 
- \frac{1}{2}\,(L_b + L_t) - \frac{1}{2}\,(z_b - z_t)\,(L_b - L_t)
+ 2 -\lambda^{1/2}(1,z_b,z_t)\,L(1,z_b,z_t).
\eqa
%--
\bqa
b_0^{\ep}(2,1,s,\mb,\mt) &=& - 4\,\Bigl\{
          - L_t + \ln(z_t - 1) + \frac{1}{2}\,b_0^{\ep}(2,1,s,\mt,\mb)
\nl
{}&+& \frac{1}{z_t-1}\,\Bigl[
          - L_t + \frac{1}{4}\,L^2_b + 2
          + \frac{1}{4}\,b_0^{\ep}(1,1,s,\mt,\mb)
          + \frac{1}{2}\,b_0^{\ep}(2,1,s,\mt,\mb)\Bigr]
          \Bigr\},
\nl
b_0^{\ep}(2,1,s,m_f,0) &=& 2\,\li{2}{\frac{1}{1-z_f}} -
\ln^2 (z_f - 1) + L^2_f.
\eqa
%--
\bei
\item[--] {\underline{Vacuum bubbles}}
\eei
%--
For vacuum bubbles, see the definition in \eqn{defTf}, we introduce variables
%--
\bq
x_i = \frac{m^2_i}{\mws}, \qquad \cL_i = \ln x_i,
\eq
%--
and derive the following results,
%--
\bqa
T_f(0,\mw,0) &=& -\frac{1}{2}-\frac{3}{2}\,\zeta(2),
\nl
T_f(0,\mz,0) &=& -\frac{1}{2}-\frac{3}{2}\,\zeta(2)+\cL_{\ssZ}-\cL_{\ssZ}^2,
\nl
T_f(0,\mt,0) &=& -\frac{1}{2}-\frac{3}{2}\,\zeta(2)+\cL_t-\cL_t^2,
\nl
T_f(0,m_f,\mw) &=& -\frac{1}{2}+\frac{1}{2}\,\zeta(2)
+\cL_f-\frac{1}{2}\,\cL_f^2,
\nl
T_f(0,m_f,\mz) &=& -\frac{1}{2}+\frac{1}{2}\,\zeta(2)
+\cL_f\,\Bigl( 1 -\cL_{\ssZ}-\frac{1}{2}\,\cL_f\Bigr)
+\frac{1}{2}\,\cL_{\ssZ}^2,
\nl
T_f(\mz,m_f,m_f) &=& -\frac{1}{2}+\frac{1}{2}\,\zeta(2)
+\cL_f\,\Bigl(1-\cL_{\ssZ}-\frac{1}{2}\,\cL_f\Bigr)
+\frac{1}{2}\,\cL_{\ssZ}^2,
\nl
T_f(\mh,m_f,m_f) &=& -\frac{1}{2}+\frac{1}{2}\,\zeta(2)
+\cL_f\,\Bigl(1-\cL_{\ssH}-\frac{1}{2}\,\cL_f\Bigr)
+\frac{1}{2}\,\cL_{\ssH}^2,
\nl
T_f(0,m_f,\mh) &=& -\frac{1}{2}+\frac{1}{2}\,\zeta(2)
+\cL_f\,\Bigl(1-\cL_{\ssH}-\frac{1}{2}\,\cL_f\Bigr)
+\frac{1}{2}\,\cL_{\ssH}^2.
\eqa
%--
Here, $f$ is a generic light fermion, $f \not= t$. For the $t - b$ doublet
we obtain instead,
%--
\bqa
T_f(\mw,\mb,\mt) &=& -\frac{1}{2}-\frac{1}{2}\,\zeta(2)
+\cL_b-\frac{1}{2}\,\cL_b^2-
                 \frac{x_t}{x_t-1}\,\Bigl[
\cL_t\,\cL_b + \frac{1}{2}\,\cL_t^2 +
                 \li{2}{1-\frac{1}{x_t}}\Bigr]
\nl
{}&-&\frac{1}{x_t-1}\,\li{2}{1-x_t},
\nl
T_f(0,\mb,\mt) &=& -\frac{1}{2}+\frac{1}{2}\,\zeta(2)+
\cL_b\,\Bigl(1-\frac{1}{2}\,\cL_b-\cL_t\Bigr)
+\frac{1}{2}\,\cL_t^2,
\label{icisknown}
\eqa
%--
and the above list includes all possible cases to be found at two loops. 
%--
\bei
\item[--] {\underline{Vertices and self-energies}}
\eei
%--
For all the diagrams at two loops where we have large values of
one of the external $|p^2|$ and small values for some of the internal masses 
it will be convenient to extract the large logarithms by defining 
{\em regular} functions. Note that individual scalar contributions may
even show power behavior: such is the case, for instance, for 
$V^{\ssM}_0$ when $p^2_{1,2}= 0$ and $m_2 = \dots = m_5 = m$ when
$m^2 \ll |P^2|$. In this case the behavior is $|P^2|/m^2$. 

In computing the vacuum polarization function we have introduced few
objects that are special in the sense that some of the masses that appear
in the argument list are zero as a consequence of the algorithm of
reduction in sub-loops. This special feature modifies the singular structure
of the function and introduces finite reminders that should not be confused
with the {\em regular} ones for which a regulator exists; something that is 
usually mispresented is that $m= 0 (\ep\,$regulated) is different from 
$m \to 0$. Introducing the collinear factor
%--
\bq
C^f_{\ssF} = -L_f+L_{\ssF}\,L_f+\frac{1}{2}\,L_f^2,
\eq
%--
we derive the following results, where $\{0\}_n$ stands for $n$ zero masses:
%--
\bqa
s^{\ssE\,;\,{\rm fin}}(s,\mz,m_f,m_f,m_f,m_f) &=& 
s^{\ssE\,;\,{\rm reg}}(s,\mz,\{0\}_4) + C^f_{\ssZ},
\nl
s^{\ssE\,;\,{\rm fin}}(s,\mz,m_f,m_f,0,m_f) &=& 
s^{\ssE\,;\,{\rm reg}}(s,\mz,\{0\}_4) + C^f_{\ssZ},
\nl
s^{\ssE\,;\,{\rm fin}}(s,0,\mw,m_f,m_f,m_f) &=& 
s^{\ssE\,;\,{\rm reg}}(s,\mw,\{0\}_4) + C^f_{\ssW},
\nl
s^{\ssE\,;\,{\rm fin}}(s,\mw,0,m_f,m_f,m_f) &=& 
s^{\ssE\,;\,{\rm reg}}(s,\mw,\{0\}_4) + C^f_{\ssW},
\nl
s^{\ssE\,;\,{\rm fin}}(s,\mw,0,m_f,0,m_f) &=& 
s^{\ssE\,;\,{\rm reg}}(s,\mw,\{0\}_4) + C^f_{\ssW},
\nl
s^{\ssE\,;\,{\rm spec}}(s,m_f,m_f,0,\mw,0) &=& 
s^{\ssE\,;\,{\rm spec}\ssR}(s,\{0\}_3,\mw,0)+
                        \frac{L_f+L^2_f}{x_{\ssW}-1}.
\eqa
%--
\bqa
s^{\ssE\,;\,{\rm fin}}(s,\mw,\muq,\md,\muq,\md) &=& 
s^{\ssE\,;\,{\rm reg}}(s,\mw,\{0\}_4) + C^d_{\ssW},
\nl
s^{\ssE\,;\,{\rm fin}}(s,\mw,\md,\muq,\md,\muq) &=& 
s^{\ssE\,;\,{\rm reg}}(s,\mw,\{0\}_4) + C^u_{\ssW},
\nl
s^{\ssE\,;\,{\rm fin}}(s,\mw,\muq,\md,\md,\md) &=& 
s^{\ssE\,;\,{\rm reg}}(s,\mw,\{0\}_4) + C^d_{\ssW},
\nl
s^{\ssE\,;\,{\rm fin}}(s,\mw,\md,\muq,\muq,\muq) &=& 
s^{\ssE\,;\,{\rm reg}}(s,\mw,\{0\}_4) + C^u_{\ssW},
\nl
s^{\ssE\,;\,{\rm fin}}(s,\mz,\muq,\muq,\md,\muq) &=& 
s^{\ssE\,;\,{\rm reg}}(s,\mz,\{0\}_4) + C^u_{\ssZ},
\nl
s^{\ssE\,;\,{\rm fin}}(s,\mz,\muq,\muq,\muq,\muq) &=& 
s^{\ssE\,;\,{\rm reg}}(s,\mz,\{0\}_4) + C^u_{\ssZ},
\nl
s^{\ssE\,;\,{\rm fin}}(s,\mz,\md,\md,\muq,\md) &=& 
s^{\ssE\,;\,{\rm reg}}(s,\mz,\{0\}_4) + C^d_{\ssZ},
\nl
s^{\ssE\,;\,{\rm fin}}(s,\mz,\md,\md,\md,\md) &=& 
s^{\ssE\,;\,{\rm reg}}(s,\mz,\{0\}_4) + C^d_{\ssZ},
\nl
s^{\ssE\,;\,{\rm spec}}(s,\muq,\muq,0,\mw,0) &=& 
s^{\ssE\,;\,{\rm spec}\ssR}(s,\{0\}_3,\mw,0)+
                      \frac{L_u+L^2_u}{x_{\ssW}-1},
\nl
s^{\ssE\,;\,{\rm spec}}(s,\md,\md,0,\mw,0) &=& 
s^{\ssE\,;\,{\rm spec}\ssR}(s,\{0\}_3,\mw,0)+
                      \frac{L_d+L^2_d}{x_{\ssW}-1}.
\eqa
%--
\bqa
s^{\ssE\,;\,{\rm fin}}(s,\mw,\mt,\mb,\mt,\mb) &=& 
s^{\ssE\,;\,{\rm reg}}(s,\mw,\mt,0,\mt,0)
\nl
{}&+& \frac{L_b}{z_t-1}\,\Bigl[
      1 - \frac{1}{2}\,L_b
     -\,\frac{1}{x_t-1}\,\lpar x_t\,L_t - L{\ssW}\rpar \Bigr],
\nl
s^{\ssE\,;\,{\rm fin}}(s,\mw,\mt,\mb,\mb,\mb) &=& 
s^{\ssE\,;\,{\rm reg}}(s,\mw,\mt,\{0\}_3)
- L_b\,\lpar 1 - L_{\ssW} - \frac{1}{2}\,L_b\rpar +
\frac{z_t}{z_t-1}\,L_b\,L_t,
\nl
s^{\ssE\,;\,{\rm fin}}(s,\mz,\mb,\mb,\mb,\mb) &=& 
s^{\ssE\,;\,{\rm reg}}(s,\mz,\{0\}_4)
- L_b\,\lpar 1 - L_{\ssZ} - \frac{1}{2}\,L_b \rpar,
\nl
s^{\ssE\,;\,{\rm fin}}(s,\mz,\mb,\mb,\mt,\mb) &=& 
s^{\ssE\,;\,{\rm reg}}(s,\mz,\{0\}_2,\mt,0) +
\frac{1}{z_t-1}\,L_b\,\lpar 1 - L_{\ssZ} - \frac{1}{2}\,L_b\rpar,
\nl
s^{\ssE\,;\,{\rm spec}}(s,\mb,\mb,0,\mw,0) &=& 
s^{\ssE\,;\,{\rm spec}\ssR}(s,\{0\}_3,\mw,0) + \frac{L_b+L^2_b}{z_t-1}.
\eqa
%--
\section{The limit $p^2 \to 0$ \label{limits}}
%--
There are several places where one needs Green functions, self-energies in
particular, evaluated at $p^2 = 0$; e.g.\ renormalization of the electric 
charge requires the vacuum polarization function subtracted at $p^2= 0$ and
corrections to $\gf$ require the $\wb-\wb$ self-energy in the same limit.

Of course, evaluation of two-loop two-point functions at $p^2 = 0$ follows
a slightly different and somehow independent approach. Every two-loop
two-point diagram at $p^2 = 0$ is a vacuum bubble that can be reduced to
one master integral by means of integration-by-part identities. However,
we have opted for a different approach where we start from the result at
arbitrary $p^2$ and perform the limit $p^2 \to 0$; in this way we have
a powerful check on the correctness of the original procedure.
%--
\subsection{One-loop functions}
%--
For most of the functions appearing in an arbitrary self-energy the limit
$s = - p^2 \to 0$ is almost trivial; nevertheless we report the corresponding
expansions:
%--
\bq
\beta^{-1}(z_m)\,L_{\beta}(z_m) = - \frac{1}{2}\,z^{-1}_m -
\frac{1}{12}\,z^{-2}_m - \frac{1}{60}\,z^{-3}_m,
\eq
%--
with $z_m = m^2/s$ and $L_m = \ln z_m$.
%--
\bq
b_0^{\ep}(2,1,s,m^2,m^2)= 
z^{-1}_m\,\Bigl[ L_m + \frac{1}{6}\,z^{-1}_m\,( L_m - 1) +
\frac{1}{30}\,z^{-2}_m\,( L_m - \frac{3}{2} ) \Bigr],
\eq
%--
where $L_m = \ln z_m$. Similarly we obtain
%--
\bqa
b_0^{\ep}(1,1,s,m^2,m^2) &=& 
- L^2_m + \frac{1}{3}\,z^{-1}_m\,L_m + \frac{1}{30}\,z^{-2}_m\,
( L_m - 1 ) + \frac{1}{210}\,z^{-3}_m\,( L_m - \frac{21}{14} ),
\nl
b_0^{\ep}(1,1,s,0,m^2) &=& 
- 2 + 2\,L_m - L^2_m - z^{-1}_m\,( \frac{3}{2} - L_m ) -
z^{-2}_m\,( \frac{17}{18} - \frac{1}{3}\,L_m ),
\nl
b_0^{\ep}(2,1,s,m^2,0) &=& 2\,z^{-1}_m\,( L_m - 1 ) +
                        z^{-2}_m\,( L_m - \frac{5}{2} ).
\eqa
%--
The expansions for $b_0^{\ep}$ with unequal masses are rather long and will not
be reported here.
%--
\subsection{Two-loop functions}
%--
For two-loop functions we rewrite scaled functions, e.g. 
$s^{\ssA\,;\,{\rm fin}}$ in terms of the original ones, e.g. $S^{\ssA}$; when 
a Taylor expansion is needed we derive it in terms of vacuum bubbles. Consider
an example; given
%--
\bqa
s^{\ssA\,;\,{\rm fin}}(s,m_1,m_2,m_3) &=& s^{-1}\,\Bigl[
       L_{\ssW}\,\lpar \frac{s}{2} - 3\,\sum_i\,m^2_i\rpar
       + 2\,L_{\ssW}\,\sum_i\,m^2_i\,L_i
       - L^2_{\ssW}\,\sum_i\,m^2_i
\nl
{}&+& {\cal P}_{\ssU\ssV}\,S^{\ssA}(s,m_1,m_2,m_3)\Bigr],
\eqa
%--
where the operator $P_{\ssU\ssV}$ acts as
%--
\bq
{\cal P}_{\ssU\ssV}\,\ep^{-n} = 0,
\qquad
{\cal P}_{\ssU\ssV}\,\DUV = 0.
\eq
%--
we use the first few terms in the expansion, e.g.
%--
\bqa
S^{\ssA}(s,m_1,m_2,m_3) &=& T_{111}(m_1,m_3,m_2) +
\Bigl[ T_{121}(m_1,m_3,m_2) - \frac{4}{n}\,T_{121}(m_1,m_3,m_2) 
\nl
{}&+& \frac{4}{n}\,m^2_3\,T_{131}(m_1,m_3,m_2) \Bigr]\,s + \ord{s^2}.
\eqa
%--
IBP techniques are then use to reduce vacuum bubbles to our master integral
$T_{121}(m_1,m_3,m_2)$. Therefore, one function suffices in deriving the final
result. This sort of results is relevant whenewer we need corrections to the 
limit $s \to 0$ or when we want to prove that vacuum bubbles give the correct
limit of the theory. For tensor integrals some additional work is needed; 
first we give an example,
%--
\bqa
\int d^n q_1\,d^n q_2\,\frac{q^{\mu}_i\,q^{\nu}_j}
{\lpar q^2_1 + m^2_1\rpar^{n_1}\,
\lpar q^2_2 + m^2_3\rpar^{n_2}\,
\lpar (q_1 - q_2)^2 + m^2_2\rpar^{n_3}} &=&
T^{ij}_{n_1,n_2,n_3}\lpar m_1,m_3,m_2\rpar\,\delta^{\mu\nu},
\eqa
%--
and derive the reduction of vacuum bubble form factors in terms of scalar
vacuum bubbles,
%--
\bqa
n\,T^{11}_{n_1,n_2,n_3} &=& T_{n_1-1,,n_2,n_3} - m^2_1\,T_{n_1,n_2,n_3},
\nl
n\,T^{12}_{n_1,n_2,n_3} &=& 
       \frac{1}{2}\,T_{n_1-1,,n_2,n_3} + \frac{1}{2}\,T_{n_1,n_2-1,n_3} 
- \frac{1}{2}\,T_{n_1,n_2,n_3-1} 
     + \frac{1}{2}\,(m^2_3-m^2_1-m^2_2)\,T_{n_1,n_2,n_3},
\nl
n\,T^{22}_{n_1,n_2,n_3} &=& 
        T_{n_1,n_2-1,n_3} - m^2_2\,T_{n_1,n_2,n_3},
\eqa
%--
where all $T$ have argument $m_1,m_2,m_3$. Suppose now that we have to compute
%--
\bqa
S^{\ssD}(\mu\,|\,0) &=& \int d^n q_1\,d^n q_2\,q^{\mu}_1
\nl
{}&\times& \Bigl[
\lpar q^2_1+m_1^2\rpar\,
\lpar (q_1+p)^2+m_1^2\rpar\,
\lpar (q_1-q_2)^2+m_2^2\rpar\,
\lpar q^2_2+m_3^2\rpar\,
\lpar (q_2+p)^2+m_3^2\rpar\Bigr]^{-1},
\eqa
%--
in the limit $p^2 \to 0$. the recipe is simple: expand the propagators, e.g.
%--
\bq
\lpar (q_1+p)^2+m_1^2\rpar^{-k} = 
\lpar q^2_1+m_1^2\rpar^{-k} -
\lpar p^2 + 2\,\spro{p}{q_1}\rpar\,\lpar q^2_1+m_1^2\rpar^{-k-1} +
\dots,
\eq
%--
to obtain
%--
\bqa
n\,S^{\ssD}_{11}(0,\{m\}) &=&
       - T_{131}(m_1,m_3,m_2)
       - 3\,T_{221}(m_1,m_3,m_2)
\nl
{}&+& ( m^2_1-m^2_2+m^2_3)\,T_{231}(m_1,m_3,m_2)
       + 2\,m^2_1\,T_{321}(m_1,m_3,m_2)
       - A_0(2,m_1)\,A_0(3,m_3),
\nl\nl
n\,S^{\ssD}_{12}(0,\{m\}) &=&
       - 3\,T_{221}(m_1,m_3,m_2)
       + 2\,m^2_3\,T_{231}(m_1,m_3,m_2)
\nl
{}&-& T_{311}(m_1,m_3,m_2)
       + (m^2_1-m^2_2+m^2_3)\,T_{321}(m_1,m_3,m_2)
       - A_0(2,m_3)\,A_0(3,m_1).
\eqa
%--
For tensor integrals of higher rank we need to define the $T^{ijkl}$ form
factors as
%--
\bqa
\int d^n q_1\,d^n q_2\,\frac{q^{\mu}_i\,q^{\nu}_j\,q^{\alpha}_k\,q^{\beta}_l}
{\lpar q^2_1 + m^2_1\rpar^{n_1}\,
\lpar q^2_2 + m^2_3\rpar^{n_2}\,
\lpar (q_1 - q_2)^2 + m^2_2\rpar^{n_3}} &=&
T^{ijkl}_{\ssA\,;\,n_1,n_2,n_3}(m_1,m_3,m_2)\,
\delta_{\mu\nu}\,\delta_{\alpha\beta} 
\nl
{}&+& T^{ijkl}_{\ssB\,;\,n_1,n_2,n_3}(m_1,m_3,m_2)\,
\delta_{\mu\alpha}\,\delta_{\nu\beta} 
\nl
{}&+& T^{ijkl}_{\ssC\,;\,n_1,n_2,n_3}(m_1,m_3,m_2)\,
\delta_{\mu\beta}\,\delta_{\nu\alpha},
\eqa
%--
with the following results:
%--
\bqa
{}&{}& n^2\,T^{1122}_{\ssA\,;\,n_1,n_2,n_3} = 
       -n\,T^{1122}_{\ssB\,;\,n_1,n_2,n_3} 
       -n\,T^{1122}_{\ssC\,;\,n_1,n_2,n_3} 
\nl
{}&{}&\qquad + T_{n_1-1,n_2-1,n_3} 
       -m^2_2\,T_{n_1-1,n_2,n_3} 
       -m^2_1\,T_{n_1,n_2-1,n_3} 
       +m^2_1\,m^2_2\,T_{n_1,n_2,n_3},
\eqa
%--
\bqa
{}&{}& (n-1)\,T^{1122}_{\ssC\,;\,n_1,n_2,n_3} = 
       (1-n^2)\,T^{1122}_{\ssB\,;\,n_1,n_2,n_3} 
       +\frac{1}{4}\,T_{n_1-2,n_2,n_3} 
\nl
{}&{}&\qquad + (\frac{1}{2}-\frac{1}{n})\,T_{n_1-1,n_2-1,n_3} 
       -\frac{1}{2}\,T_{n_1-1,n_2,n_3-1} 
       +\Bigl[(\frac{1}{n}-\frac{1}{2})\,m^2_2 + \frac{1}{2}\,(m^2_3-m^2_1)
        \Bigr]\,T_{n_1-1,n_2,n_3} 
\nl
{}&{}&\qquad + \frac{1}{4}\,T_{n_1,n_2, n_3-2} 
       +\frac{1}{2}\,(m^2_1+m^2_2-m^2_3)\,T_{n_1,n_2,n_3-1} 
       +\Bigl[ \frac{n-1}{n}\,m^2_1 m^2_2 + \frac{1}{4}\,
               \lambda(m^2_1,m^2_2,m^2_3)\Bigr]\,T_{n_1,n_2,n_3}   
\nl
{}&{}&\qquad + \frac{1}{4}\,T_{n_1,n_2-2,n_3} 
       -\frac{1}{2}\,T_{n_1,n_2-1,n_3-1} 
       +\Bigl[(\frac{1}{n}-\frac{1}{2})\,m^2_1 + \frac{1}{2}\,(m^2_3-m^2_2)
        \Bigr]\,T_{n_1,n_2-1,n_3} 
\eqa
%--
\bqa
{}&{}& n\,(n-1)\,(n+2)\,T^{1122}_{\ssB\,;\,n_1,n_2,n_3} =
       \frac{1}{4}\,n\,T_{n_1-2,n_2,n_3} 
       -(1-\frac{1}{2}\,n)\,T_{n_1-1,n_2-1,n_3} 
\nl
{}&{}&\qquad - \frac{1}{2}\,n\,T_{n_1-1,n_2,n_3-1} 
       -\Bigl[(\frac{1}{2}\,n-1)\,m^2_2+\frac{1}{2}\,(m^2_1-m^2_3)\Bigr]\,
         T_{n_1-1,n_2,n_3} 
       +\frac{1}{4}\,n\,T_{n_1,n_2-2,n_3} 
       -\frac{1}{2}\,n\,T_{n_1,n_2-1,n_3-1} 
\nl
{}&{}&\qquad - \Bigl[(\frac{1}{2}\,n-1)\,m^2_1+
    \frac{1}{2}\,(m^2_2-m^2_3)\Bigr]\,T_{n_1,n_2-1,n_3} 
       +\frac{1}{4}\,n\,T_{n_1,n_2, n_3-2} 
       -\frac{1}{2}\,n\,(m^2_3-m^2_1-m^2_2)\,T_{n_1,n_2,n_3-1} 
\nl
{}&{}&\qquad - \Bigl[(1-n)\,m^2_1 m^2_2-\frac{1}{4}\,n\,
           \lambda(m^2_1,m^2_2,m^2_3)\Bigr]\,T_{n_1,n_2,n_3}, 
\eqa
%--
\bqa
{}&{}& n^2\,T^{1222}_{\ssA\,;\,n_1,n_2,n_3} =
       -n\,T^{1222}_{\ssB\,;\,n_1,n_2,n_3} 
       -n\,T^{1222}_{\ssC\,;\,n_1,n_2,n_3} 
\nl
{}&{}&\qquad + \frac{1}{2}\,T_{n_1-1,n_2-1,n_3} 
       -\frac{1}{2}\,m^2_2\,T_{n_1-1,n_2,n_3} 
       +\frac{1}{2}\,T_{n_1,n_2-2,n_3} 
       -\frac{1}{2}\,T_{n_1,n_2-1,n_3-1} 
\nl
{}&{}&\qquad + \frac{1}{2}\,(m^2_3-2\,m^2_2-m^2_1)\,T_{n_1,n_2-1,n_3} 
       +\frac{1}{2}\,m^2_2\,T_{n_1,n_2,n_3-1} 
       +\frac{1}{2}\,m^2_2\,(m^2_1+m^2_2-m^2_3)\,T_{n_1,n_2,n_3} 
\eqa
%--
\bqa
&{}& T^{1222}_{\ssC\,;\,n_1,n_2,n_3} = 
       -(n+1)\,T^{1222}_{\ssB\,;\,n_1,n_2,n_3} 
       +\frac{1}{2\,n}\,T_{n_1-1,n_2-1,n_3} 
\nl
{}&{}&\qquad - \frac{1}{2\,n}\,m^2_2\, T_{n_1-1,n_2,n_3} 
       +\frac{1}{2\,n}\,T_{n_1,n_2-2,n_3} 
       -\frac{1}{2\,n}\,T_{n_1,n_2-1,n_3-1} 
\nl
{}&{}&\qquad + \frac{1}{n}\,\Bigl[\frac{1}{2}\,(m^2_3-m^2_1)-m^2_2\Bigr]\, 
         T_{n_1,n_2-1,n_3} 
\nl
{}&{}&\qquad + \frac{1}{2\,n}\,m^2_2\,T_{n_1,n_2,n_3-1} 
       -\frac{1}{2\,n}\,m^2_2\,(m^2_3-m^2_1-m^2_2)\,
       T_{n_1,n_2,n_3},  
\eqa
%--
\bqa
{}&{}& n\,(n+2)\,T^{1222}_{\ssB\,;\,n_1,n_2,n_3} =
        \frac{1}{2}\,T_{n_1-1,n_2-1,n_3} 
       -\frac{1}{2}\,m^2_2\,T_{n_1-1,n_2,n_3} 
\nl
{}&{}&\qquad + \frac{1}{2}\,T_{n_1,n_2-2,n_3} 
       -\frac{1}{2}\,T_{n_1,n_2-1,n_3-1} 
       -\Bigl[ \frac{1}{2}\,m^2_1+m^2_2-\frac{1}{2}\,m^2_3\Bigr]\,
         T_{n_1,n_2-1,n_3} 
\nl
{}&{}&\qquad + \frac{1}{2}\,m^2_2\,T_{n_1,n_2,n_3-1} 
       -\frac{1}{2}\,m^2_2\,(m^2_3-m^2_1-m^2_2)\,
         T_{n_1,n_2,n_3},
\eqa
%--
\bqa
{}&{}& n^2\,T^{2222}_{\ssA\,;\,n_1,n_2,n_3} = 
       -n\,T^{2222}_{\ssB\,;\,n_1,n_2,n_3} 
       -n\,T^{2222}_{\ssC\,;\,n_1,n_2,n_3} 
\nl
{}&{}&\qquad + T_{n_1,n_2-2,n_3} 
       -2\,m^2_2\,T_{n_1,n_2-1,n_3} 
       +m^4_2\,T_{n_1,n_2,n_3},
\eqa
%--
\bqa
{}&{}& T^{2222}_{\ssC\,;\,n_1,n_2,n_3} = 
       -(n+1)\,T^{2222}_{\ssB\,;\,n_1,n_2,n_3} 
       +\frac{1}{n}\,T_{n_1,n_2-2,n_3} 
\nl
{}&{}&\qquad - 2\,\frac{1}{n}\,m^2_2\,T_{n_1,n_2-1,n_3} 
       +\frac{1}{n}\,m^4_2\,T_{n_1,n_2,n_3},
\eqa
%--
\bqa
{}&{}& n\,(n+2)\,T^{2222}_{\ssB\,;\,n_1,n_2,n_3} =
       T_{n_1,n_2-2,n_3} 
       -2\,m^2_2\,T_{n_1,n_2-1,n_3} 
       +m^4_2\,T_{n_1,n_2,n_3}.
\eqa
%--
For instance, we obtain
%--
\bqa
S^{\ssE}_{221}(0,\{m\}) &=& 4\,(\frac{1}{n}-\frac{1}{n+2})\,\Bigl[
        T_{131} - 2\,m^2_3\,T_{141} + 
        m^4_3\,T_{151}\Bigr].
\eqa
%--
\bei
\item[--] {\underline{Laurent expansion}}
\eei
%--
There are cases, however, where a Taylor expansion is not allowed. Consider
$S^{\ssC}(s,m_1,m_2,\{0\}_2)$ which contains propagator factors $q^2_2$ and
$(q_2+p)^2$. The limit $p^2 \to 0$ is IR divergent and the corresponding
singularity should not be regularized by $\ep$, a fact that would change the
pole structure of the diagram, but a Laurent expansion around $p^2 = 0$ is
needed. In the following list, where we define
%--
\bq
L_{\ssW} = \ln\frac{\mws}{s}, \quad
\cL_m= \ln\frac{m^2}{\mws}, \quad
\cL_{m_i} = \ln\frac{m^2_i}{\mws},
\eq
%--
and where $L_{\rm an}$ is given in \eqn{deflambda}, we present the relevant 
limits:
%--
\bqa
S^{\ssC\,;\,{\rm fin}}(s,0,m,\{0\}_2) &=&
        L_{\ssW}\,L_{\rm an} 
 - \lpar L_{\ssW} + L_{\rm an} \rpar \, \lpar \frac{23}{6} - \cL_m \rpar
       - \frac{1}{2} \, L_{\ssW}^2 
       - \frac{107}{12}
       + \frac{9}{2}\,\zeta(2)
\nl
{}&+& 2\,\cL_m + \frac{1}{2}\,\cL^2_m + \ord{s},
\nl
S^{\ssC\,;\,{\rm fin}}(s,m_1,m_2,\{0\}_2) &=&
        L_{\ssW}\,L_{\rm an} 
       - L_{\ssW} \, \Bigl[
           \frac{23}{6}
          - \cL_{m_1}
          - \lpar \cL_{m_1} + \cL_{m_2} \rpar \frac{x_2}{x_1-x_2} \Bigr]
       - \frac{1}{2} \, L_{\ssW}^2 
\nl
{}&+& L_{\rm an} \, \Bigl[
          \frac{23}{6}
          - \cL_{m_1}
          - \lpar \cL_{m_1} - \cL_{m_2} \rpar \, \frac{x_2}{x_1-x_2} \Bigr]
          - \frac{113}{12}
          + 3\,\zeta(2)
          + 3\,\cL_{m_1}
\nl
{}&-& \frac{1}{2}\,\cL^2_{m_1}
          - T^{\rm fin}_{121}(0,m_1,m_2)
+ \frac{x_2}{x_1-x_2}\,\Bigl[
- \cL_{m_1}\,\cL_{m_2}
          + 4\,\cL_{m_1}
          - 2\,\cL_{m_2}
          - \cL^2_{m_1}
\nl
{}&-& \zeta(2)
          - 1
          - 2\,T^{\rm fin}_{121}(0,m_1,m_2)\Bigr] + \ord{s}.
\eqa
%--
Note that these functions are of $\ord{\ln^2 s}$ for $s \to 0$. Furthermore,
%--
\bqa
s\,S^{\ssE\,;\,{\rm fin}}(s,0,m,\{0\}_3) &=& 
        L_{\ssW}\,L_{\rm an} 
       - \lpar L_{\ssW} - L_{\rm an} \rpar \, \lpar
          \frac{7}{3} - L_{\ssW 2}
          \rpar
       - \frac{1}{2}\,\Bigl( L_{\ssW}^2 + L_{\rm an}^2 + \cL^2_m\Bigr)
       + \frac{11}{6}\,\cL_m
       - \frac{49}{12},
\nl
s\,S^{\ssE\,;\,{\rm fin}}(s,m,m,\{0\}_3) &=& 
        L_{\ssW}\,L_{\rm an} 
       - \lpar L_{\ssW} - L_{\rm an} \rpar \, \lpar
          \frac{4}{3} - \cL_m
          \rpar
       - \frac{1}{2}\,\Bigl( L_{\ssW}^2 + L_{\rm an}^2 + \cL^2_m\Bigr)
       + \frac{5}{6}\,\cL_m
       - \frac{9}{4},
\nl
s\,S^{\ssE\,;\,{\rm fin}}(s,m_1,m_2,\{0\}_3) &=& 
        L_{\ssW}\,L_{\rm an} 
       - \lpar L_{\ssW} - L_{\rm an} \rpar \, \lpar
          \frac{7}{3} - \cL_{m_1} \rpar
       + \frac{x_2}{x_1-x_2}\,\lpar \cL_{m_1} - \cL_{m_2} \rpar
\nl
{}&-& \frac{1}{2}\,\Bigl( L_{\ssW}^2 + L_{\rm an}^2 + \cL^2_{m_1}\Bigr)
       - \frac{49}{12}
       + \frac{11}{6}\,\cL_{m_1}
\nl
{}&+& \frac{x_2}{x_1-x_2}\,\Bigl[
        \frac{11}{6}\,\Bigl( \cL_{m_1} -  \cL_{m_2}\Bigr)
       - \frac{1}{2}\,\Bigl( \cL^2_{m_1} - \cL^2_{m_2}\Bigr) \Bigr].
\eqa
%--
\subsection{Limit $m_f \to 0$ of $p^2 \to 0$}
%--
After performing the limit $p^2 \to 0$ we still would like to set light
fermion masses to zero, whenever possible. Such is the situation for our
main traget, the inclusion of $\alpha$ and of $\gf$ in the IPS. Functions for 
which a Laurent expansion is needed include all cases where we have 
a vacuum bubble $T_{n_1 n_2 n_3}(0,m,M)$, or permutations of it, and 
$n_1 \ge 2$. For expanding these functions we have developed a special 
technique, based on the idea of writing differential equations for 
diagrams~\cite{Kotikov:1990kg}.
This technique works in all cases where we know how to compute integration 
constants. 
%--
\bei
\item[--] {\underline{$T_{121}(m,M,m)$ with $m \to 0$}}
\eei
%--
Consider the following example: $T_{121}(m,M,m)$ with $m \to 0$. We write the 
equation
%--
\bq
\frac{d}{d m^2}\,T_{121}(m,M,m) = - T_{221}(m,M,m) - T_{122}(m,M,m).
\label{Lexp}
\eq
%--
We use IBP identities to express the r.h.s of \eqn{Lexp} in terms of 
$T_{121}(m,M,m)$. Using \eqn{defTf} we make the ans{\"a}tze,
%--
\bqa
T_f(m,M,m) &=& 
\sum_{n=0} a_n\,\lpar \frac{m^2}{M^2}\rpar^n
+ \cL_m\,\sum_{n=0} b_n\,\lpar \frac{m^2}{M^2}\rpar^n +
\cL^2_m\,\sum_{n=0} c_n\,\lpar \frac{m^2}{M^2}\rpar^n,
\eqa
%--
insert into \eqn{Lexp} and derive
%-
\bq
a_0 = T_f(0,M,0) =
- \frac{1}{2} - \frac{3}{2}\,\zeta(2) + \cL_{\ssM} - \cL^2_{\ssM},
\eq
%--
where $\cL_m= \ln m^2/\mws$ and $\cL_{\ssM}= \ln M^2/\mws$. Here, the 
integration constant is known, see \eqn{icisknown}. Furthermore, we have
%--
\bqa
a_1 &=& 2 + 2\,\cL_{\ssM}, \qquad
a_2= - \frac{1}{2}\,\Bigl[
   - 2\,\zeta(2)
   - 5
   - 2\,\cL_{\ssM}
   - 2\,\cL^2_{\ssM}
   - 4\,a_0\Bigr], 
\nl
a_3 &=& - \frac{1}{3}\,\Bigl[
   - 12\,\zeta(2)
   - \frac{92}{3}
   - 8\,\cL_{\ssM}
   - 12\,\cL^2_{\ssM}
   - 24\,a_0\Bigr]
\nl
b_0 &=& 0, \qquad b_1 = - 2, \qquad b_2 = - 3 + 2\,\cL_{\ssM}, \qquad
b_3 = - \frac{32}{3} + 8\,\cL_{\ssM},
\nl
c_0 &=& 0, \qquad c_1 = 0, \qquad c_2 = - 1, \qquad c_3 = - 4,
\eqa
%--
etc. 
%--
\bei
\item[--] {\underline{$T_f(m,M,0)$ with $m \to 0$}}
\eei
%--
Similarly to the previous case, we obtain
%--
\bqa
T_f(m,M,0) &=& 
\sum_{n=0} a_n\,\lpar \frac{m^2}{M^2}\rpar^n
+ \ln \frac{m^2}{\mws}\,\sum_{n=0} b_n\,\lpar \frac{m^2}{M^2}\rpar^n,
\eqa
%--
with coefficients $a_0 = T_f(0,M,0)$ again given in \eqn{icisknown} and
%--
\bqa
a_1 &=& 1 + \cL_{\ssM}, \qquad
a_2 = \frac{1}{2}\,\lpar \frac{1}{2} + \cL_{\ssM} \rpar, \qquad
a_3 = \frac{1}{3}\,\lpar \frac{1}{3} + \cL_{\ssM} \rpar, 
\nl
b0 &=& 0, \qquad b_1 = - 1, \qquad b_2 = - \frac{1}{2}, \qquad 
b_3 = - \frac{1}{3}.
\eqa
%--
\bei
\item[--] {\underline{small, unequal, masses}}
\eei
%--
For small, unequal, masses we write two differential equations,
%--
\bq
\frac{d}{d m^2_1}\,T_{121}(m_1,M,m_2) = - T_{221}(m_1,M,m_2),
\quad
\frac{d}{d m^2_2}\,T_{121}(m_1,M,m_2) = - T_{122}(m_1,M,m_2),
\eq
%--
and introduce the ansatz
%--
\bqa
T_f(m_1,M,m_2) &=& 
a_0 + a_1\,\frac{m^2_1+m^2_2}{M^2} + a_2\,\frac{m^4_1+m^4_2}{M^4} +
a_3\,\frac{m^2_1\,m^2_2}{M^4} +
a_4\,\frac{m^6_1+m^6_2}{M^6} +
a_5\,m^2_1\,m^2_2\,\frac{m^2_1+m^2_2}{M^6} 
\nl
{}&+& \ln \frac{m^2_1}{\mws} \,\Bigl[
b_1\,\frac{m^2_1}{M^2} + b_2\,\frac{m^2_1 m^2_2}{M^4} +
b_3\,\frac{m^4_1}{M^4} +
b_4\,\frac{m^2_1 m^4_2}{M^6} +
b_5\,\frac{m^4_1 m^2_2}{M^6} +
b_6\,\frac{m^6_1}{M^6}\Bigr] 
\nl
{}&+& \ln \frac{m^2_2}{\mws} \,\Bigl[
b_1\,\frac{m^2_2}{M^2} + b_2\,\frac{m^2_2 m^2_1}{M^4} +
b_3\,\frac{m^4_2}{M^4} +
b_4\,\frac{m^2_2 m^4_1}{M^6} +
b_5\,\frac{m^4_2 m^2_1}{M^6} +
b_6\,\frac{m^6_2}{M^6}\Bigr]
\nl
{}&+& \ln \frac{m^2_1}{\mws}\,\ln \frac{m^2_2}{\mws}\,\Bigl[
c_1\,\frac{m^2_1 m^2_2}{M^4} +
c_2\,m^2_1\,m^2_2\,\frac{m^2_1+m^2_2}{M^6}\Bigr]. 
\eqa
%-- 
The solution is $a_0 = T_f(0,M,0)$ (\eqn{icisknown}) and
%--
\bqa
a_1 &=& 1 + \cL_{\ssM}, \qquad
a_2 = \frac{1}{2}\,\lpar \frac{1}{2} + \cL_{\ssM} \rpar, \qquad
a_3 = 2 + \zeta(2) + \cL^2_{\ssM} + 2\,a_0, 
\nl
a_4 &=& \frac{1}{3}\,\lpar \frac{1}{3} + \cL_{\ssM} \rpar, \qquad
a_5 = 5 + 2\,\zeta(2) + \cL_{\ssM} + 2\,\cL^2_{\ssM} + 4\,a_0,
\nl
b_1 &=& - 1, \qquad
b_2 = - 1 + \cL_{\ssM}, \qquad
b_3 = - \frac{1}{2}, \qquad
b_4 = - 1 + 2\,\cL_{\ssM}, \qquad
b_5 = - 4 + 2\,\cL_{\ssM}, \qquad
b_6 = - \frac{1}{3},
\nl
c_1 &=& - 1, \qquad 
c_2 = - 2.
\eqa
%--
\bei
\item[--] {\underline{one small mass}}
\eei
%--
For one small mass, $m \ll m_1, m_2$, we obtain
%--
\bqa
T_f(m,m_1,m_2) &=& 
\sum_{n=0} a_n\,\lpar \frac{m^2}{\mws}\rpar^n
+ \cL_m\,\sum_{n=1} b_n\,\lpar \frac{m^2}{\mws}\rpar^n,
\eqa
%--
with a solution $a_0 = T_f(0,m_1,m_2)$ and
%--
\bqa
a_1 &=& \cL_1\,\cL_2 \, \frac{x_2}{X^2}
       + \cL_1 \, \lpar
          - \frac{x_2}{X^2}
          + \frac{1}{X}
          \rpar
       + \cL_1^2 \, \frac{x_2}{X^2}
       - \cL_2 \, \frac{x_2}{X^2}
+ \zeta(2)\,\frac{x_2}{X^2}
       + \frac{x_2}{X^2}
       + \frac{1}{X}
       + 2\,a_0 \, \frac{x_2}{X^2},
\nl
b_1 &=& \cL_1 \, \frac{x_2}{X^2}
       - \cL_2 \, \frac{x_2}{X^2}
       - \frac{1}{X},
\nl
a_2 &=& \cL_1\,\cL_2 \, \lpar
           3\,\frac{x^2_2}{X^4}
          + 2\,\frac{x_2}{X^3}
          \rpar
       + \cL_1 \, \lpar
          - \frac{5}{2}\,\frac{x^2_2}{X^4}
          + \frac{1}{2}\,\frac{1}{X^2}
          \rpar
       + \cL_1^2 \, \lpar
           3\,\frac{x^2_2}{X^4}
          + 2\,\frac{x_2}{X^3}
          \rpar
\nl
{}&+& \cL_2 \, \lpar
          - \frac{7}{2}\,\frac{x^2_2}{X^4}
          - \frac{x_2}{X^3}
          \rpar
       + 3\,\zeta(2)\,\frac{x^2_2}{X^4}
       + 2\,\zeta(2)\,\frac{x_2}{X^3}
       + 3\,\frac{x^2_2}{X^4}
       + \frac{9}{2}\,\frac{x_2}{X^3}
       + \frac{1}{4}\,\frac{1}{X^2}
       + a_0 \, \lpar
           6\,\frac{x^2_2}{X^4}
          + 4\,\frac{x_2}{X^3}
          \rpar,
\nl
b_2 &=& \cL_1 \, \lpar
           3\,\frac{x^2_2}{X^4}
          + 2\,\frac{x_2}{X^3}
          \rpar
       + \cL_2 \, \lpar
          - 3\,\frac{x^2_2}{X^4}
          - 2\,\frac{x_2}{X^3}
          \rpar
       - 3\,\frac{x_2}{X^3}
       - \frac{1}{2}\,\frac{1}{X^2},
\nl
a_3 &=& \cL_1\,\cL_2 \, \lpar
           10\,\frac{x^3_2}{X^6}
          + 12\,\frac{x^2_2}{X^5}
          + 3\,\frac{x_2}{X^4}
          \rpar
       + \cL_1 \, \lpar
          - \frac{23}{3}\,\frac{x^3_2}{X^6}
          - 4\,\frac{x^2_2}{X^5}
          + 2\,\frac{x_2}{X^4}
          + \frac{1}{3}\,\frac{1}{X^3}
          \rpar
\nl
{}&+& \cL_1^2 \, \lpar
           10\,\frac{x^3_2}{X^6}
          + 12\,\frac{x^2_2}{X^5}
          + 3\,\frac{x_2}{X^4}
          \rpar
+ \cL_2 \, \lpar
          - \frac{37}{3}\,\frac{x^3_2}{X^6}
          - 10\,\frac{x^2_2}{X^5}
          - \frac{x_2}{X^4}
          \rpar
\nl
{}&+& 10\,\zeta(2)\,\frac{x^3_2}{X^6}
       + 12\,\zeta(2)\,\frac{x^2_2}{X^5}
       + 3\,\zeta(2)\,\frac{x_2}{X^4}
       + 10\,\frac{x^3_2}{X^6}
       + \frac{59}{3}\,\frac{x^2_2}{X^5}
       + \frac{49}{6}\,\frac{x_2}{X^4}
       + \frac{1}{9}\,\frac{1}{X^3}
\nl
{}&+& a_0 \, \lpar
           20\,\frac{x^3_2}{X^6}
          + 24\,\frac{x^2_2}{X^5}
          + 6\,\frac{x_2}{X^4}
          \rpar,
\nl
b_3 &=& \cL_1 \, \lpar
           10\,\frac{x^3_2}{X^6}
          + 12\,\frac{x^2_2}{X^5}
          + 3\,\frac{x_2}{X^4}
          \rpar
       + \cL_2 \, \lpar
          - 10\,\frac{x^3_2}{X^6}
          - 12\,\frac{x^2_2}{X^5}
          - 3\,\frac{x_2}{X^4}
          \rpar
       - 10\,\frac{x^2_2}{X^5}
       - 7\,\frac{x_2}{X^4}
       - \frac{1}{3}\,\frac{1}{X^3},
\eqa
%--
where we have introduced
%--
\bq
x_i = \frac{m^2_i}{\mws}, \quad X = x_1 - x_2, \quad
\cL_i = \ln x_i, \quad
\cL_m= \ln\frac{m^2}{\mws}.
\eq
%--
\section{Finite mass renormalization in $\Pi^{\ssR}$}
%--
As we have discussed in the previous sections there is full control upon 
finite parts at arbitrary scale. In the following $M_0 = M/\ctw$, $M$ being 
the renormalized $W$ boson mass; furthermore, $\mh$ (etc) is the $H$ boson 
(etc) renormalized mass.
Consider, for example, the {\em subtracted} two-loop photon self-energy, 
as defined in sect.~6 of I,
%--
\bq
\Pi^{\ssR}(s) = \Pi^{(2)}_{\ssQ\ssQ\ext}(s) - \Pi^{(2)}_{\ssQ\ssQ\ext}(0) =
\frac{1}{\ctws}\,F^{\ssA\ssA}_1(s) + F^{\ssA\ssA}_2(s) + 
\stws\,F^{\ssA\ssA}_3(s).
\eq
%--
Before using $\Pi^{\ssR}(s)$ for physical predictions all renormalized masses 
must be replaced by on-shell masses, using 
$M_{\ssR} = M_{\ssO\ssS} + \ord{g^2}$ in one-loop terms and
$M_{\ssR} = M_{\ssO\ssS}$ in two-loop terms (see \sect{FiniteR}, 
\eqnst{OSf}{OSW}{OSH}). For instance, the effect of finite mass renormalization
on $\Pi(s)= \Pi^{\ssR}(s) - \Pi^{\ssR}(0)$ is as follows; define 
UV finite factors
%--
\bq
M^2 = \mws\,\lpar 1 + \frac{g^2}{\pi^2}\,\delta M^2\rpar,
\eq
%--
etc. In the massless limit for fermions we get
%--
\bq
\Pi^{(1)}(s) \to \Pi^{(1)}(s) + \frac{g^4}{\pi^4}\,\stws\,\Delta\,\Pi(s),
\eq
%--
\bqa
\Delta\,\Pi(s) &=& \frac{1}{12}\,\sum_l\,\delta m^2_l
+ \frac{1}{8}\,x_{\ssW}\lpar 1 + 12\,x_{\ssW}\rpar\,\delta M^2\,
b^{\rm fin}_0(2,1,s,\mw,\mw)
\nl
{}&-& \frac{4}{3}\,x^2_{\ssT}\,\Delta m^2_t\,
b^{\rm fin}_0(2,1,s,\mt,\mt)
- \frac{3}{4}\,\lpar x_{\ssW} + \frac{1}{4}\rpar\,\delta M^2
+ \frac{2}{3}\,\lpar x_{\ssT} + \frac{1}{6}\rpar\,\delta m^2_t,
\eqa
%--
with $x_i= m^2_i/s$.
%--
The complete expression for $\Pi(s)$ is too long to be reported here and
can be found at http://www.to.infn.it/\~{}giampier/REN/pis.ps.
%--
%%%%%%%%%%%%%%%%%%%%%%%%%%%%%%%%%%%%%%%%%%%%%%%%%%%%%%%%%%%%%%%%%%%%%%%%%%%%%

\end{document}